\documentclass[journal=jacsat,manuscript=article]{achemso}

\usepackage[version=3]{mhchem} 

\usepackage{graphicx}
\usepackage{threeparttable}
\usepackage{array}
\usepackage{amsfonts}
\usepackage{amssymb}
\usepackage{mathtext}
\usepackage{amsmath}
\usepackage{multirow}
\usepackage{rotating}
\usepackage{ulem}

\SectionNumbersOn

\author{Sergei F. Chekmarev}
\email{chekmarev@itp.nsc.ru}%
\affiliation[Institute of Thermophysics, SB RAS] {Institute of Thermophysics, SB RAS, 630090 Novosibirsk, Russia}%
\alsoaffiliation[Novosibirsk State University] {Department of Physics, Novosibirsk State University, 630090
Novosibirsk, Russia}
\title[An \textsf{achemso} demo]{First-Passage Time Distributions in Two-State Protein Folding Kinetics: Exploring the Native-Like States vs Overcoming the Free Energy Barrier}


\begin{document}
\begin{abstract}
Using a $\beta$-hairpin protein as a representative example of two-state folders, we studied how the exploration of native-like states affects the folding kinetics. It has been found that the first-passage time (FPT) distributions are essentially single-exponential not only for the times to overcome the free energy barrier that separates unfolded and native-like states but also for the times to find the native state among the native-like ones. If the protein explores native-like states for a time much longer than the time to overcome the free energy barrier, which was found to be characteristic of high temperatures, the resulting FPT distribution to reach the native state remains close to exponential but the mean FPT (MFPT) is determined not by the height of the free energy barrier but by the time to explore native-like states. The mean time to overcome the free energy barrier is found to be in reasonable agreement with the Kramers rate formula and generally far shorter than the MFPT to reach the native state. The time to find the native state among native-like ones increases with temperature, which explains the known U-shape dependence of the MFPTs on temperature.  
\end{abstract}


\section{Introduction}
Most of small, single-domain globular proteins (approximately to one hundred residues) fold in a two-state cooperative manner  \cite{jackson1991folding,schindler1995extremely,jackson1998how,plaxco1998contact,plaxco2000topology,grantcharova2001mechanisms,ivankov2003contact,akmal2004nature,oliveberg2005experimental,gelman2014fast}. The process of folding represents a transition from an unfolded state of the protein to its functional (native) state over a free energy barrier \cite{sali1994how,zwanzig1995simple,onuchic1997theory,karlpus1997levinthal,dobson1998understanding,pande1998pathways,kubelka2004protein,naganathan2005direct,barrick2009what,berezhkovskii2011peptides,lane2013probing}. The barrier is created due to an interplay between energy and entropy, i.e., while the energy directs the protein towards the native state, the entropy returns it back towards numerous unfolded states \cite{sali1994how,onuchic1997theory,dobson1998understanding}. On a free energy landscape, the unfolded and folded states form basins of attraction separated by the free energy  barrier \cite{onuchic1997theory,dobson1998understanding,shea2001from,gruebele2002protein,henry2004combinatorial,das2006low,best2013native}. The first-passage time (FPT) to reach the native state has a single-exponential distribution with the mean FPT (MFPT) associated with the height of the free energy barrier. If observed, such a distribution allows one to suggest that only two protein states are essentially populated - for unfolded and folded protein conformations. Along with these well-documented properties of two-state folders, one issue requires clarification. According to the Anfinsen principle \cite{anfinsen1973principles}, the native state of a protein is characterized by the minimum in free energy. This implies that the native state represents an ensemble of conformationally close structures (``native ensemble'') rather than a unique structure \cite{anfinsen1973principles,mc_cammon1977dynamics,shortle1998clustering,kay2005nmr,lindorff_larsen2005simultaneous,shehu2006modeling,%
best2006relation,shehu2007characterization,shehu2009multiscale,dubay2015fluctuations,munoz2016limited}; the solution NMR experiments give some examples of these structures \cite{palmer2004nmr,kay2005nmr,lindorff_larsen2005simultaneous,best2006relation,andersen2006minimization}. The flexibility of the native state is considered to be an essential property of a protein to perform its function \cite{privalov1979stability,ishima2000protein,taverna2002why,eisenmesser2005intrisic,baldwin2009nmr}. The structures that form the free energy basin of folded states are generally native-like. Among them, the native ensemble structures, which correspond to the minimum of the basin, constitute a tiny fraction. Therefore, when a protein comes to the basin of native-like states, it does not necessarily reach the native state immediately, but may dwell, and typically does, in this basin exploring native-like conformations until it finds a native one. As a result, the MFPT may not be determined solely by the height of the free energy barrier but be affected by the protein dynamics in the native-like basin. So, why the exploration of native-like states does not change the single-exponential FPT distributions, which are typically observed for two-state folders and are considered to be an intrinsic indicator of two-state folding? 

To gain insight into this issue, we perform molecular dynamics (MD) simulations of folding of a representative example of two-state folders - a $\beta$-hairpin protein. We show that along with the single-exponential FPT distribution to come to the basin of native-like states by overcoming the free energy barrier, the distribution of times to reach the native state within this basin is also single-exponential. If the protein explores native-like states for a time much longer than the time to overcome the free energy barrier, the resulting FPT distribution to reach the native state remains close to an exponential distribution, but the MFPT is largely determined not by the height of the free energy barrier but by the time to find the native state among the native-like ones. This time increases with temperature, which explains the known U-shape dependence of the MFPT on temperature. We also use the Kramers rate formula to estimate the transition times from the basin of unfolded states to the basin of native-like states, and find that these times are in reasonable agreement with the corresponding times obtained in the simulations and typically much shorter than the MFPTs to reach the native state. 

The paper is organized as follows. Section \ref{sec:2} describes the system and methods, Sect. \ref{sec:3} presents the results and their discussion, and Sec. \ref{sec:4} summarizes the results and gives some concluding
remarks. 

\section{System and Simulation Method}
\label{sec:2}%
The $\beta$-hairpin protein we study is a 12-residue protein with the sequence KTWNPATGKWTE (2evq.pdb) \cite{andersen2006minimization}. Since a larger number of folding trajectories was required to have well-converged FPT distributions (ten to twenty-five thousand trajectories were run), a coarse-grained simulation similar to that in the previous work  \cite{chekmarev2013protein} was employed. It included a C$_{\alpha}$-bead protein model and G\={o}-type interaction potential \cite{go1983theoretical}. The  C$_{\alpha}$-bead model was constructed on the basis of the NMR solution structures of the protein \cite{andersen2006minimization}. Specifically, the first structure among the 43 structures in the NMR ensemble was taken as a reference native structure. The G\={o}-type potential consisted of three terms, which accounted for the rigidity of the backbone and the contributions of native and non-native contacts in the form of the Lennard-Jones potential \cite{hoang2000molecular}. Two C$_{\alpha}$-beads were considered to be in native contact if they were not the nearest neighbors along the protein chain and had the inter-bead distance not longer than $d_{\mathrm{cut}}=7.5\mathrm{\AA}$, which was found to be suitable for the correct formation of the native structure.  In this case, the number of native contacts is $N_{\mathrm{nat}}=N_{\mathrm{nat}}^{\mathrm{NAT}}=27$. The simulations were performed with a constant-temperature molecular dynamics (MD) based on the coupled set of Langevin equations \cite{biswas1986simulated}. The time-step was $\Delta t=0.0125\tau$, where $\tau$ is the characteristic time. At the length scale $l=7.5\mathrm{\AA}$ and the attractive energy $\epsilon=2.2$ kcal/mol \cite{miyazawa1996residue}, $\tau=(Ml^2/\epsilon)^{1/2} \approx 2.6$ ps, where $M=110$ Da is the average mass of the residue. The friction constant $\gamma=M/\tau$ in the Langevin equations, which determined the protein friction against the surroundings, was varied from $\gamma=3M/\tau$ to $\gamma=50M/\tau$, where the upper bound corresponds to water solution at room temperature \cite{klimov1997viscosity}. In what follows, the temperature is measured in the units of $\epsilon$, i.e., the Boltzmann constant is set to unity. 

Folding trajectories were initiated at partially folded states of the protein, which were obtained by equilibration of a fully extended conformation for $10^3$ time steps. Starting from these states, the trajectories were continued until the root-mean-square-deviation (RMSD) from the reference structure, $\sigma_{\mathrm{nat}}$, was less than some RMSD threshold $\sigma_{\mathrm{nat}}^{\mathrm{thrh}}$. Specifically, $\sigma_{\mathrm{nat}}^{\mathrm{thrh}}$ was taken to be equal to 1.0 {\AA}, which is slightly higher than the maximum value of the pairwise C$_{\alpha}$ RMSD for the 43-member ensemble of the NMR solution protein structures ($\approx 0.65\mathrm{\AA}$) \cite{andersen2006minimization}. The structures corresponding to the end points of the trajectories were considered to form the native ensemble. 

\section{Results and Discussion}
\label{sec:3}%
\begin{figure}\centering%
\resizebox{0.49\linewidth}{!}{ \includegraphics*{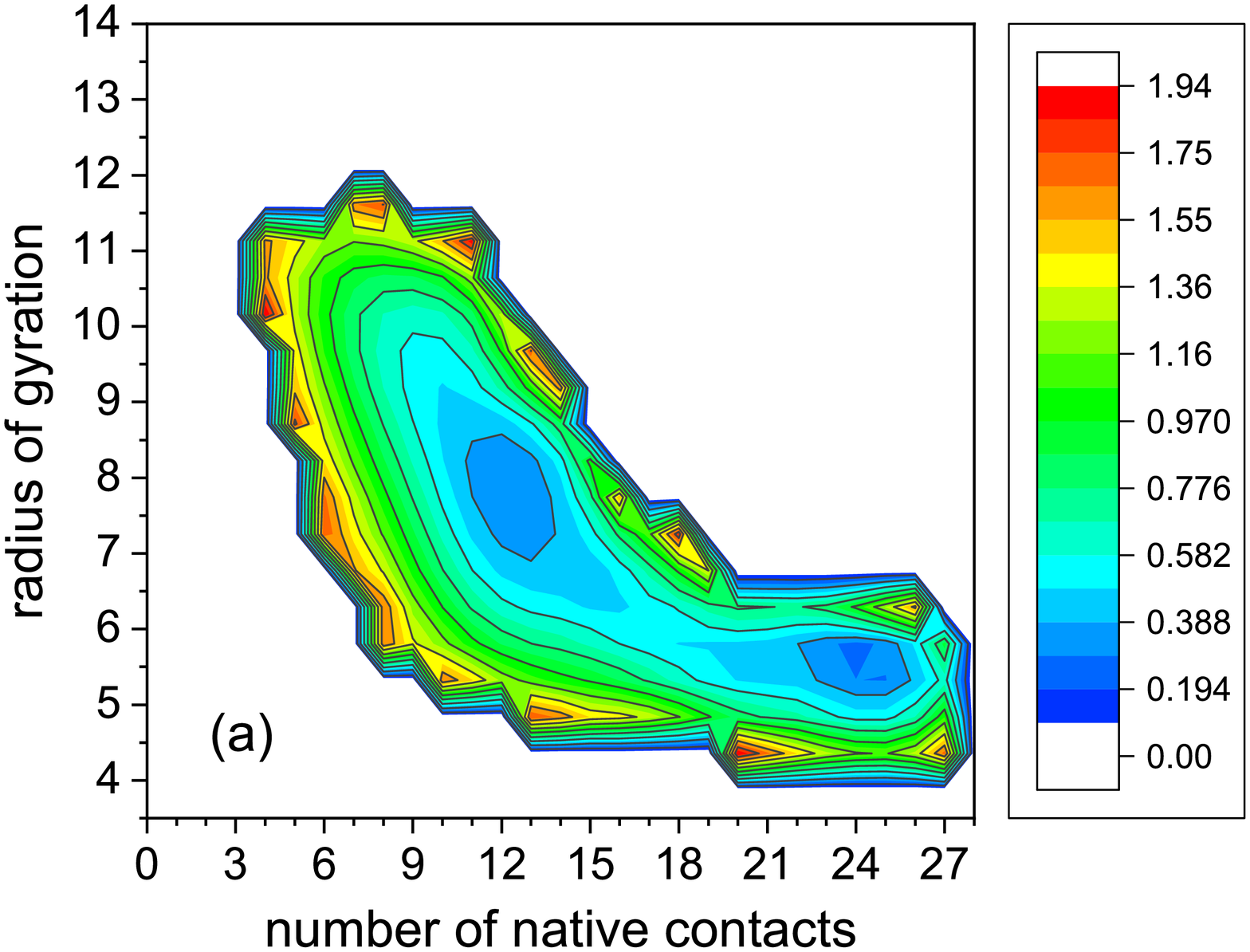}}%
\hfill
\resizebox{0.49\linewidth}{!}{ \includegraphics*{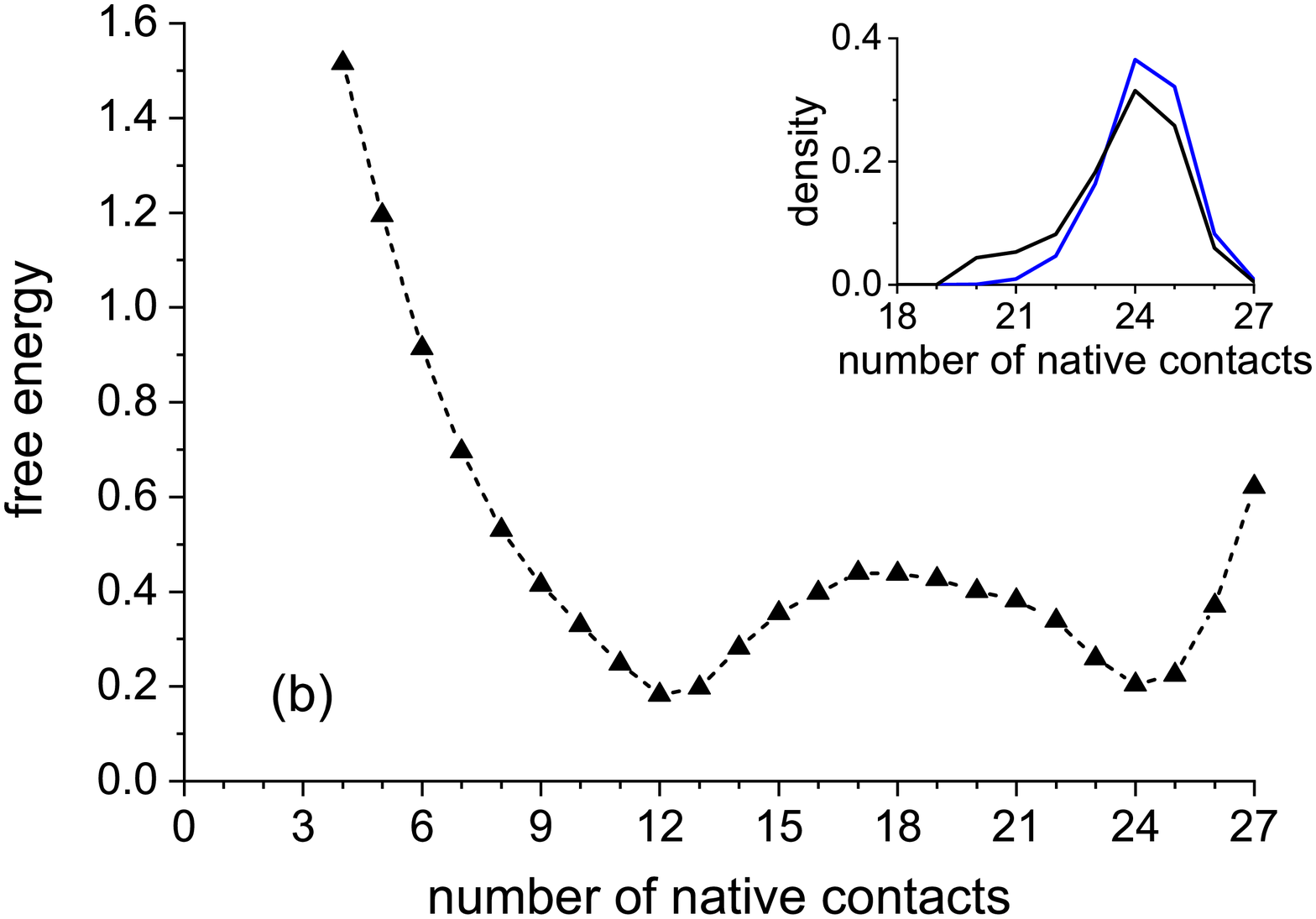}}%
\hfill
\resizebox{0.49\linewidth}{!}{ \includegraphics*{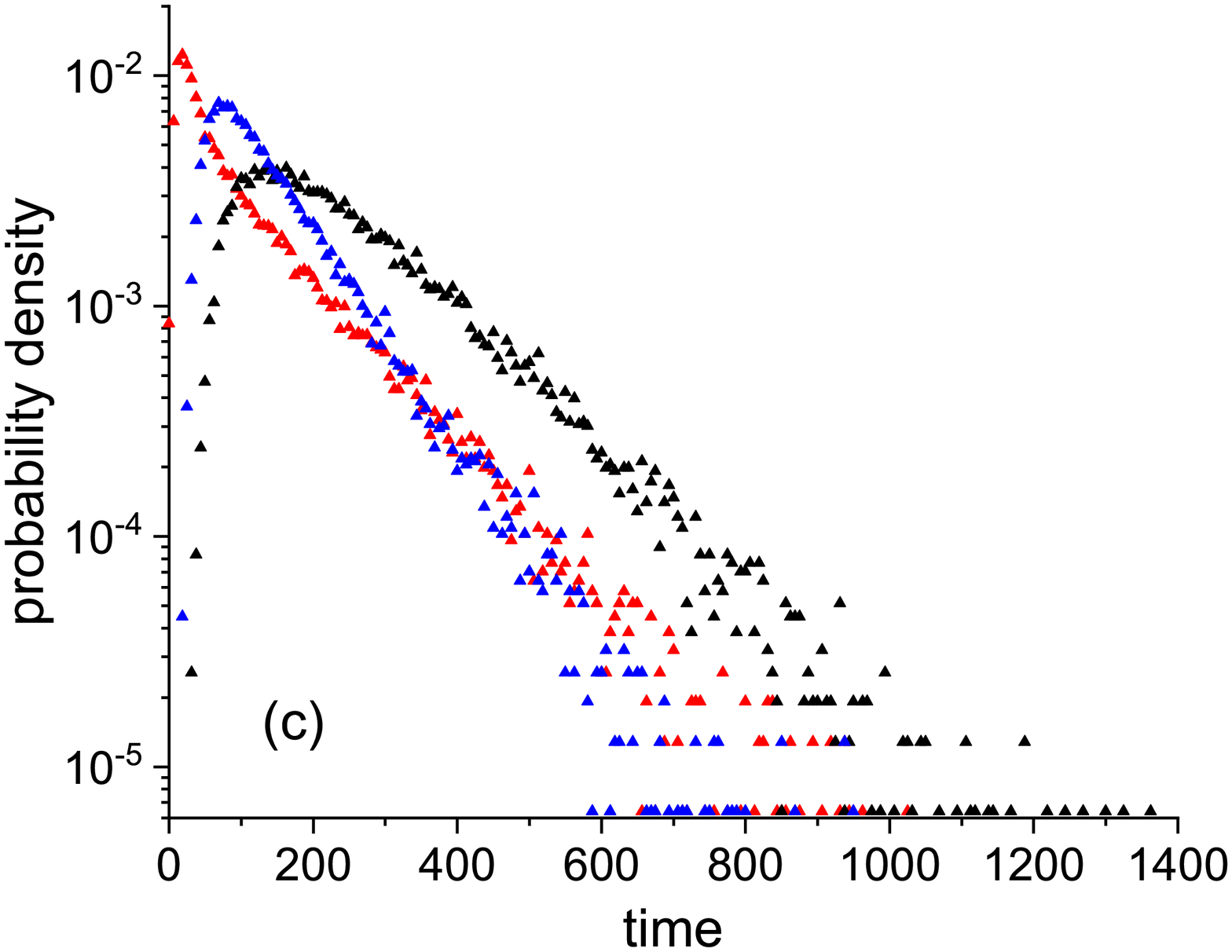}}%
\hfill
\resizebox{0.49\linewidth}{!}{ \includegraphics*{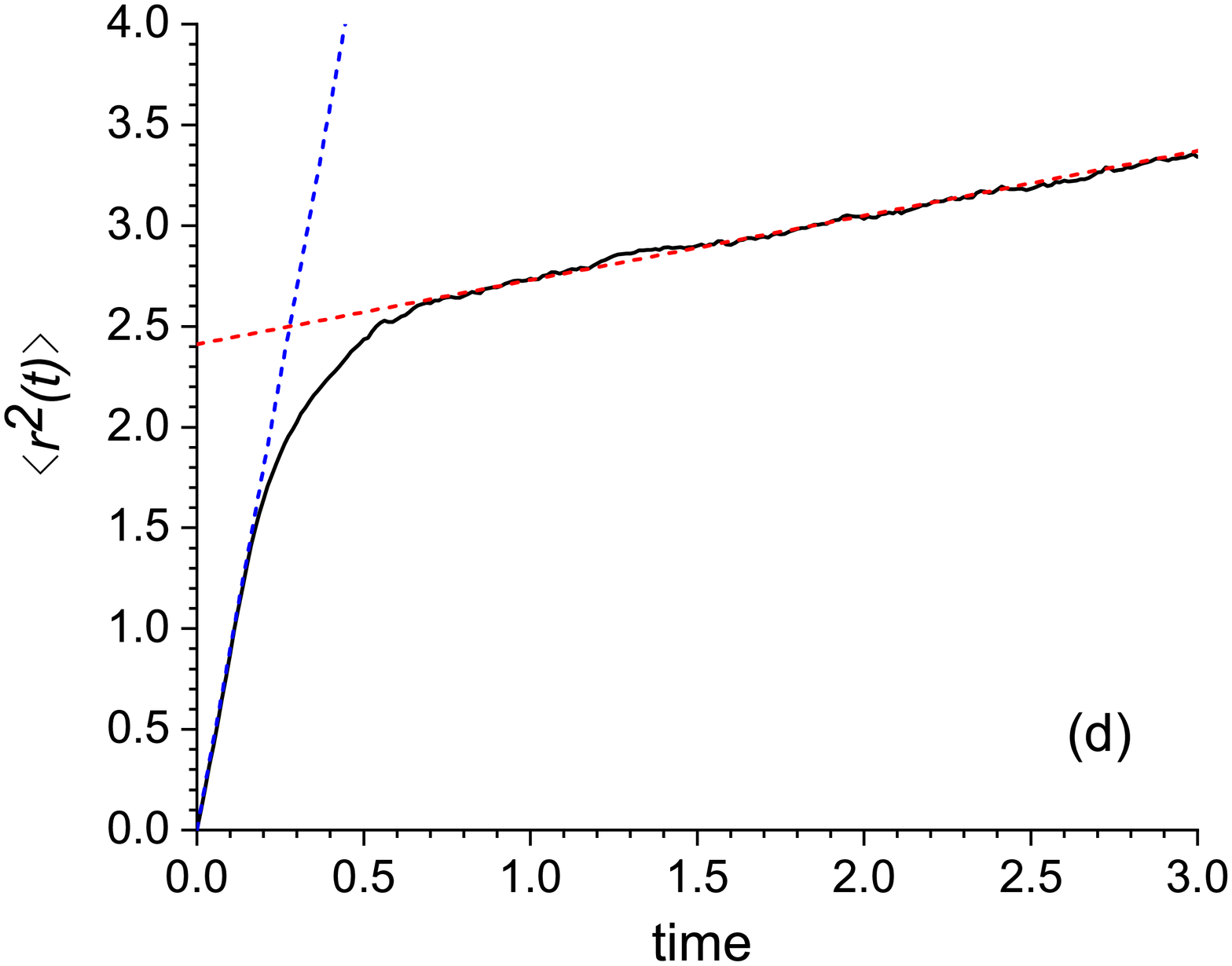}}%
\caption{$T=0.1$. ({\bf{a}}) The free energy surface $F(N_{\mathrm{nat}},R_{\mathrm{g}})$. ({\bf{b}}) Free energy profile $F(N_{\mathrm{nat}})$. The insert shows the normalized distributions of the protein states in the native-state ensemble (blue curve) and in the native-like basin (black curve). ({\bf{c}}) First-passage time distributions: the $\mathrm{U} \rightarrow \mathrm{NL}$ trajectories (blue triangles), the $\mathrm{NL} \rightarrow \mathrm{N}$ trajectories (red), and the $\mathrm{U} \rightarrow \mathrm{N}$ trajectories (black). ({\bf{d}}) The mean-square deviation of the number of natives contacts $N_{\mathrm{nat}}$ from that at the transition state $N_{\mathrm{nat}}^{\mathrm{TS}}$ (black curve); the blue and red dashed lines are the linear fits to the curve for short and long times, respectively.}   
\label{all_0_1}
\end{figure}

\begin{figure}\centering%
\resizebox{0.49\linewidth}{!}{ \includegraphics*{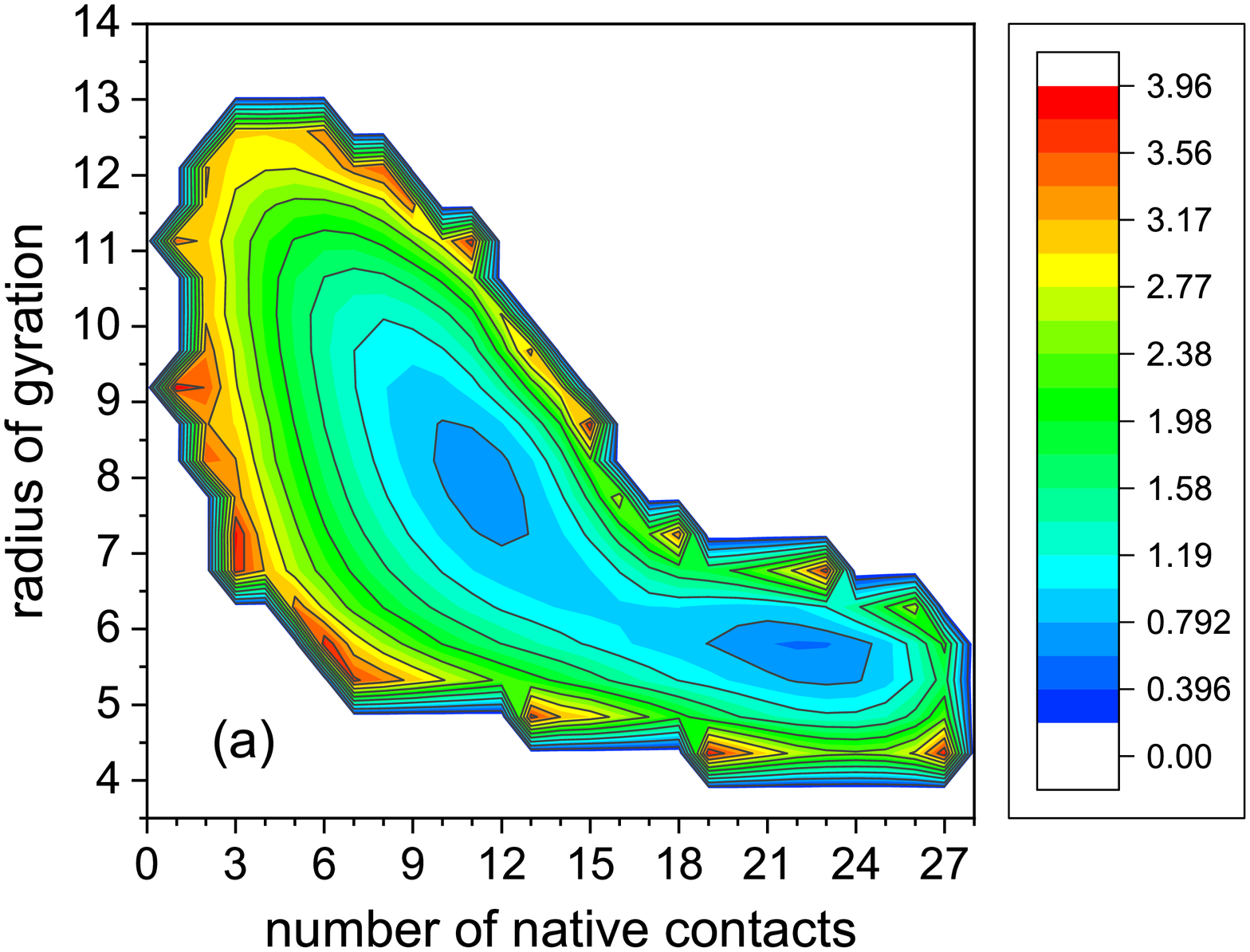}}%
\hfill
\resizebox{0.49\linewidth}{!}{ \includegraphics*{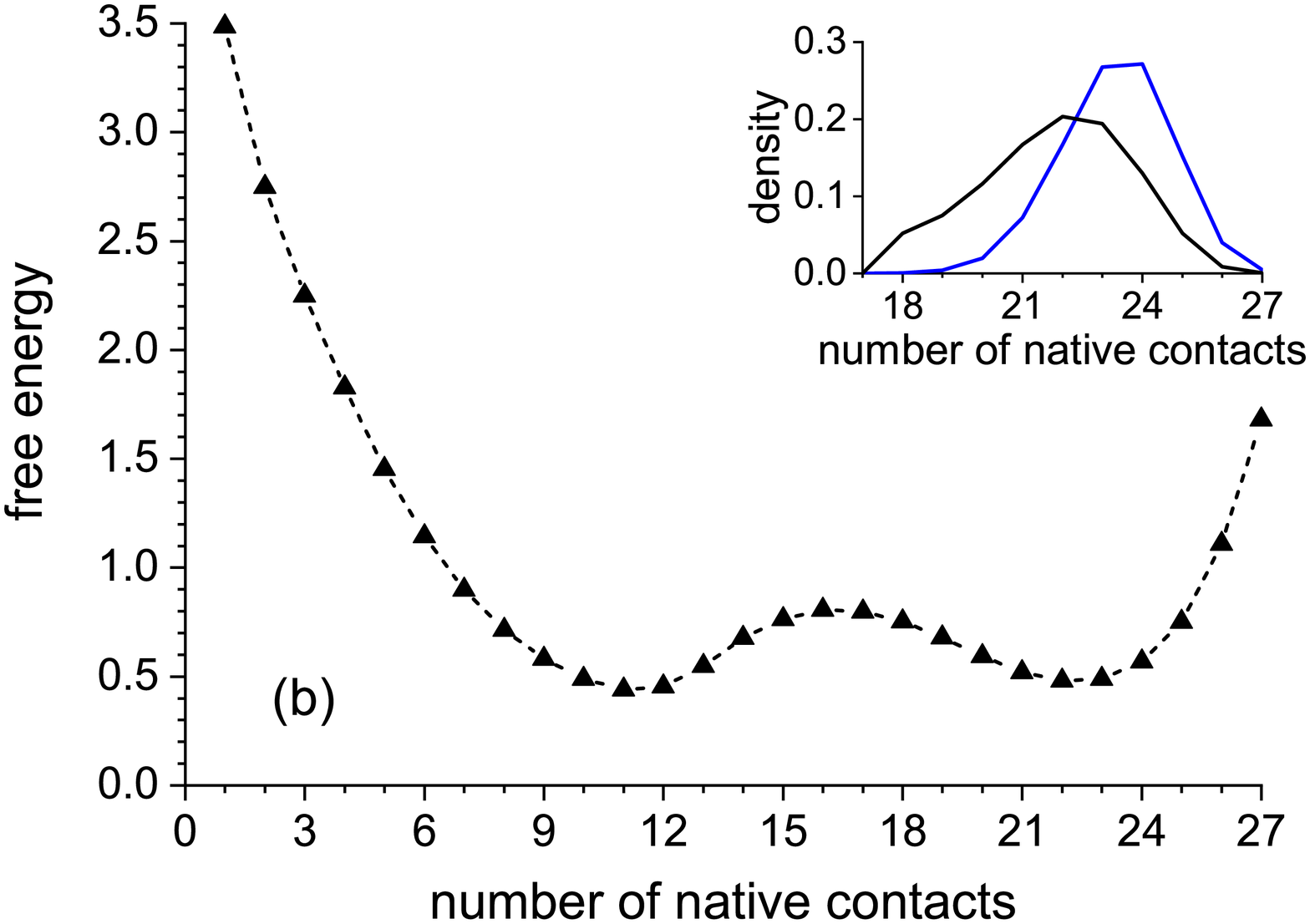}}%
\hfill
\resizebox{0.49\linewidth}{!}{ \includegraphics*{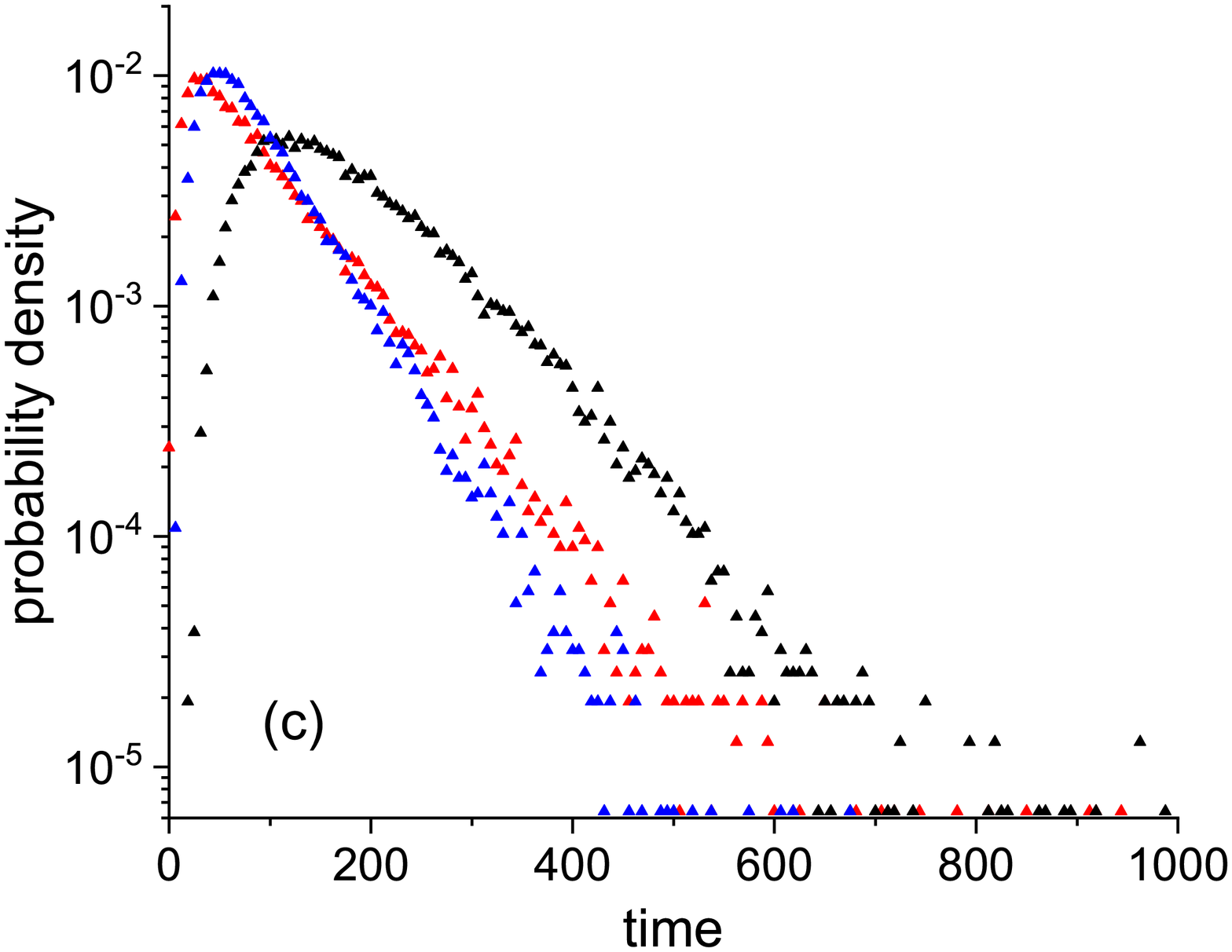}}%
\hfill
\resizebox{0.49\linewidth}{!}{ \includegraphics*{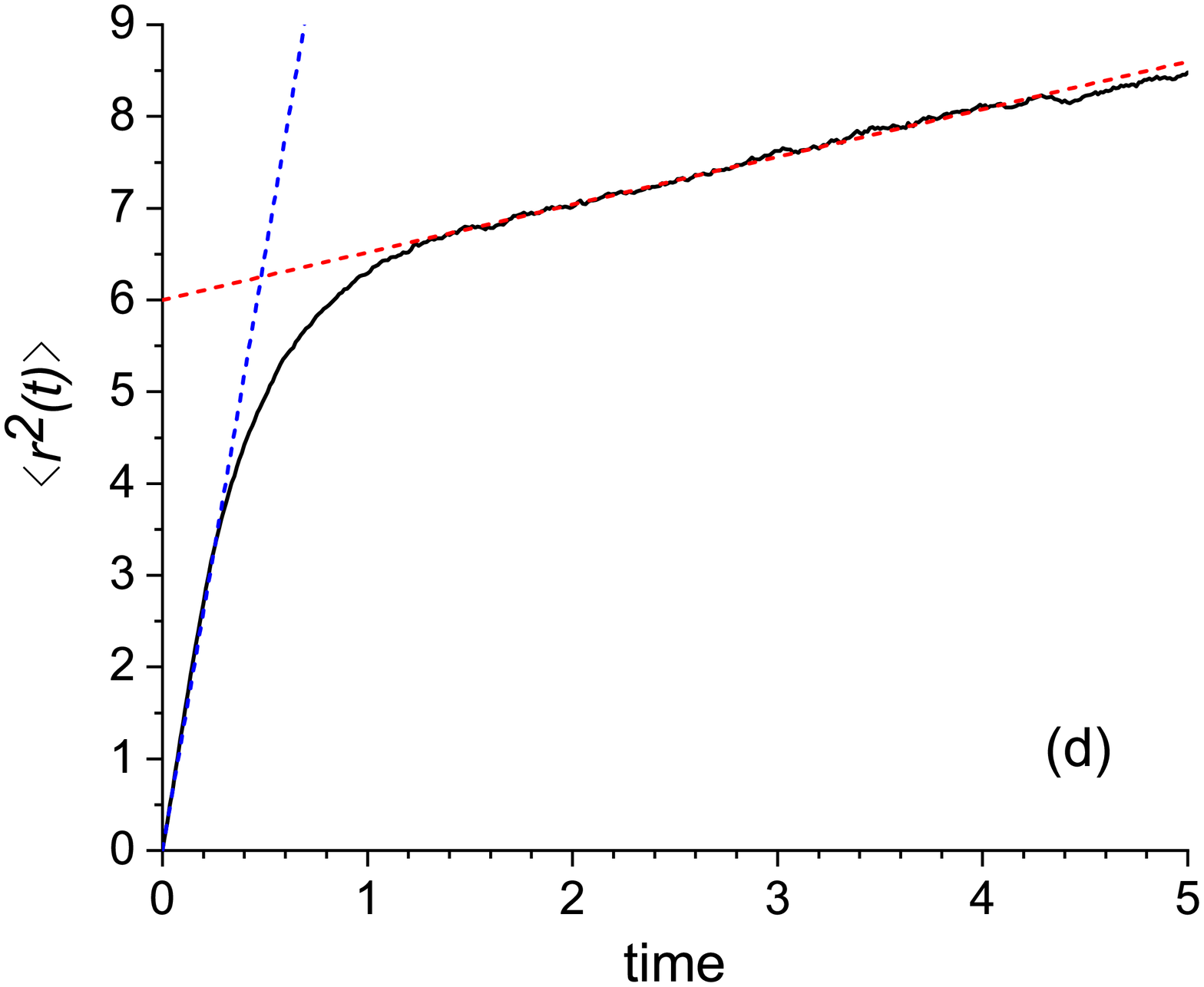}}%
\caption{$T=0.2$. The notations are as in Fig. \ref{all_0_1}.}   
\label{all_0_2}
\end{figure}

\begin{figure}\centering%
\resizebox{0.49\linewidth}{!}{ \includegraphics*{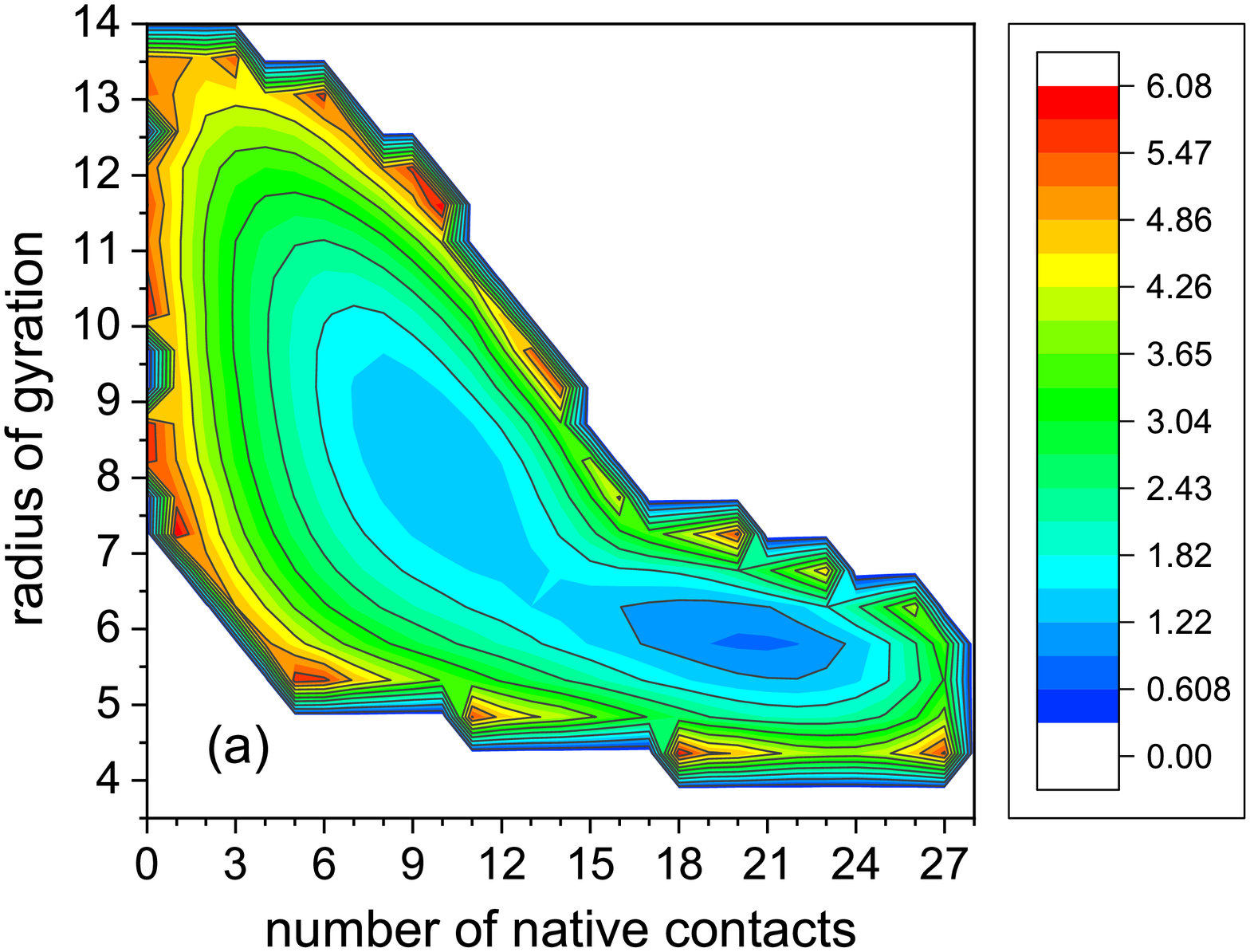}}%
\hfill
\resizebox{0.49\linewidth}{!}{ \includegraphics*{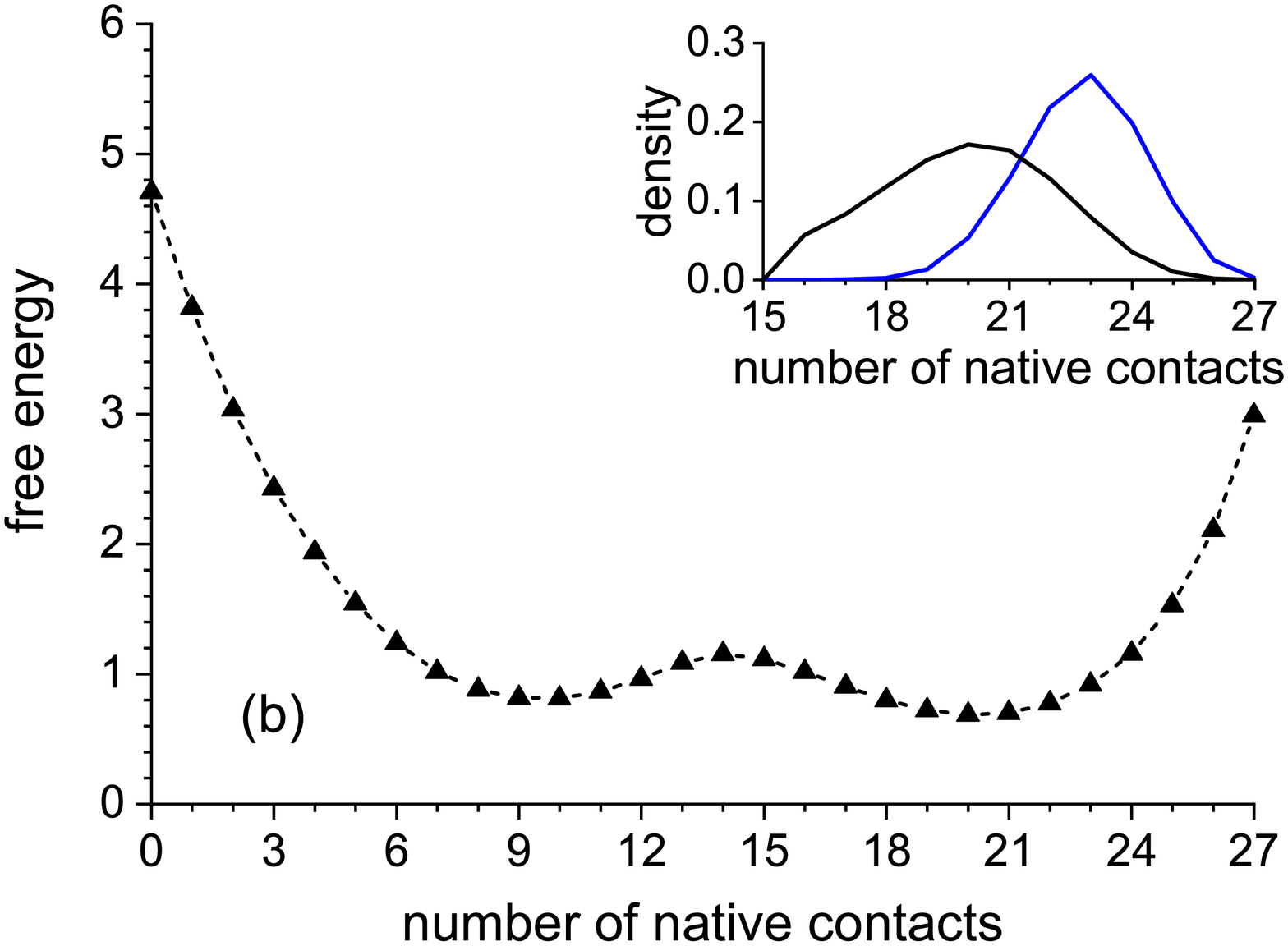}}%
\hfill
\resizebox{0.49\linewidth}{!}{ \includegraphics*{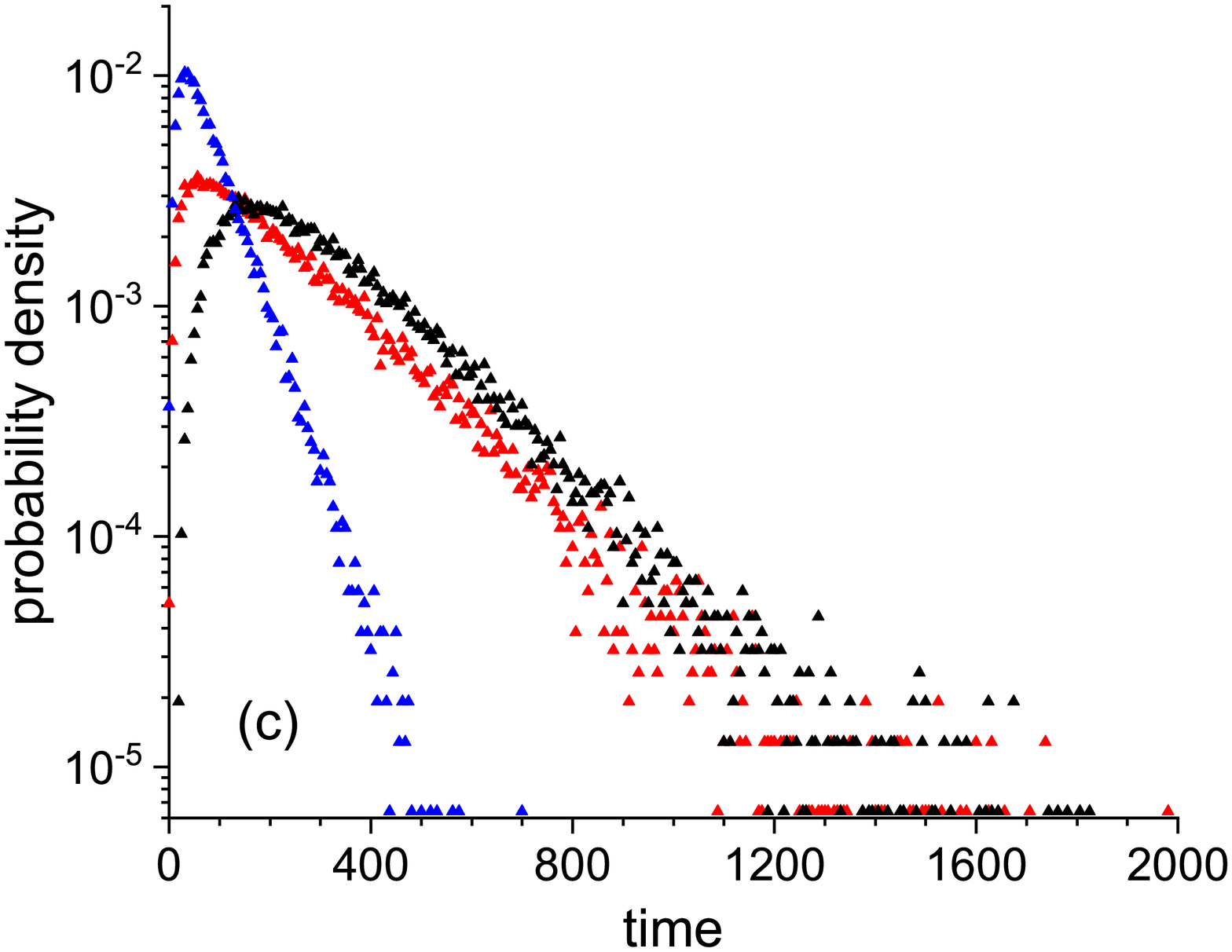}}%
\hfill
\resizebox{0.49\linewidth}{!}{ \includegraphics*{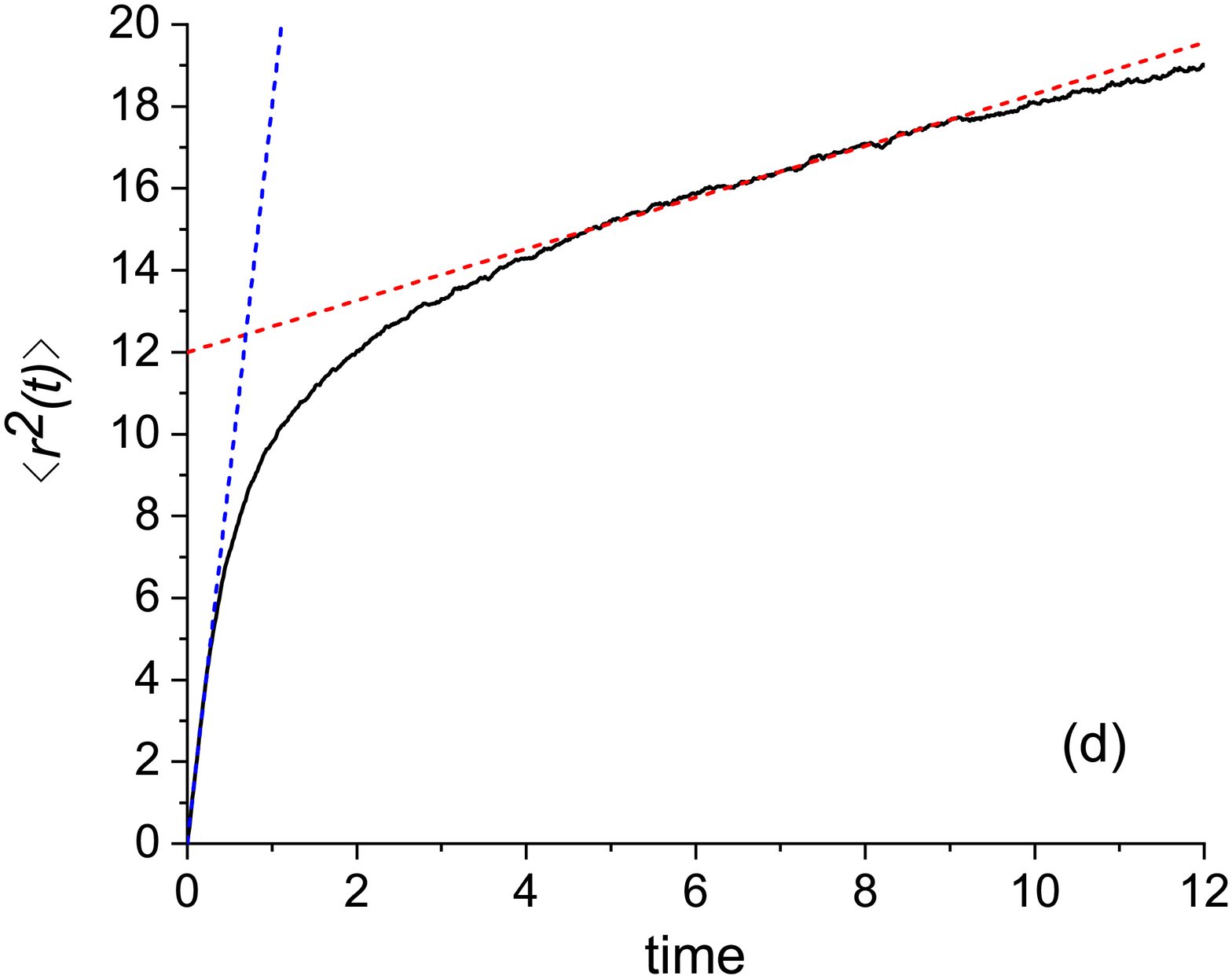}}%
\caption{$T=0.3$. The notations are as in Fig. \ref{all_0_1}.}   
\label{all_0_3}
\end{figure}

The simulations were performed for five temperatures ranging from $T=0.1$ to $T=0.3$. Figures \ref{all_0_1}, \ref{all_0_2} and \ref{all_0_3} present the results for $T=0.1$, $T=0.2$ and $T=0.3$, respectively. The friction constant is $\gamma=10M/\tau$. Twenty five thousand of folding trajectories were run for each temperature. Figure \ref{all_0_1} shows the results for $T=0.1$. The distribution of protein states is presented in Fig. \ref{all_0_1}{\bf{a}} and Fig. \ref{all_0_1}{\bf{b}} as a free energy surface (FES) and a free energy profile (FEP), respectively. The number of native contacts,  $N_{\mathrm{nat}}$, and the radius of gyration, $R_{\mathrm{g}}$, were used as collective variables; the former characterized the protein proximity to the native state and the latter the protein compactness. The number of native contacts is commonly employed as a reaction coordinate \cite{sali1994how,socci1996diffusive,onuchic1997theory,dobson1998understanding,best2010coordinate} and, as has been recently shown, captures remarkably well the transition states (TSs) for a variety of proteins \cite{best2013native}. Since the MD trajectories were terminated upon reaching the native state, i.e., ''nonequilibrium'' conditions were simulated, the present FES and FEP represent the distributions of probabilities of protein states rather the true free energy landscapes. For the FES, the free energy was calculated as $F(N_{\mathrm{nat}},R_{\mathrm{g}})=-T\ln P(N_{\mathrm{nat}},R_{\mathrm{g}})$, where $P(N_{\mathrm{nat}},R_{\mathrm{g}})$ is the probability to find the protein in a state with the given number of native contacts and radius of gyration. For the FEP, the free energy was calculated as $F(N_{\mathrm{nat}})=-T\ln P(N_{\mathrm{nat}})$, where the probability for the protein to have  $N_{\mathrm{nat}}$ contacts, $P(N_{\mathrm{nat}})$, was calculated by the summation of protein states at the current value of $N_{\mathrm{nat}}$. In agreement with previous studies of $\beta$-hairpin folding  \cite{chekmarev2013protein,munoz1997folding,dinner1999understanding,zhou2001free,zagrovic2001hairpin,bolhuis2003transition,krivov2004hidden,bussi2006free}, the FES (Fig. \ref{all_0_1}{\bf{a}}) and FEP (Fig. \ref{all_0_1}{\bf{b}}) reveal two basins of attraction - one for partially folded conformations (smaller values of $N_{\mathrm{nat}}$), and the other for native-like states (larger values of $N_{\mathrm{nat}}$). The basins are separated by a free energy barrier at the TS at  $N_{\mathrm{nat}} = N_{\mathrm{nat}}^{\mathrm{TS}} \approx 18$. The insert in Fig. \ref{all_0_1}{\bf{b}} also shows the normalized distribution of the protein states in the native ensemble, i.e., of the states in which the folding trajectories were terminated (blue curve). It is centered at the point corresponding to the minimum of the FEP ($N_{\mathrm{nat}}=24$), in agreement with Anfinsen principle \cite{anfinsen1973principles}, and is close to the distribution of the states in the native-like basin (black curve). The essential difference between these distributions is that they represent drastically different numbers of states, specifically, $2.5 \times 10^4$ states in the former case (which is the number of simulated folding trajectories) and $\approx 2 \times 10^8$ states in the latter case. Accordingly, approximately $2 \times 10^8/2.5 \times 10^4=8 \times 10^3$ native-like states are required to pass through in order to reach the native state.

To estimate the contribution of native-like states to the overall FPT distribution, each MD trajectory from the unfolded to the native state ($\mathrm{U} \rightarrow \mathrm{N}$) was divided into two parts, i.e., the trajectory to come from the unfolded state to the basin of native-like states ($\mathrm{U} \rightarrow \mathrm{NL}$), and the continuation of this trajectory in the basin of native-like states, until the protein finds the native state ($\mathrm{NL} \rightarrow \mathrm{N}$). The $\mathrm{U} \rightarrow \mathrm{NL}$ trajectories are somewhat similar to the transition-path trajectories \cite{chung2012single,jacobs2018accurate}. To divide the $\mathrm{U} \rightarrow \mathrm{N}$ trajectory, here and in all other cases we studied, we chose the points $N_{\mathrm{nat}}=N_{\mathrm{nat}}^{\mathrm{TS}}+2$, where the height of the basin of native-like states on the TS side typically was $\approx 90\%$. It has been found that at the given choice of dividing point, the $\mathrm{U} \rightarrow \mathrm{NL}$ trajectories did not return from the basin of native-like states to the unfolded basin, except for small deviations from the TS (Supporting Information, Fig. S1). Figure \ref{all_0_1}{\bf{c}} shows the FPT distributions for the $\mathrm{U} \rightarrow \mathrm{NL}$ (blue), $\mathrm{NL} \rightarrow \mathrm{N}$ (red) and $\mathrm{U} \rightarrow \mathrm{N}$ (black) trajectories. It is seen that not only the $\mathrm{U} \rightarrow \mathrm{NL}$ distribution is essentially single-exponential, which is implied by two-state kinetics, but the NL-N distribution is also approximately single-exponential. A steep rise in the  $\mathrm{U} \rightarrow \mathrm{NL}$ distribution at small times reflects the times required to come to the bottom of the unfolded basin from a less folded state rather than the times to overcome the free energy barrier \cite{chekmarev2005folding}. According to the Poisson law of zero-order (the waiting times for the first events), the single-exponential FPT distribution for the $\mathrm{U} \rightarrow \mathrm{NL}$ trajectories suggests that in the basin of native-like states the protein explores equally probable and accessible states (see also a simple illustration of this in the Supporting Information, Fig. S2). In the present case, the $\mathrm{U} \rightarrow \mathrm{NL}$ and $\mathrm{NL} \rightarrow \mathrm{N}$ trajectories contribute to the overall MFPT approximately equally: the MFTPs for these trajectories are $\langle t_{\mathrm{U} \rightarrow \mathrm{NL}}\rangle \approx 144$ and $\langle t_{\mathrm{NL} \rightarrow \mathrm{N}}\rangle \approx 106$, which, in sum, gives for the overall MFPT $\langle t_{\mathrm{U} \rightarrow \mathrm{N}}\rangle \approx 250$. The $\mathrm{U} \rightarrow \mathrm{N}$ distribution is approximately exponential but mostly for the times longer than the MFTPs for the $\mathrm{U} \rightarrow \mathrm{NL}$ and $\mathrm{NL} \rightarrow \mathrm{N}$ trajectories. 

\begin{figure}\centering%
\resizebox{0.49\linewidth}{!}{ \includegraphics*{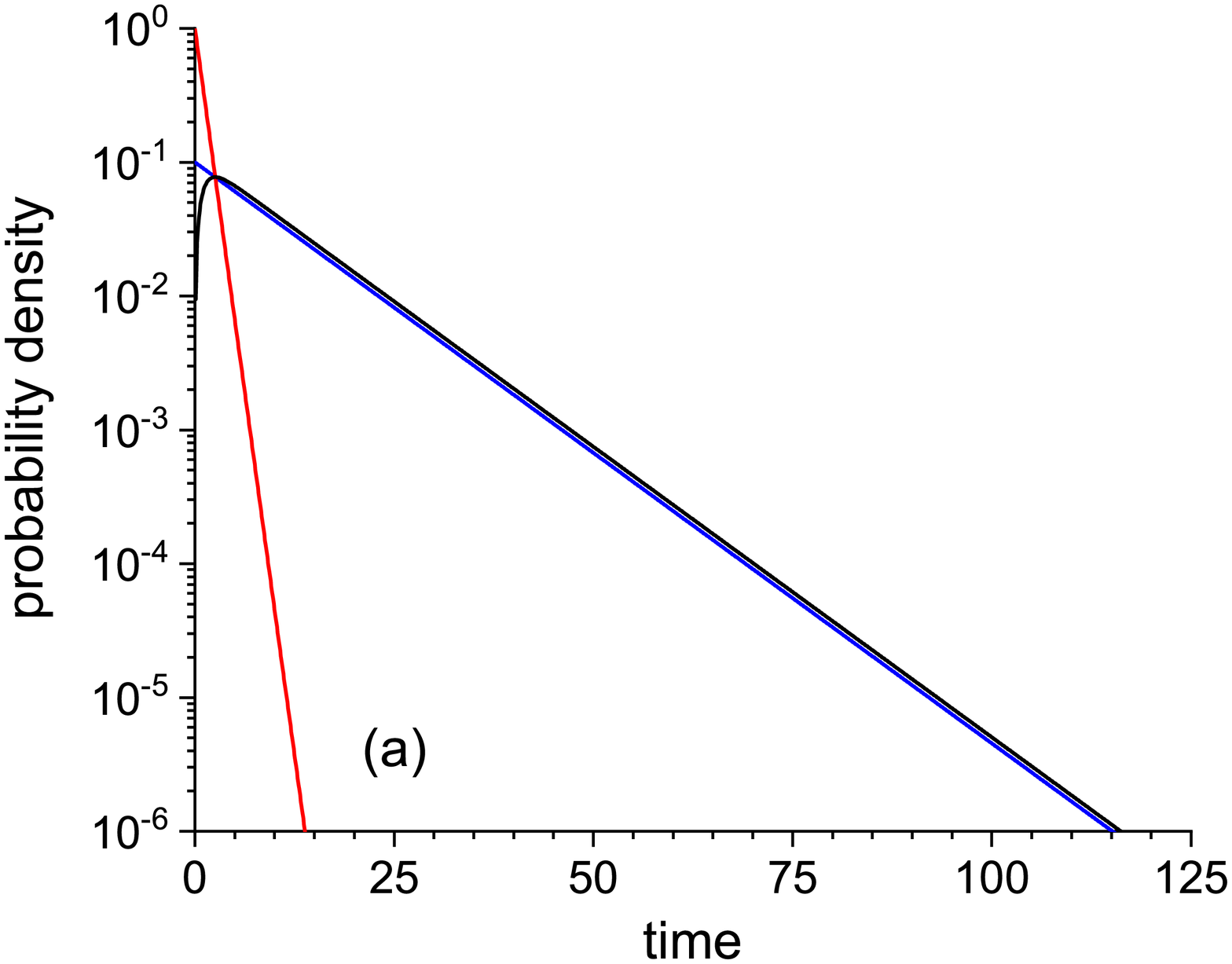}}%
\hfill
\resizebox{0.49\linewidth}{!}{ \includegraphics*{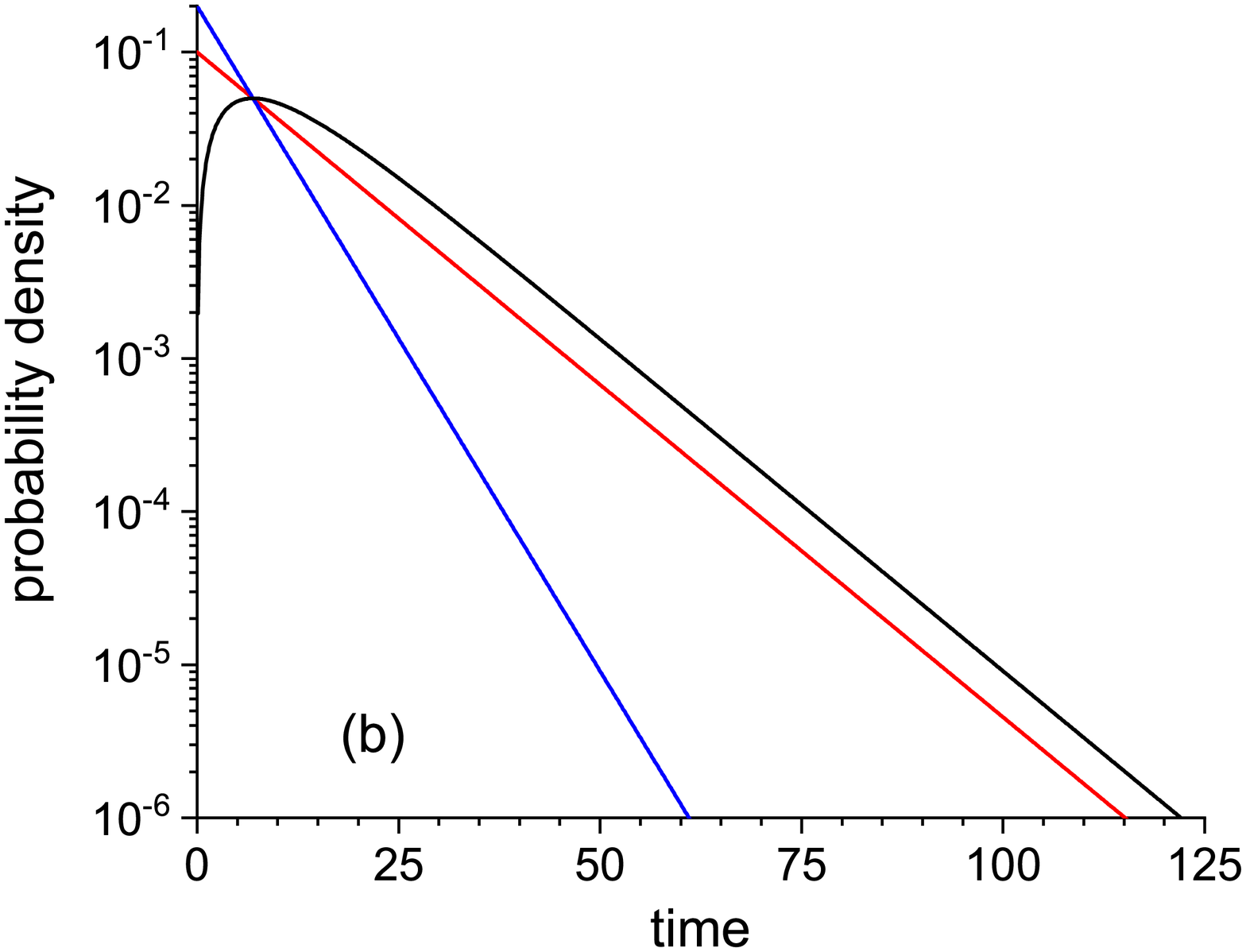}}%
\hfill
\resizebox{0.49\linewidth}{!}{ \includegraphics*{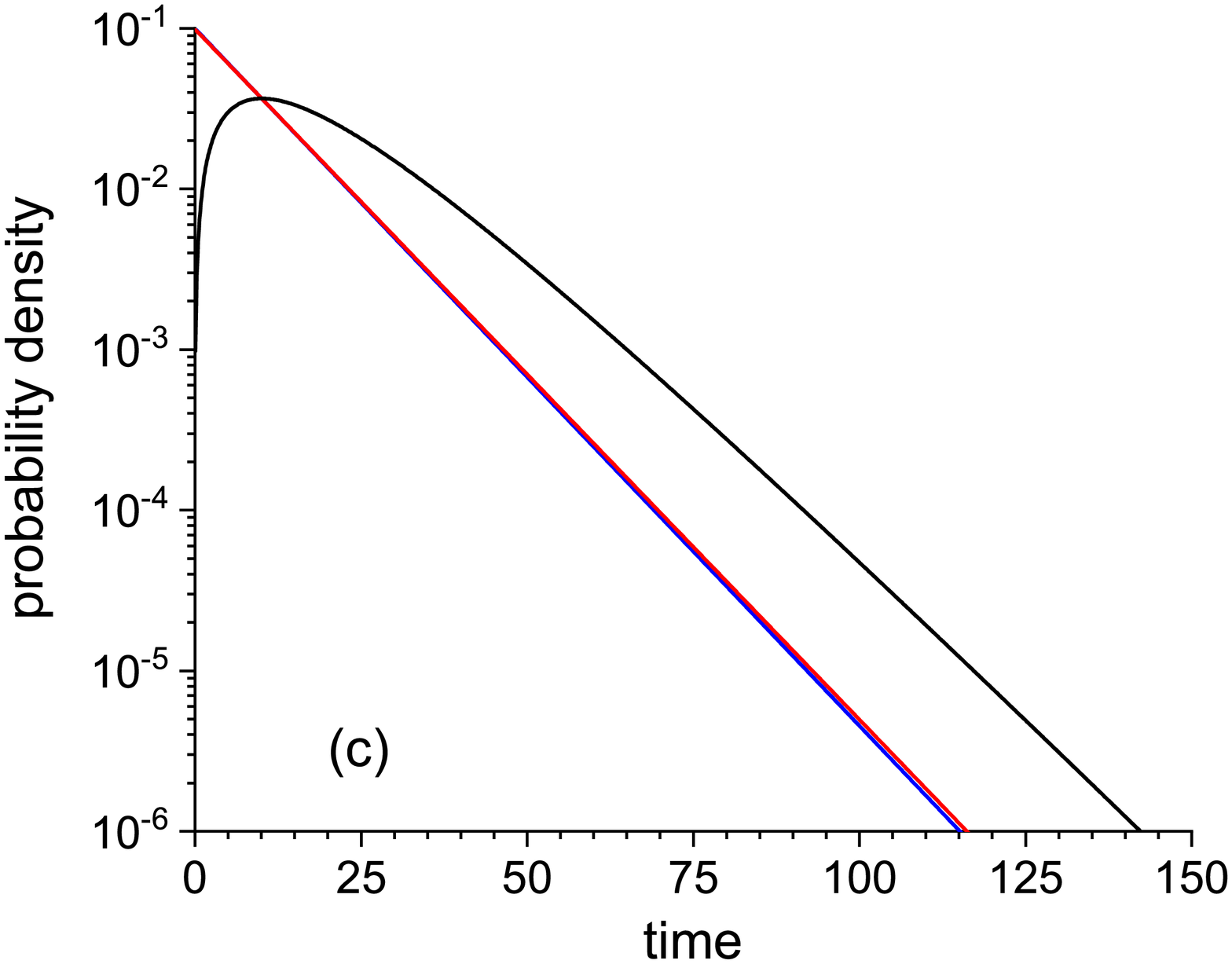}}%
\hfill
\resizebox{0.49\linewidth}{!}{ \includegraphics*{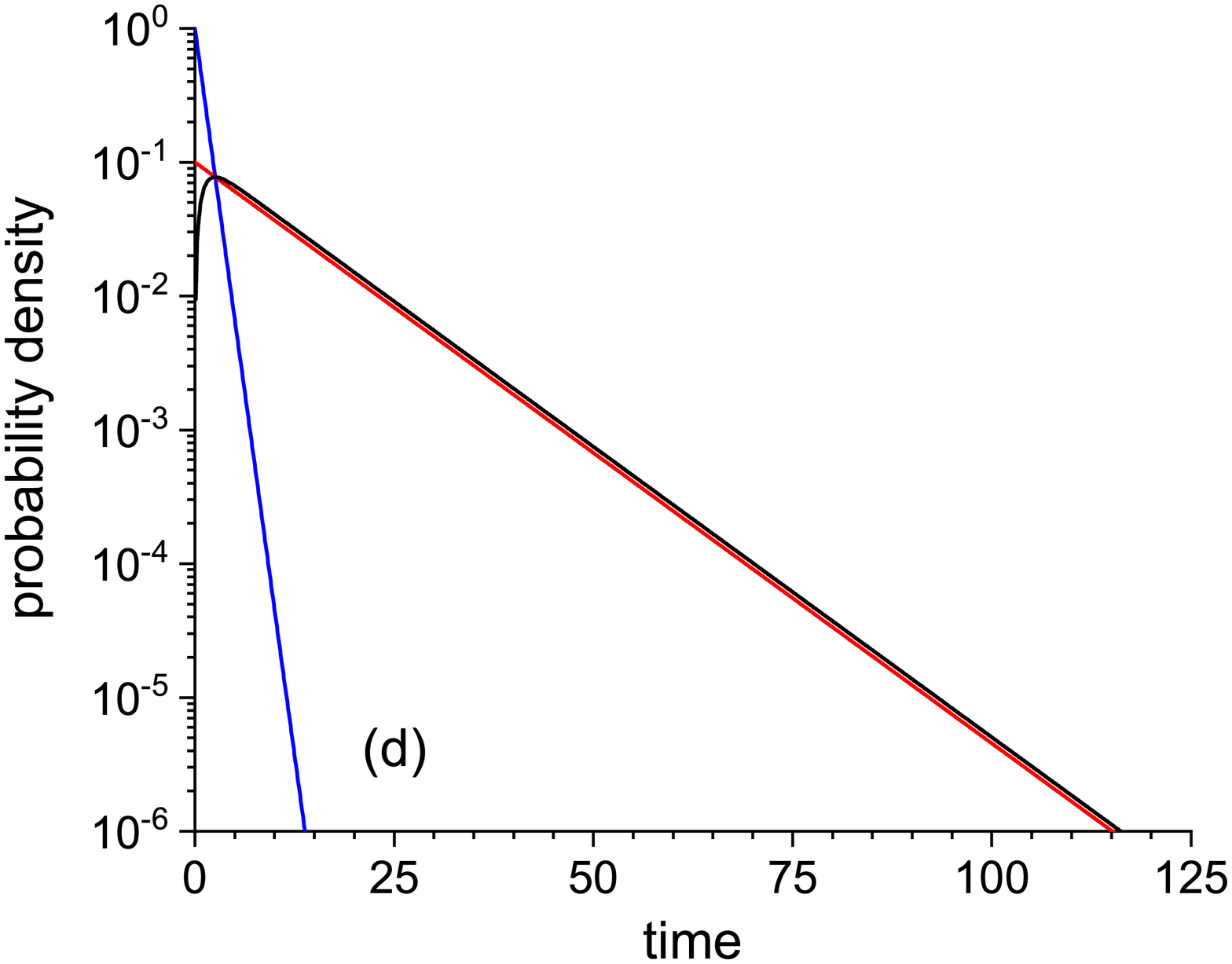}}%
\caption{The evolution of theoretical first-passage time distribution: ({\bf a}) $\langle t_{\mathrm{\mathrm{U} \rightarrow \mathrm{NL}}}\rangle=10.0$ and $\langle t_{\mathrm{NL} \rightarrow \mathrm{N}} \rangle=1.0$, ({\bf b}) $\langle t_{\mathrm{\mathrm{U} \rightarrow \mathrm{NL}}}\rangle=5.0$ and $\langle t_{\mathrm{NL} \rightarrow \mathrm{N}} \rangle=10.0$, ({\bf c}) $\langle t_{\mathrm{U} \rightarrow \mathrm{NL}}\rangle=10.0$ and $\langle t_{\mathrm{NL} \rightarrow \mathrm{N}} \rangle=10.1$, and ({\bf d}) $\langle t_{\mathrm{U} \rightarrow \mathrm{NL}}\rangle=1.0$ and $\langle t_{\mathrm{NL} \rightarrow \mathrm{N}} \rangle=10.0$. The distributions for the $\mathrm{U} \rightarrow \mathrm{NL}$ trajectories are shown in blue, for the $\mathrm{NL} \rightarrow \mathrm{N}$ trajectories in red, and for the $\mathrm{U} \rightarrow \mathrm{N}$ trajectories in black.}   
\label{fpt_theor}
\end{figure}

Theoretically, the $\mathrm{U} \rightarrow \mathrm{N}$ FPT distribution is determined as $$p_{\mathrm{U} \rightarrow \mathrm{N}}(t)=\int_{0}^{t}p_{\mathrm{NL} \rightarrow \mathrm{N}}(t_{1})p_{\mathrm{U} \rightarrow \mathrm{NL}}(t-t_{1})dt_{1}$$ In the case when the $\mathrm{U} \rightarrow \mathrm{NL}$ and  $\mathrm{NL} \rightarrow \mathrm{N}$ distributions are single-exponential, i.e., $p_{\mathrm{U} \rightarrow \mathrm{NL}}(t)=(1/\langle t_{\mathrm{U} \rightarrow \mathrm{NL}} \rangle) \exp(-t/\langle t_{\mathrm{U} \rightarrow \mathrm{NL}} \rangle)$ and  $p_{\mathrm{NL} \rightarrow \mathrm{N}}(t)=(1/\langle t_{\mathrm{NL} \rightarrow \mathrm{N}} \rangle) \exp(-t/\langle t_{\mathrm{NL} \rightarrow \mathrm{N}} \rangle)$, we have 
\begin{equation}\label{eq1}%
p_{\mathrm{U} \rightarrow \mathrm{N}}(t)=\frac{1}{\langle t_{\mathrm{U} \rightarrow \mathrm{NL}}\rangle-\langle t_{\mathrm{NL} \rightarrow \mathrm{N}} \rangle} [e^{-t/\langle t_{\mathrm{U} \rightarrow \mathrm{NL}} \rangle}-e^{-t/\langle t_{\mathrm{NL} \rightarrow \mathrm{N}} \rangle}]
\end{equation}
If $\langle t_{\mathrm{U} \rightarrow \mathrm{NL}}\rangle \gg \langle t_{\mathrm{NL} \rightarrow \mathrm{N}} \rangle$ or $\langle t_{\mathrm{NL} \rightarrow \mathrm{N}}\rangle \gg \langle t_{\mathrm{U} \rightarrow \mathrm{NL}} \rangle$, $p_{\mathrm{U} \rightarrow \mathrm{N}}(t)$ transforms into the corresponding single-exponential distribution, i.e., the $\mathrm{U} \rightarrow \mathrm{NL}$ or $\mathrm{NL} \rightarrow \mathrm{N}$ distribution, respectively. However, if $\langle t_{\mathrm{U} \rightarrow \mathrm{NL}} \rangle$ and $\langle t_{\mathrm{NL} \rightarrow \mathrm{N}}\rangle$ are compatible, $p_{\mathrm{U} \rightarrow \mathrm{N}}(t)$ may deviate from a single-exponential distribution considerably, revealing a steep rise at the initial times. The most significant deviation occurs at $\langle t_{\mathrm{NL} \rightarrow \mathrm{N}}\rangle \approx \langle t_{\mathrm{U} \rightarrow \mathrm{NL}} \rangle$. Then, $p_{\mathrm{U} \rightarrow \mathrm{N}}(t) \approx t/\tau^{2} \exp^{-t/\tau}$, where $\tau=\langle t_{\mathrm{NL} \rightarrow \mathrm{N}}\rangle \approx \langle t_{\mathrm{U} \rightarrow \mathrm{NL}} \rangle$. In this case, first, the rise at the initial times becomes more pronounced, and, second, the distribution approaches to a single-exponential one only at $t/\ln{t} \gg \tau$. Figure \ref{fpt_theor} shows the evolution of the FPT distributions with the ratio of the $\mathrm{NL} \rightarrow \mathrm{N}$ and $\mathrm{U} \rightarrow \mathrm{NL}$ times $\alpha= \langle t_{\mathrm{NL} \rightarrow \mathrm{N}}\rangle/\langle t_{\mathrm{U} \rightarrow \mathrm{NL}}\rangle$ from $\alpha \ll 1$ to $\alpha \gg 1$; at the lowest bound of $\alpha$, the MFPT is largely determined by the transition over the free energy barrier, and at the upper bound, by the exploration of native-like states. 

\begin{figure}\centering%
\resizebox{0.49\linewidth}{!}{ \includegraphics*{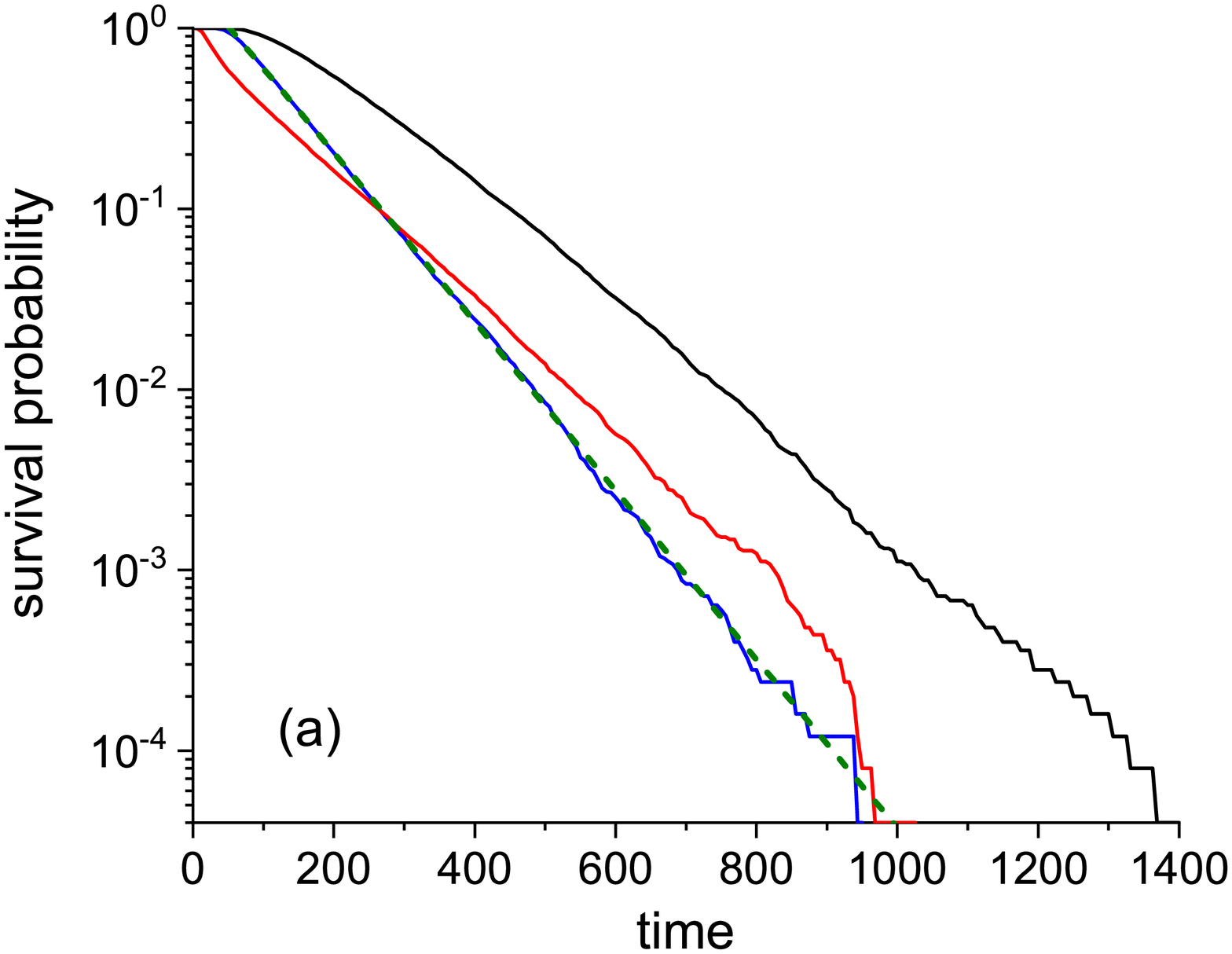}}%
\hfill
\resizebox{0.49\linewidth}{!}{ \includegraphics*{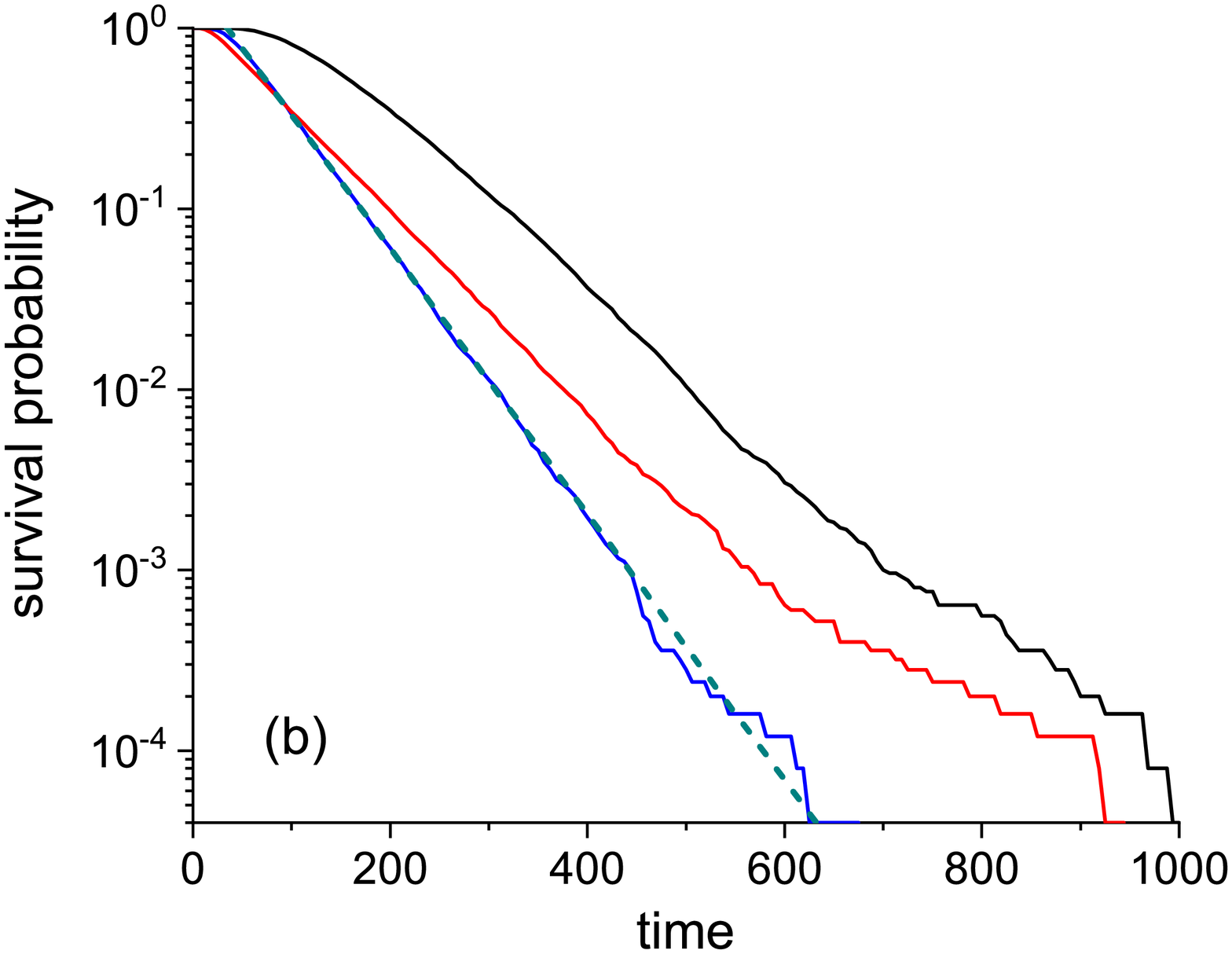}}%
\hfill
\resizebox{0.49\linewidth}{!}{ \includegraphics*{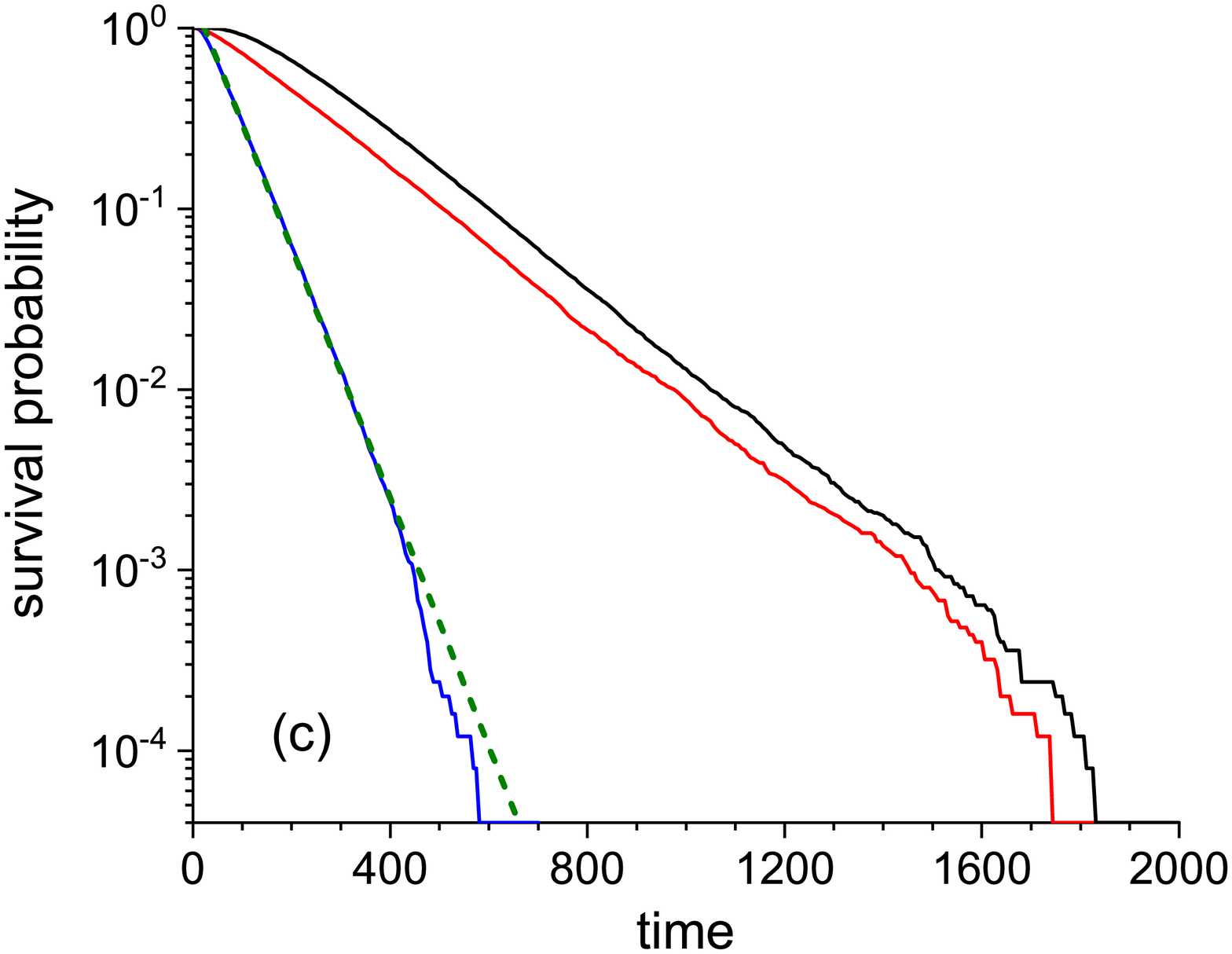}}%
\caption{The first-passage time distributions as the survival probabilities: ({\bf a}) T=0.1, ({\bf b}) T=0.2, and ({\bf c}) T=0.3. The $\mathrm{U} \rightarrow \mathrm{NL}$ distributions are shown in blue, the $\mathrm{NL} \rightarrow \mathrm{N}$ distributions in red, and the $\mathrm{U} \rightarrow \mathrm{N}$ distributions in black. The dashed green lines are the exponential fits to the the $\mathrm{U} \rightarrow \mathrm{NL}$ distributions. }   
\label{surv_all}
\end{figure}

As temperature increases, the TS slightly shifts towards the unfolded states, extending the basin of native-like states, Figs. \ref{all_0_2}{\bf a,b} and \ref{all_0_3}{\bf a,b}. The FPT distributions remain similar to those for $T=0.1$ (Figs. \ref{all_0_2}{\bf c} and \ref{all_0_3}{\bf c}), i.e., the $\mathrm{U} \rightarrow \mathrm{NL}$ and $\mathrm{NL} \rightarrow \mathrm{N}$ distributions are essentially single-exponential, whereas the $\mathrm{U} \rightarrow \mathrm{N}$ distribution has an exponential decay only at relatively long times. Figure \ref{surv_all}{\bf a-c} also shows these distributions in the form of survival probabilities, including the approximation of the $\mathrm{U} \rightarrow \mathrm{NL}$ distributions by exponential functions. The contribution of the $\mathrm{NL} \rightarrow \mathrm{N}$ trajectories to the overall $\mathrm{U} \rightarrow \mathrm{N}$ FPT distribution becomes dominant with temperature, so that both the $\mathrm{U} \rightarrow \mathrm{N}$ distribution and its MFPT are determined by the exploration of native-like states rather than by overcoming the free energy barrier; for example, at $T=0.3$, the MFPTs are $\langle t_{\mathrm{U} \rightarrow \mathrm{NL}}\rangle \approx 85$,  $\langle t_{\mathrm{NL} \rightarrow \mathrm{N}}\rangle \approx 235$, and  $\langle t_{\mathrm{U} \rightarrow \mathrm{N}}\rangle \approx 320$. The distributions of the states in the native ensembles (the inserts in Figs. \ref{all_0_2}{\bf c}-\ref{all_0_3}{\bf c}) remain close to that at $T=0.1$ (Fig. \ref{all_0_1}{\bf c}), except that they are not well-centered at the minima of the native-like basins as in Fig. \ref{all_0_1}{\bf c}, because the minima shift towards the unfolded states along with the TSs.

\begin{figure}\centering%
\resizebox{0.7\linewidth}{!}{ \includegraphics*{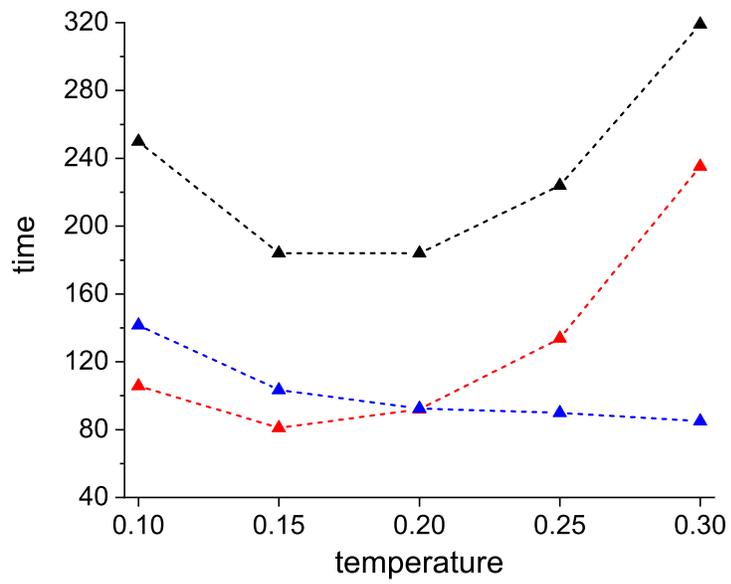}}%
\caption{The MFPTs for the simualted $\mathrm{U} \rightarrow \mathrm{NL}$ (blue), $\mathrm{NL} \rightarrow \mathrm{N}$ (red), and $\mathrm{U} \rightarrow \mathrm{N}$ (black) trajectories. The dashed lines are to guide the eye.}  
\label{u_shape}
\end{figure}

Figure \ref{u_shape} presents the temperature-dependent MFPTs for the simulated $\mathrm{U} \rightarrow \mathrm{NL}$, $\mathrm{NL} \rightarrow \mathrm{N}$ and $\mathrm{U} \rightarrow \mathrm{N}$ trajectories. The $\mathrm{U} \rightarrow \mathrm{NL}$ time decreases across the entire temperature range, while the $\mathrm{NL} \rightarrow \mathrm{N}$ time first decreases and then rapidly grows. As a result, the overall MFPT $\langle t_{\mathrm{U} \rightarrow \mathrm{N}}\rangle=\langle t_{\mathrm{U} \rightarrow \mathrm{NL}}\rangle+\langle t_{\mathrm{NL} \rightarrow \mathrm{N}}\rangle$, as a function of temperature, becomes U-shaped, as is found in theoretical models \cite{socci1996diffusive,karlpus1997levinthal,chekmarev2013protein} and experiments \cite{oliveberg1995negative}. From this follows that the ascending, high-temperature branch of the U-curve should be associated with the increase of the $\mathrm{NL} \rightarrow \mathrm{N}$ time the protein spends among native-like states (Fig. \ref{surv_all}{\bf a-c}).  

It is interesting to compare the simulated times with the times predicted by reaction-rate theory on the basis of calculated FEPs. Specifically, we can use the Kramers rate formula in the strong friction limit \cite{kramers1940brownian,berne1988classical}, which has been previously employed to calculate folding times \cite{kubelka2004protein,kubelka2006sub}. Our analysis is somewhat similar to that for the folding of a 27-bead lattice protein \cite{socci1996diffusive}. The mean time of transitions over the free energy barrier is determined as
\begin{equation}\label{eq2}%
\langle t_{\mathrm{U} \rightarrow \mathrm{NL}}\rangle=\frac{2 \pi T}{D_{\mathrm{TS}} (F_{\mathrm{U}}^{''} F_{\mathrm{TS}}^{''})^{1/2}}\exp(\Delta F/T)
\end{equation}
where $F_{\mathrm{U}}^{''}$ and $F_{\mathrm{TS}}^{''}$ are the second order derivatives of the free energy with respect to $N_{\mathrm{nat}}$ at the bottom of the basin of unfolded states and the top of the TS barrier, respectively, $D_{\mathrm{TS}}$ is the diffusion coefficient at the TS, and $\Delta F$ is the height of the TS barrier measured from the bottom of the unfolded state basin. The diffusion coefficient was calculated directly, although the autocorrelation time of $N_{\mathrm{nat}}$ could also be employed \cite{socci1996diffusive}. Specifically, as the MD trajectory reached the TS ($N_{\mathrm{nat}}=N_{\mathrm{nat}}^{\mathrm{TS}}$), the time-dependent square deviation from the TS, $R^{2}(t)=[N_{\mathrm{nat}}(t)-N_{\mathrm{nat}}^{\mathrm{TS}}]^2$, was calculated, and the diffusion coefficient was determined as $D=(1/2)d \langle R^{2}(t) \rangle/dt$, where the angle brackets stand for the ensemble average. In general, the value of the diffusion coefficient is position-dependent in protein folding\cite{best2010coordinate}. As can be seen from Figs. \ref{all_0_1}{\bf{d}}, \ref{all_0_2}{\bf{d}} and \ref{all_0_3}{\bf{d}}, there are two time intervals where the $\langle R^{2} (t) \rangle$ changes with time linearly, and thus the diffusion coefficient can be considered constant for each of these time intervals. At short times ($t < 0.2$), $D_{\mathrm{TS}} \sim 10$, but at longer times ($t > 1$ in Fig. \ref{all_0_1}{\bf{d}} and $t > 4$ in Fig. \ref{all_0_3}{\bf{d}}), it is one order of magnitude smaller (by $\approx 30$ times). At short times, the deviation from TS, $\Delta N_{\mathrm{nat}}=\langle R^{2} (t) \rangle^{1/2}$, is 2 units or less, i.e., the protein does not leave a close vicinity of the TS. In contrast, at longer times, $\Delta N_{\mathrm{nat}}$  can be as large as 4 units, which indicates that the protein moves away from the TS towards the bottom of one of the basins (see Figs. \ref{all_0_1}{\bf{b}}, \ref{all_0_2}{\bf{b}} and \ref{all_0_3}{\bf{b}}). The simulations show that in this case the protein mostly explores the basin of native-like states rather than moves around the TS (Supporting Information, Fig. S1). This suggests that the linear behavior of $\langle R^{2} (t) \rangle$ at longer times should be associated with an intra-basin diffusion rather than with the transitions over the TS barrier, i.e., with an inter-basin diffusion. Consequently, the value of the diffusion coefficient at small times was employed as the $D_{\mathrm{TS}}$ in Eq. (\ref{eq2}). Because of the discrete nature of the reaction coordinate, we used two methods to calculate the derivative $F_{\mathrm{U}}^{''}$ and $F_{\mathrm{TS}}^{''}$. In one method, the derivatives were obtained by the approximation of the $F(N_{\mathrm{nat}})$ with the second order polynomials in the vicinity of the points corresponding to the bottom of the unfolded basin ($F_{\mathrm{U}}^{''}$) and the top of the TS barrier ($F_{\mathrm{TS}}^{''}$). In the other method, the derivatives were calculated directly, as the three-point finite differences of $F(N_{\mathrm{nat}})$ at those points. The calculated parameters for Eq. (\ref{eq2}) are tabulated in Table  \ref{tbl:parameters}, including those for two intermediate temperatures, $T=0.15$ and $T=0.25$ (Supporting Information, Figs. S3 - S4). 

\begin{table}
  \caption{Parameters to calculate the $\mathrm{U} \rightarrow \mathrm{NL}$ transition time with the Kramers formula}
  \label{tbl:parameters}
  \begin{tabular}{llllll}
    \hline
     $T$ & $0.1$ & $0.15$ & $0.2$ & $0.25$ & $0.3$\\
    \hline
$\Delta F$ & 0.26 & 0.33 & 0.36 & 0.38 & 0.34\\
$F_{\mathrm{U}}^{''}\textsuperscript{\emph{a}}$ & 0.27 & 0.29 & 0.27 & 0.23 & 0.24\\
$F_{\mathrm{U}}^{''}\textsuperscript{\emph{b}}$ & 0.29 & 0.29 & 0.25 & 0.23 & 0.23\\
$F_{\mathrm{TS}}^{''}\textsuperscript{\emph{a}}$ & 0.14 & 0.19 & 0.21 & 0.26 & 0.32\\
$F_{\mathrm{TS}}^{''}\textsuperscript{\emph{b}}$ & 0.10 & 0.19 & 0.22 & 0.28 & 0.32\\
$D_{\mathrm{TS}}$ & 4.5 & 6.0 & 6.5 & 8.0 & 9.0\\
    \hline
  \end{tabular}

\textsuperscript{\emph{a}} from the polynomial approximation.\\
\textsuperscript{\emph{b}} calculated as the three-point finite difference.
\end{table}

\begin{figure}\centering%
\resizebox{0.7\linewidth}{!}{ \includegraphics*{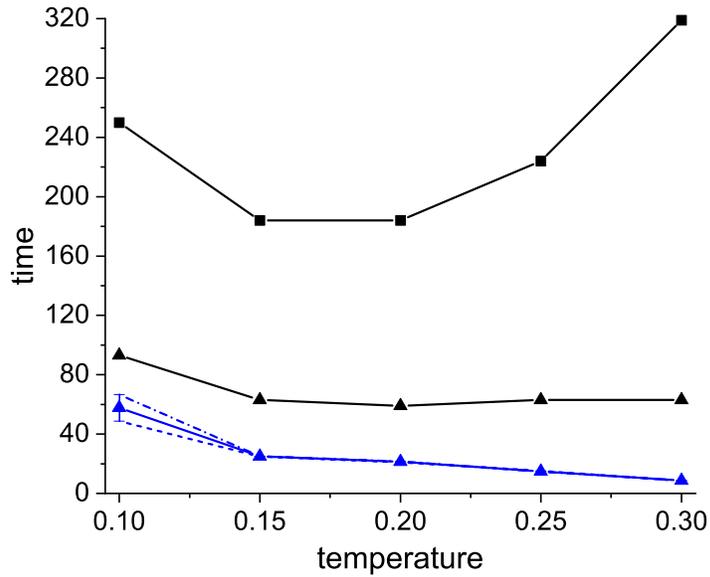}}%
\caption{The comparison of the Kramers times, Eq. (\ref{eq2}), with the simulated times. The black squares are for the $\langle t_{\mathrm{U} \rightarrow \mathrm{N}} \rangle$ times from simulations, the black triangles denote the $\langle t_{\mathrm{U} \rightarrow \mathrm{NL}} \rangle$ times calculated from the slopes of the simulated $\mathrm{U} \rightarrow \mathrm{NL}$ distributions, and the blue triangles are for $\langle t_{\mathrm{U} \rightarrow \mathrm{NL}} \rangle$ times from Eq. (\ref{eq2}) with the average values of $F_{\mathrm{U}}^{''}$ and $F_{\mathrm{TS}}^{''}$ (the dashed and dash-dotted blue lines indicate the results for $F_{\mathrm{U}}^{''}$ and $F_{\mathrm{TS}}^{''}$ obtained by the polynomial approximation of the FEP and calculated by finite-differences, respectively). In all cases, the lines are to guide the eye.}  
\label{times_cmp}
\end{figure}

Figure \ref{times_cmp} compares the calculated Kramers times with the MFPTs obtained in the simulations. The steep rise in the $\mathrm{U} \rightarrow \mathrm{NL}$ distributions (Figs. \ref{all_0_1}{\bf{c}} - \ref{all_0_3}{\bf{c}}) should be discarded, because, as has been mentioned, it does not represent the times to overcome the free energy barrier. Therefore, the $\mathrm{U} \rightarrow \mathrm{NL}$ MFPTs were determined as the rates of decay, which are inverse MFPTs, in the exponential fits to the $\mathrm{U} \rightarrow \mathrm{NL}$ distributions, as is shown in Fig. \ref{surv_all}{\bf a-c}. From Fig. \ref{times_cmp} it is seen that although the difference between the Kramers and simulated $\mathrm{U} \rightarrow \mathrm{NL}$ times is considerable (from approximately 2 times at $T=0.1$ to 8 times at $T=0.3$), these times are much shorter than the $\mathrm{U} \rightarrow \mathrm{N}$ times and exhibit a similar temperature dependence. It can also be noticed that the difference decreases as one of the basic conditions to derive Eq. (2), $\Delta F/T \gg 1$ \cite{kramers1940brownian,berne1988classical}, is better fulfilled ($\Delta F/T=2.5$ at $T=0.1$ to $\Delta F/T=1.1$ at $T=0.3$). A similar behavior of the simulated and Kramers times is observed for the other values of the friction constant, except that the difference in the absolute values of the times is smaller ($\gamma=3M/\tau$, Supporting Information, Figs. S9) or larger ($\gamma=50M/\tau$, Supporting Information, Figs. S14). 

In general, the overall folding kinetics of the present protein are found to be quite robust to the variation of factors that govern the kinetics. In agreement with earlier results for $\alpha$-helical and $\beta$-sheet proteins \cite{klimov1997viscosity}, the increase of the friction constant just slows the rate of the process (Supporting Information, Figs. S5 - S8 for $\gamma=3M/\tau$, and Figs. S10 - S13 for $\gamma=50M/\tau$). A smaller  RMSD threshold $\sigma_{\mathrm{nat}}^{\mathrm{thrh}}$ to terminate folding trajectories, such as the pairwise C$_{\alpha}$ RMSD for the ensemble of the NMR solution structures ($\approx 0.65\mathrm{\AA}$) \cite{andersen2006minimization}, mostly increases the time the protein spends in the basin of native-like states (Supporting Information, Fig. S16). A similar effect is observed when the trajectories are alternatively terminated by the condition that the current number of native contacts $N_{\mathrm{nat}}$ should be equal to that in the native state $N_{\mathrm{nat}}^{\mathrm{NAT}}=27$ because the probability to find a state with $N_{\mathrm{nat}}=27$ turns to be smaller than that to find a state with the $\sigma_{\mathrm{nat}}< 1.0 \mathrm{\AA}$ (Supporting Information, Figs. S17). All other characteristic properties of folding are well-conserved, i.e., the $\mathrm{U} \rightarrow \mathrm{NL}$ and $\mathrm{NL} \rightarrow \mathrm{N}$ distributions remain essentially single-exponential, the overall $\mathrm{U} \rightarrow \mathrm{N}$ distribution is close to exponential, and,  as temperature increases, the $\mathrm{U} \rightarrow \mathrm{N}$ MFPT is largely determined by the $\mathrm{U} \rightarrow \mathrm{NL}$ time. 

\section{Conclusion}
\label{sec:4}%
Using a coarse-grained protein model, we have performed an extensive MD simulation of folding of a $\beta$-hairpin protein, a benchmark two-state folder. Each MD trajectory to reach the protein native state from an unfolded state was divided into two parts - one to pass from the unfolded state to a native-like state by overcoming the free energy barrier that separates these states, and the other to explore the basin of native-like states until the native state is achieved. It has been found that the first-passage time (FPT) distributions for both segments of the trajectories are essentially single-exponential. The resulting FTP distribution to reach the native state generally has a steep rise at short times and an exponential decay at longer times. The deviation of this distribution from a single-exponential one is determined by the relation between the MFPTs for the constituting trajectories, i.e., the smaller one of the MFPTs is in comparison to the other, the closer the resulting distribution is to the exponential distribution for the trajectories with the longer MFPT. Accordingly, if a protein explores native-like states for a time much longer than the time to overcome the free energy barrier, the resulting FPT distribution can have exponential decay but the MFPT will be largely determined by the time to find the native state among the native-like ones rather than by the height of the free energy barrier. It has been found that this effect is characteristic of high temperatures and becomes more pronounced as temperature increases. The time to overcome the free energy barrier decreases across the entire temperature range, while the time to find the native state among native-like ones first decreases and then rapidly grows. This explains the well-known U-shape dependence of the MFPT on temperature that is found in theoretical models and experiments. Based on the free energy profiles constructed from the simulated MD trajectories, the mean times to pass from the basin of unfolded states to the basin of native-like states has been calculated using the Kramers rate formula. It has been found that these times are in reasonable agreement with the corresponding times obtained by the simulation and are far shorter than the MFPTs to reach the native state. All findings are robust to the variation of factors that govern the kinetics, such as the condition to determine the native state and the strength of protein friction against the surroundings.          

\section{Supporting Information}
Time-dependent deviations from the TS; FPT distribution for the model system; FESs, FEPs, FPT distributions, time-dependent mean-square deviations from the TS, temperature-dependent simulated MFPTs and Kramers times for different values of the friction constant; FESs, FEPs, FPT distributions and time-dependent mean-square deviations from the TS for different conditions to terminate folding trajectories.  

\begin{acknowledgement}
I thank Dmitriy Chekmarev for valuable comments on the manuscript. A support from the Russian Ministry of Education and Science is acknowledged.
\end{acknowledgement}

\clearpage

\begin{figure}\centering%
\resizebox{0.49\linewidth}{!}{ \includegraphics*{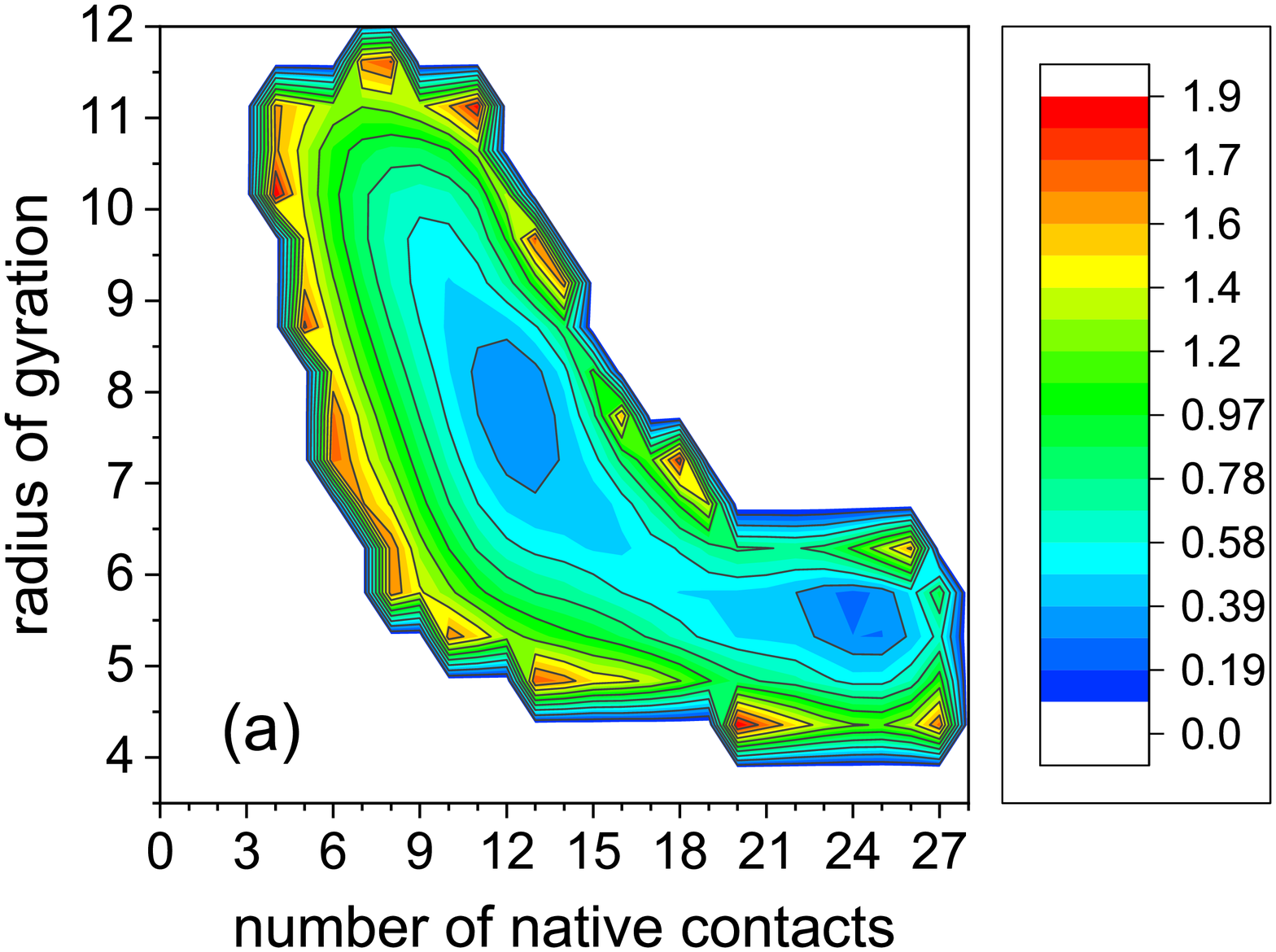}}%
\hfill
\resizebox{0.49\linewidth}{!}{ \includegraphics*{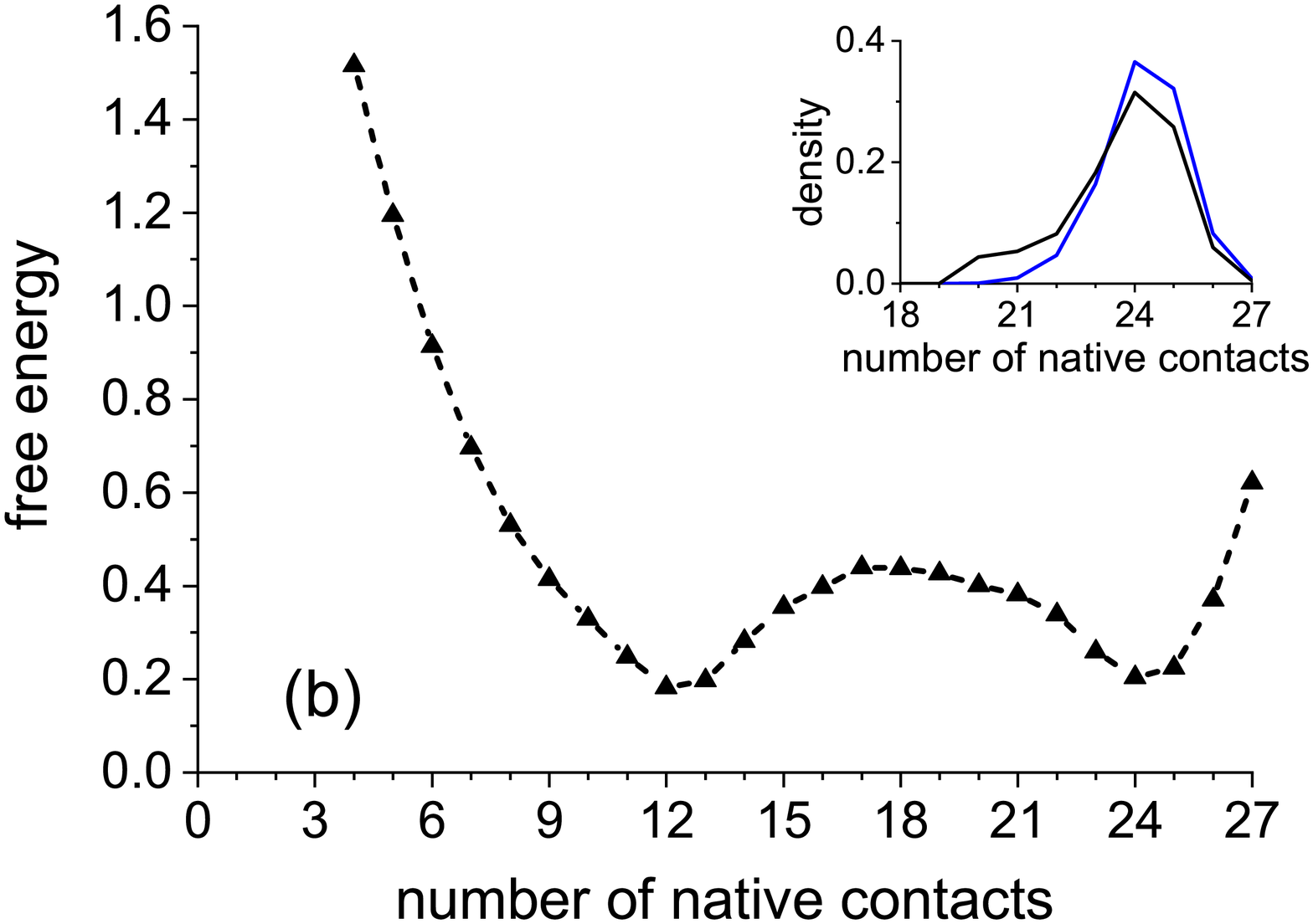}}%
\hfill
\resizebox{0.49\linewidth}{!}{ \includegraphics*{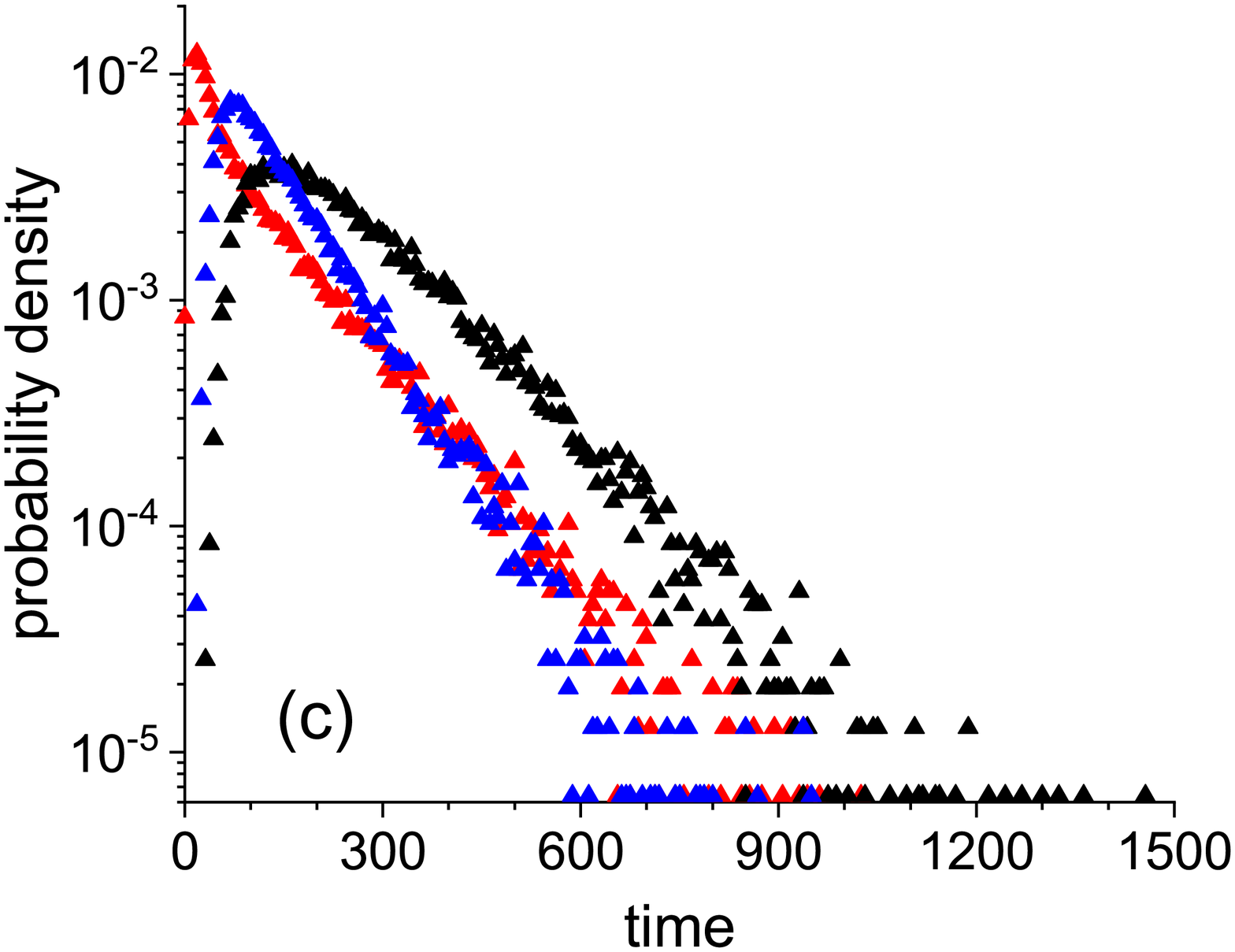}}%
\hfill
\resizebox{0.49\linewidth}{!}{ \includegraphics*{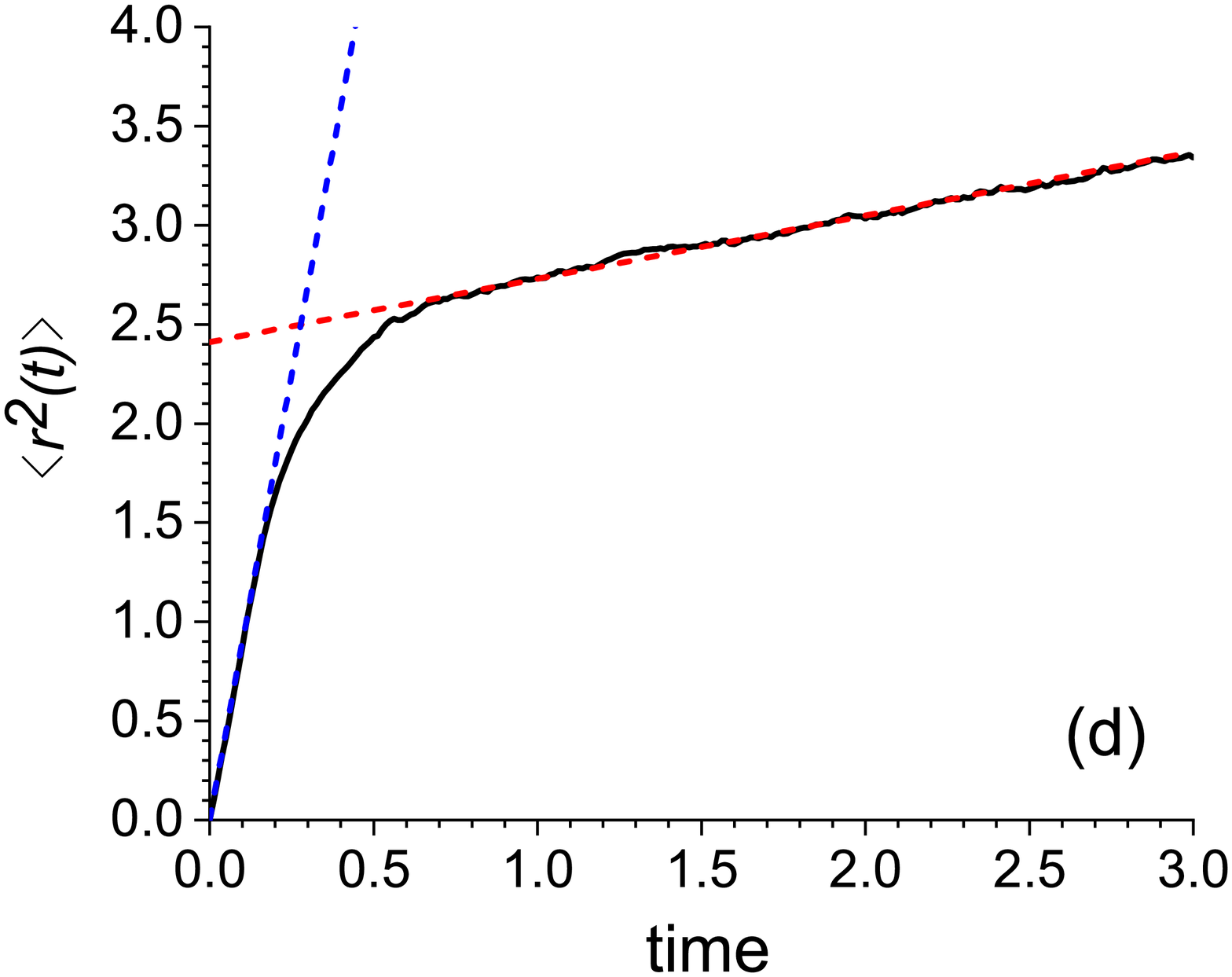}}%
\caption{TOC graphic. }   
\label{toc}
\end{figure}

\end{document}


\clearpage

{\centering \section{Deviation from the Transition State}}
\begin{figure}\centering%
\resizebox{0.75\linewidth}{!}{ \includegraphics*{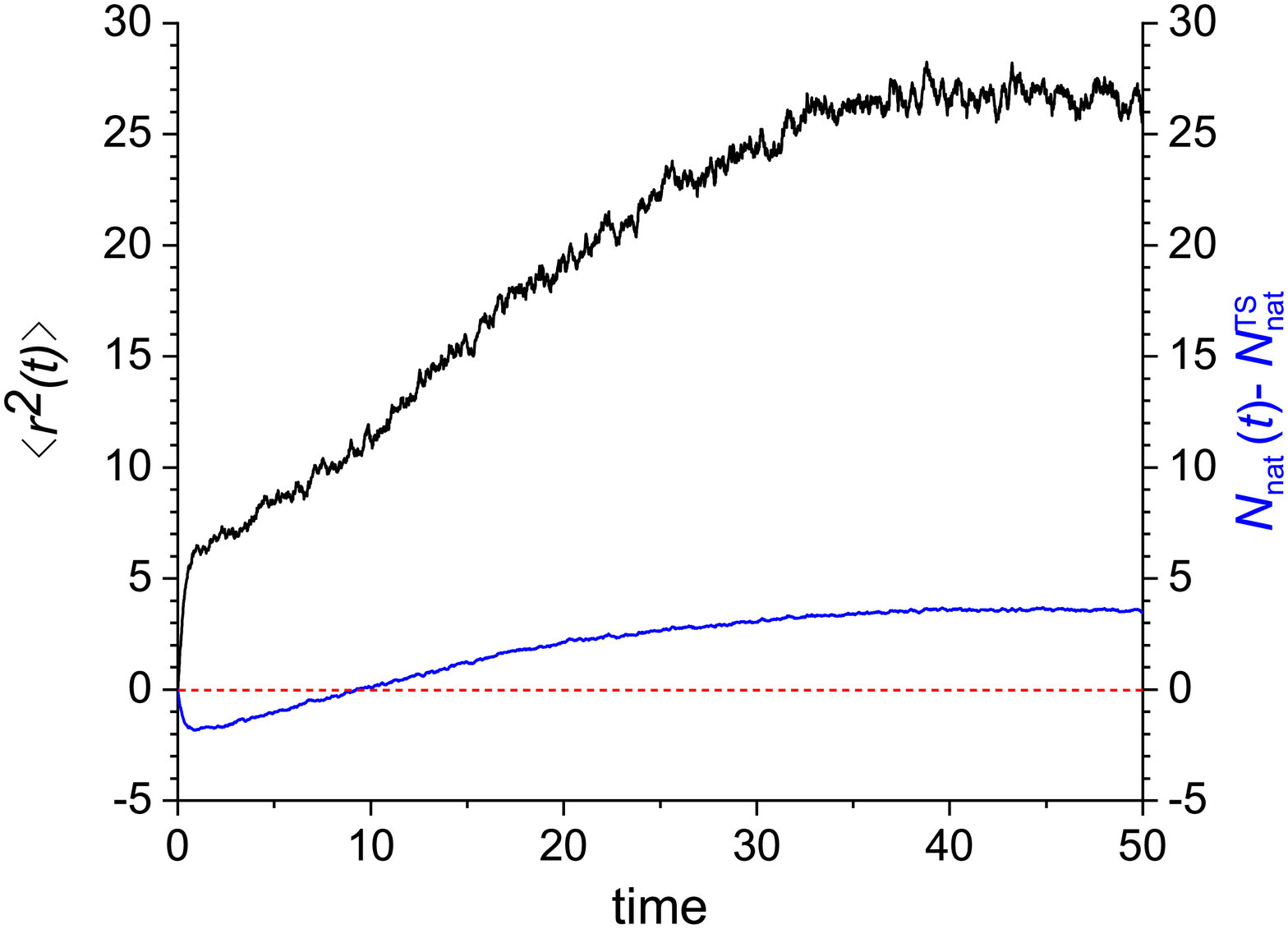}}%
\caption{$T=0.2$,  $\gamma=10M/\tau $, and ten thousand MD trajectories. The mean square (black curve) and mean (blue curve) deviations from the transition state.}   
\label{g10_r2_dn_av}
\end{figure}
\clearpage

{\centering \section{A Simple Model for Single-Exponential First-Passage Time Distribution}}
\begin{figure}\centering%
\resizebox{0.75\linewidth}{!}{ \includegraphics*{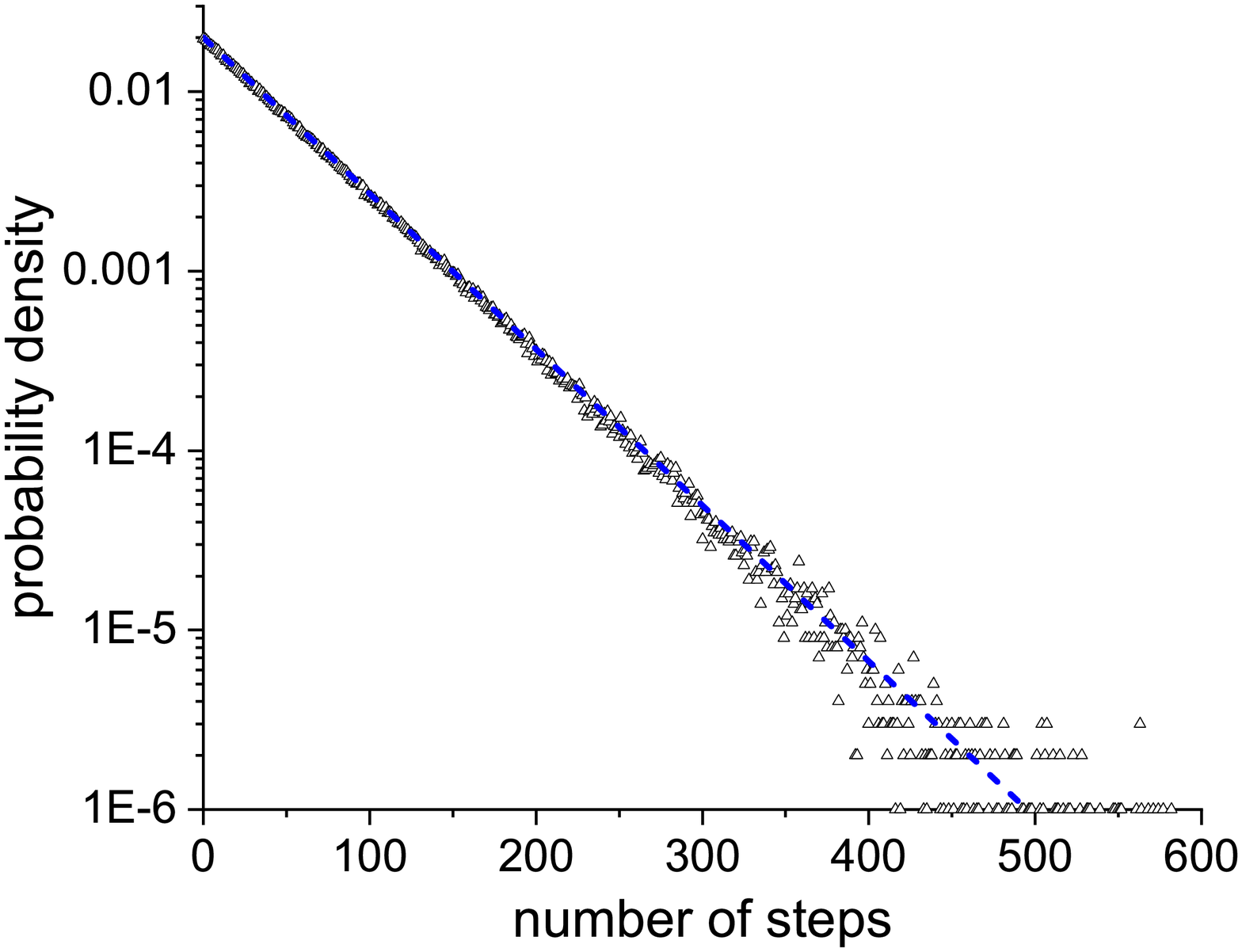}}%
\caption{A simulated distribution of first-passage times. Random number generator with a uniform distribution of the numbers between 0 and 1 was used. In the ensemble of $10^6$ trajectories, each trajectory was started from a random number and proceeded through the numbers until the value of $0.7 \pm 0.01$ was achieved. The label corresponds to the simulated trajectories, and the blue dashed line shows an exponential fit to the simulate distribution with the decay rate of 50.0.}  
\label{fpt_random}
\end{figure}
\clearpage

{\centering \section{Friction Constant $\gamma=10M/\tau $}}

\begin{figure}\centering%
\resizebox{0.49\linewidth}{!}{ \includegraphics*{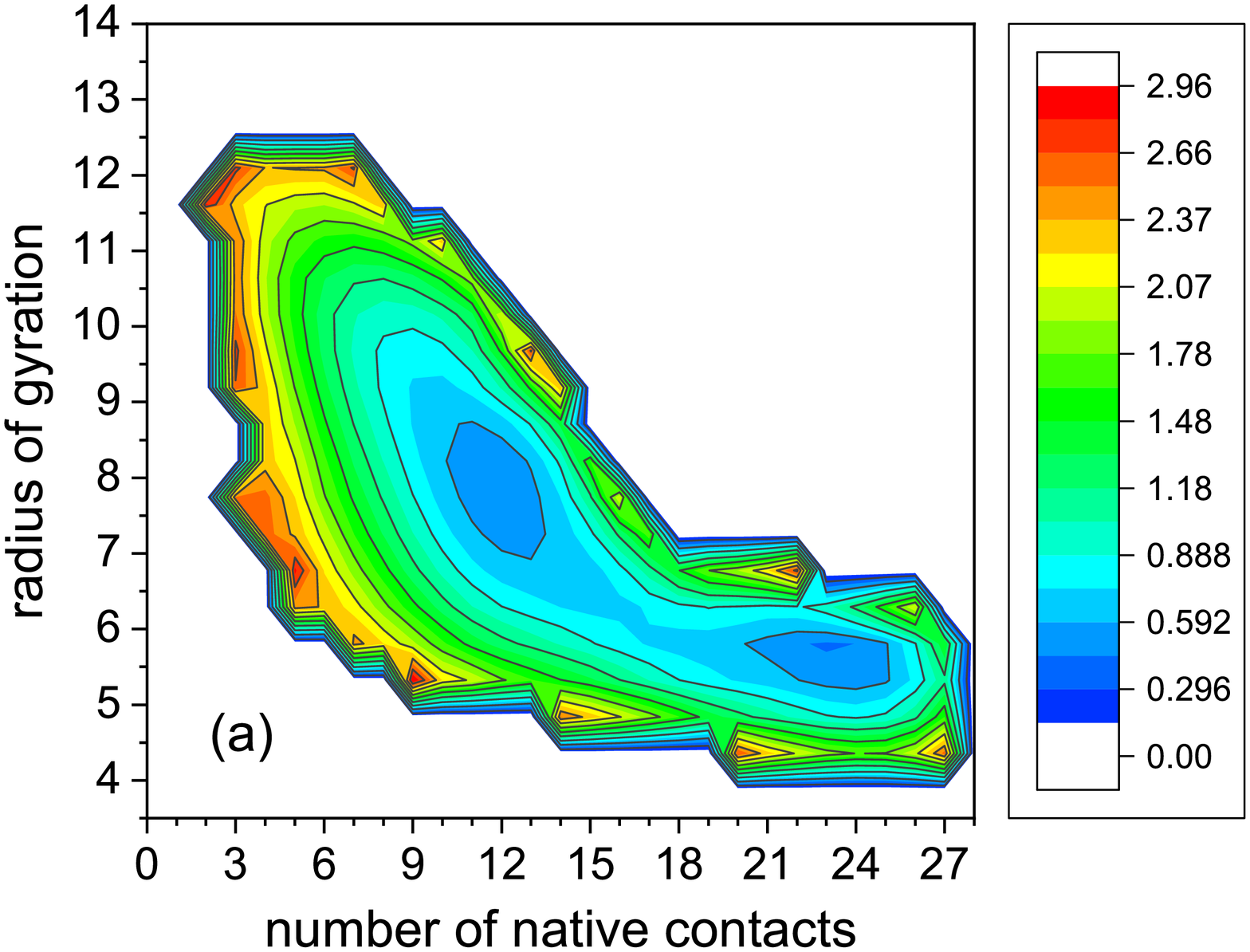}}%
\hfill
\resizebox{0.49\linewidth}{!}{ \includegraphics*{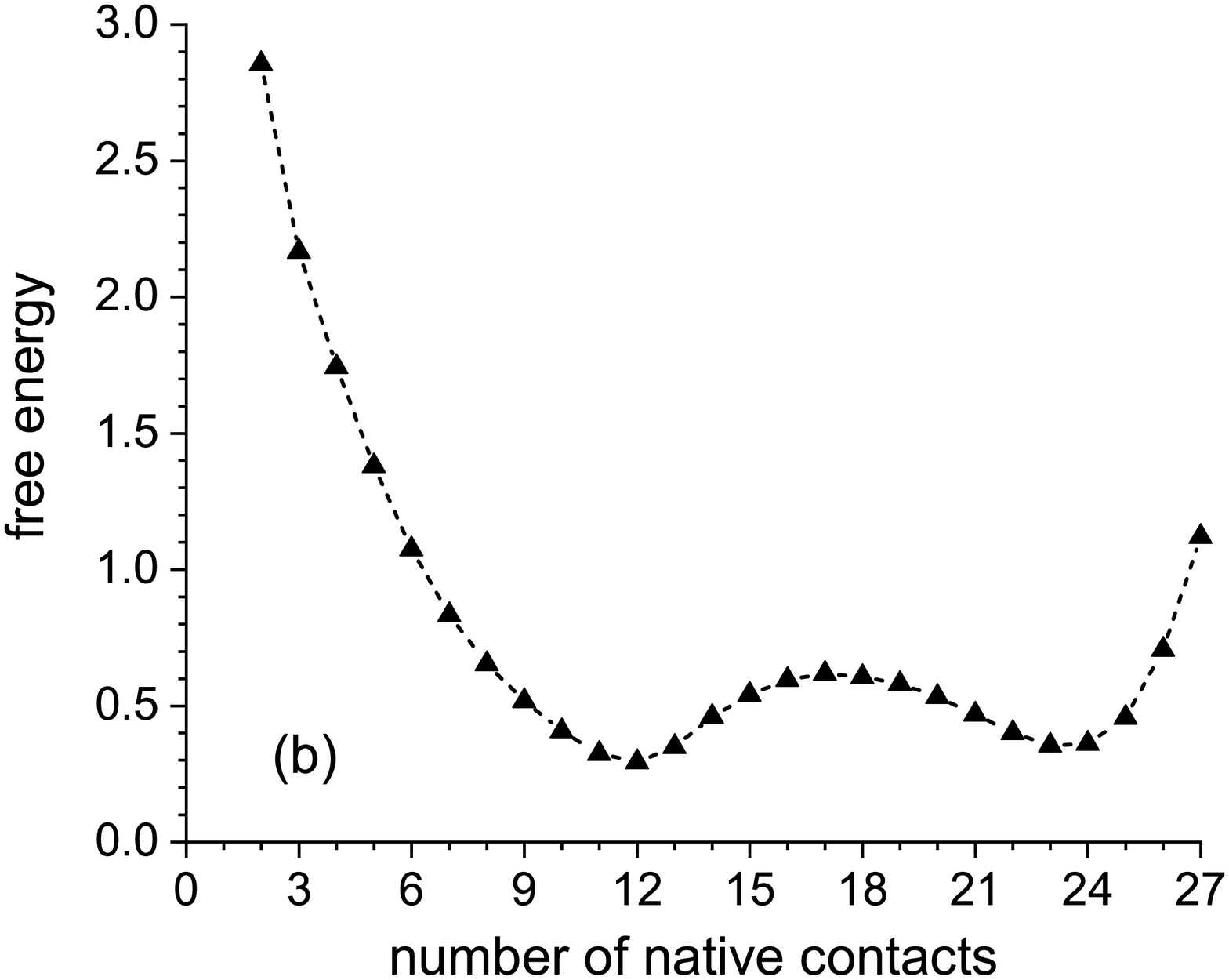}}%
\hfill
\resizebox{0.49\linewidth}{!}{ \includegraphics*{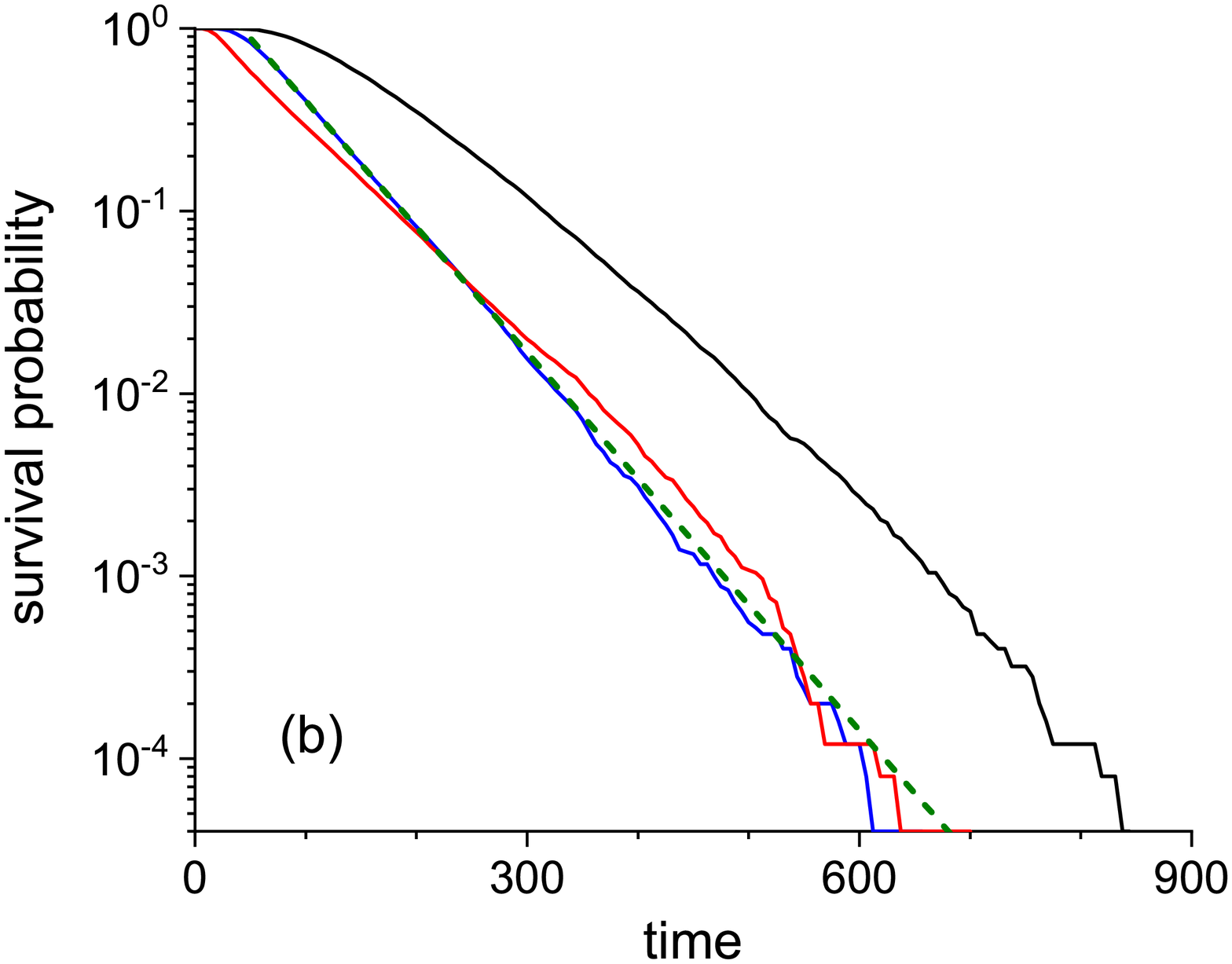}}%
\hfill
\resizebox{0.49\linewidth}{!}{ \includegraphics*{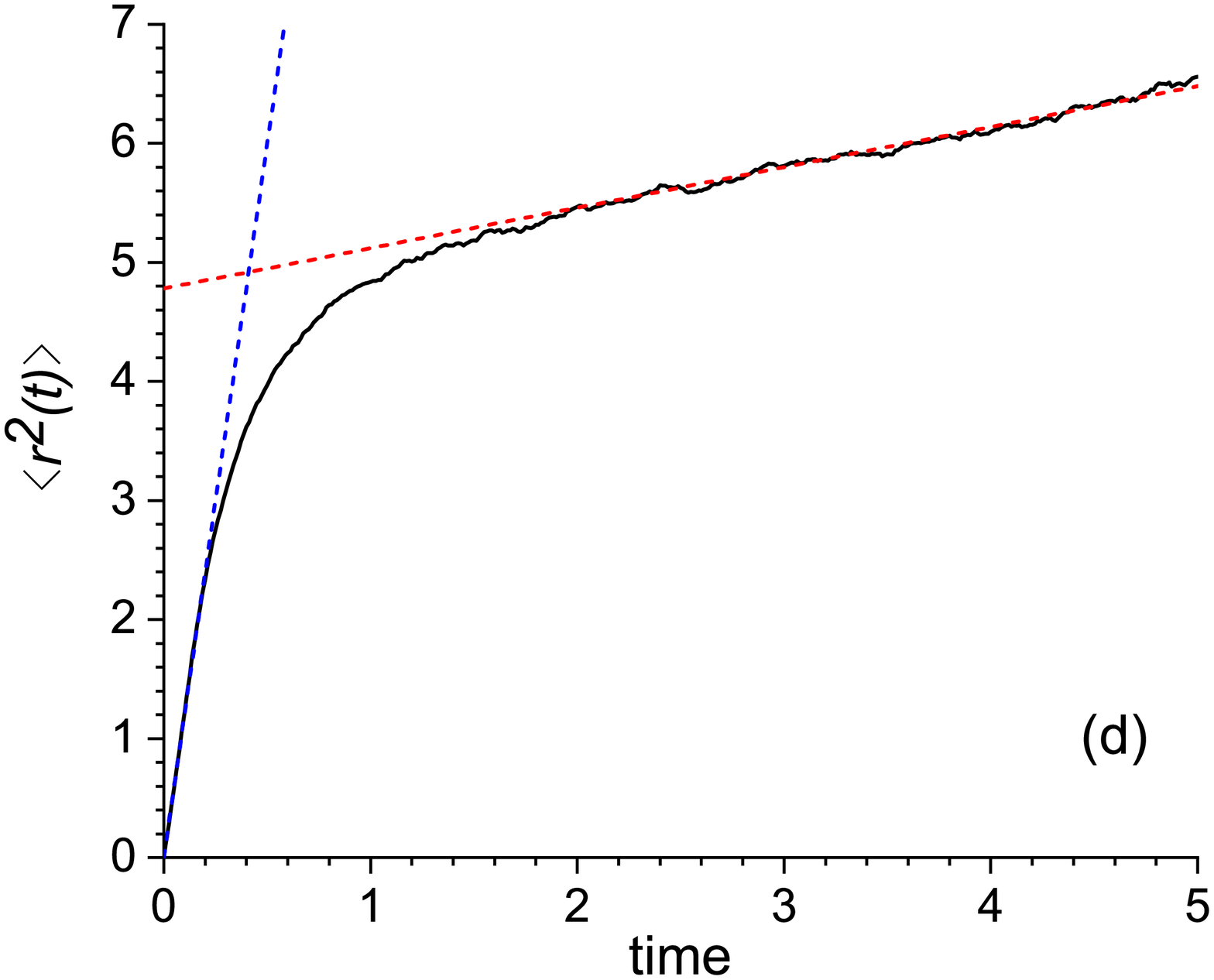}}%
\caption{$T=0.15$. ({\bf{a}}) The free energy surface $F(N_{\mathrm{nat}},R_{\mathrm{g}})$, and ({\bf{b}}) free energy profile $F(N_{\mathrm{nat}})$. ({\bf{c}}) First-passage time distributions in the form of survival probabilities: the $\mathrm{U} \rightarrow \mathrm{NL}$ trajectories (blue), the $\mathrm{NL} \rightarrow \mathrm{N}$ trajectories (red), and the U-N trajectories (black); the dashed green line denotes an exponential fit to the $\mathrm{U} \rightarrow \mathrm{NL}$ distribution. ({\bf{d}}) The time-dependent mean-square deviation from the transition state in the number of native contacts (black curve); the blue and red dashed lines are the linear fits to the curve for short and long times, respectively.}   
\label{g10_all_0_15}
\end{figure}

\begin{figure}\centering%
\resizebox{0.49\linewidth}{!}{ \includegraphics*{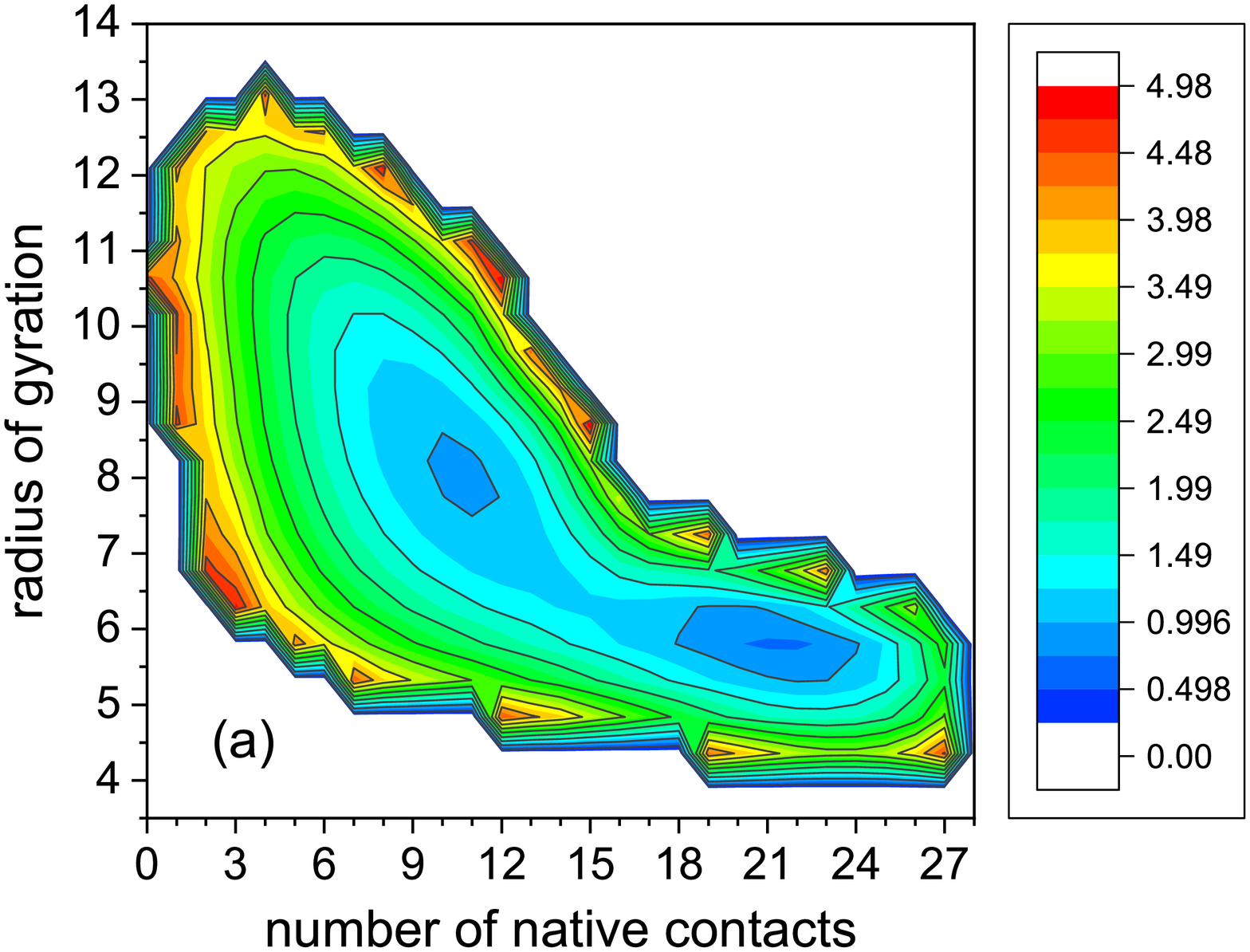}}%
\hfill
\resizebox{0.49\linewidth}{!}{ \includegraphics*{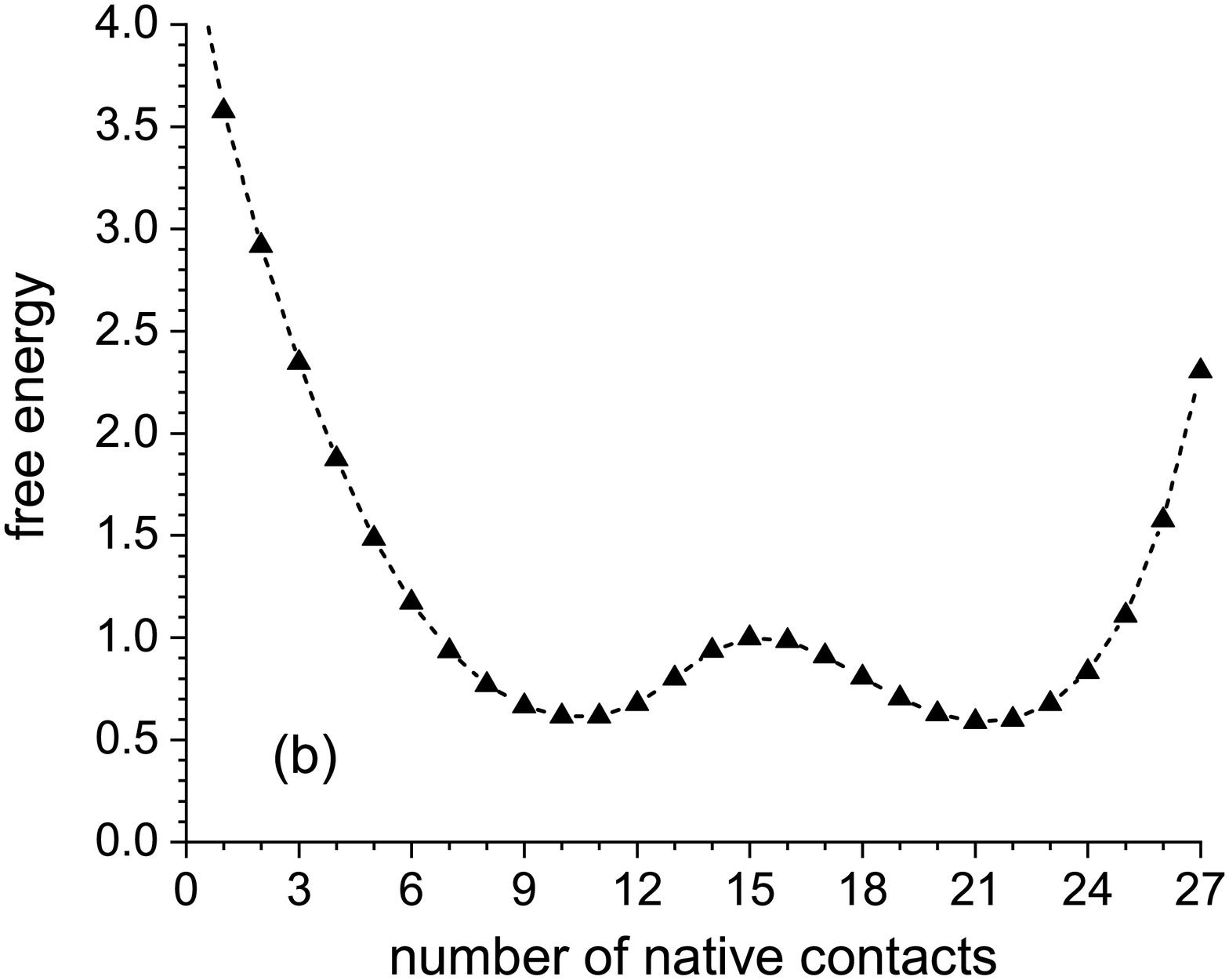}}%
\hfill
\resizebox{0.49\linewidth}{!}{ \includegraphics*{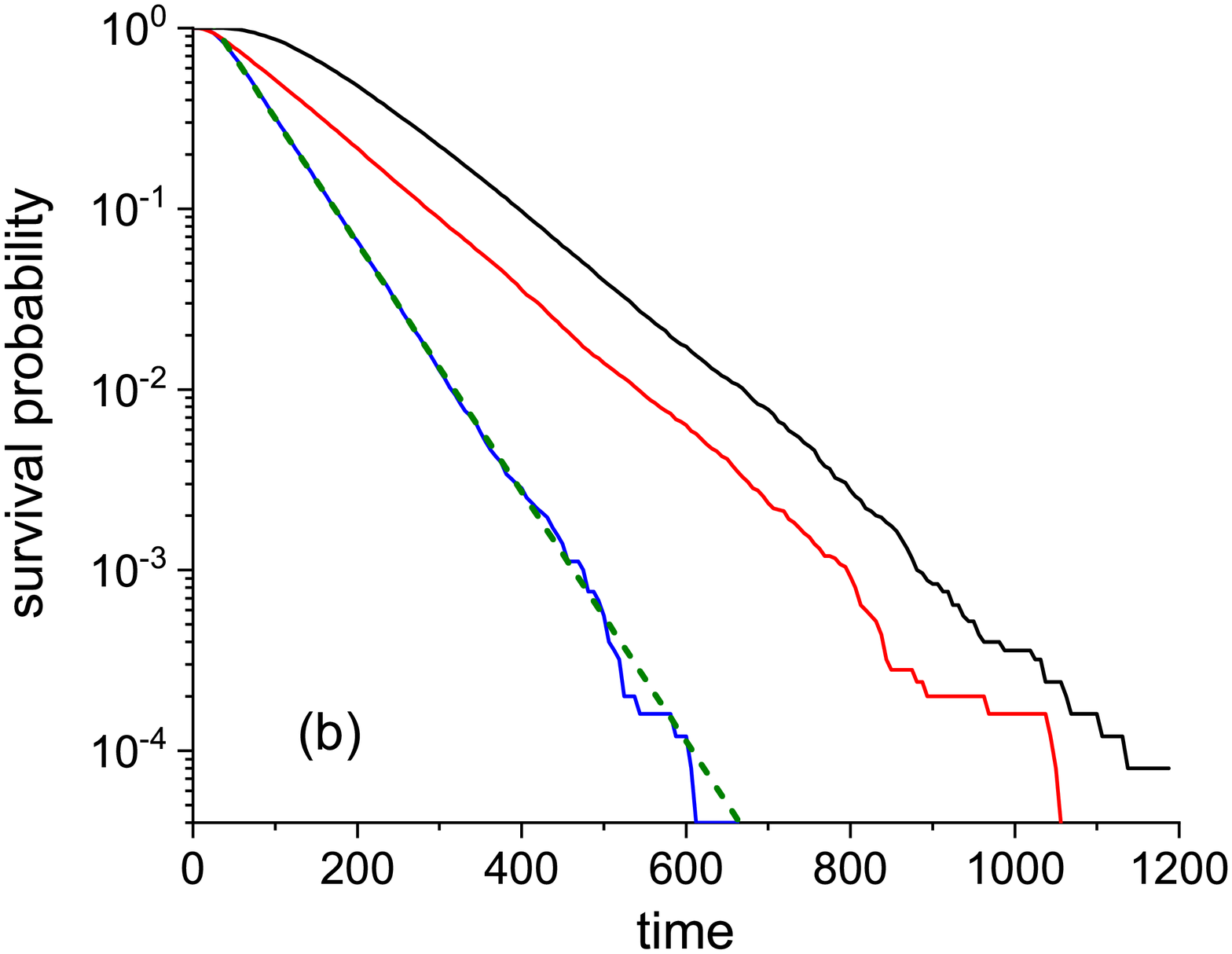}}%
\hfill
\resizebox{0.49\linewidth}{!}{ \includegraphics*{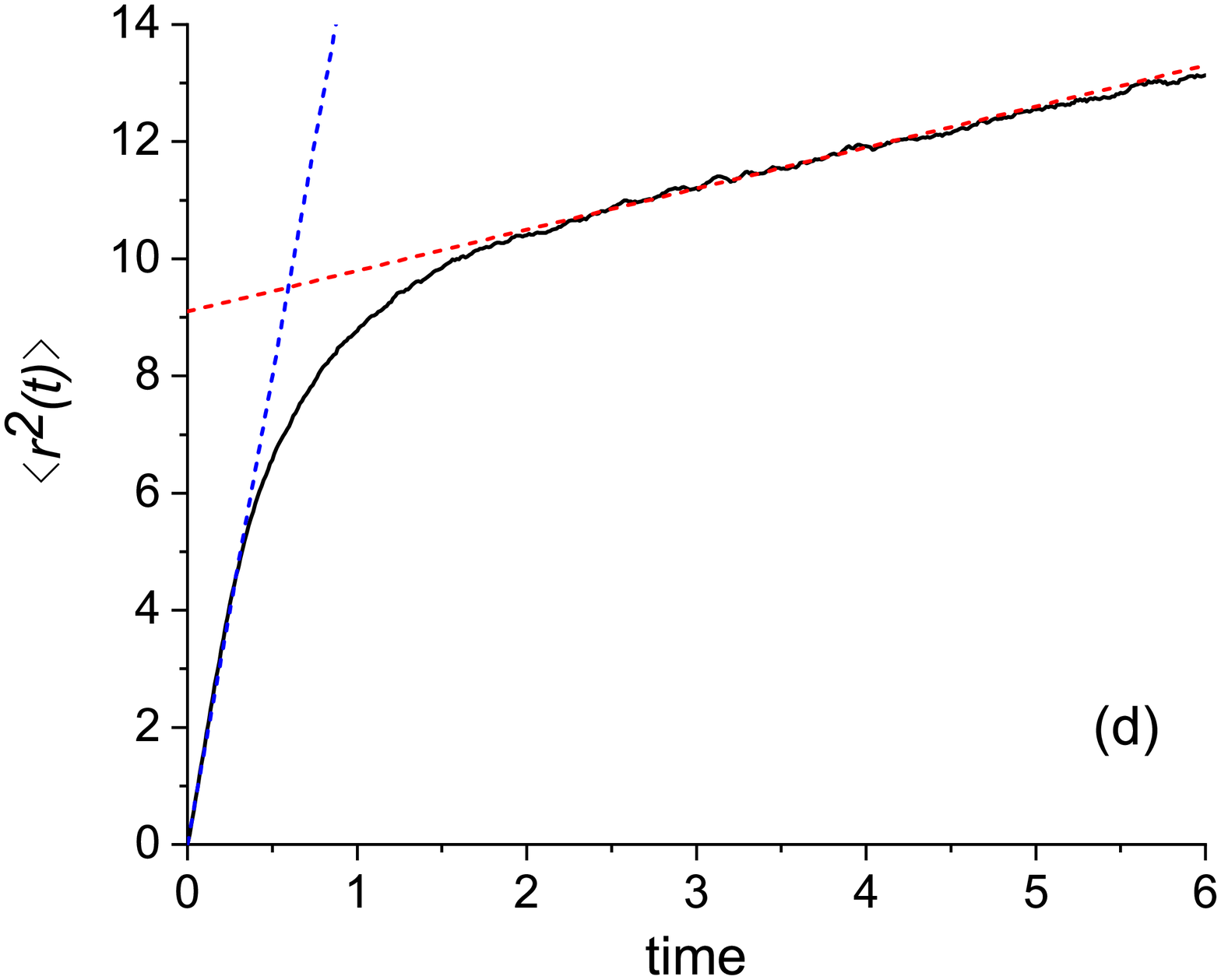}}%
\caption{$T=0.25$. The notations are as in Fig. \ref{g10_all_0_15}.}   
\label{g10_all_0_25}
\end{figure}

\clearpage

\begin{table}
  \caption{Parameters to calculate the $\mathrm{U} \rightarrow \mathrm{NL}$ transition time with the Kramers formula}
  \label{tbl:parameters}
  \begin{tabular}{llllll}
    \hline
     $T$ & $0.1$ & $0.15$ & $0.2$ & $0.25$ & $0.3$\\
    \hline
$\Delta F$ & 0.26 & 0.33 & 0.36 & 0.38 & 0.34\\
$F_{\mathrm{U}}^{''}\textsuperscript{\emph{a}}$ & 0.27 & 0.29 & 0.27 & 0.23 & 0.24\\
$F_{\mathrm{U}}^{''}\textsuperscript{\emph{b}}$ & 0.29 & 0.29 & 0.25 & 0.23 & 0.23\\
$F_{\mathrm{TS}}^{''}\textsuperscript{\emph{a}}$ & 0.14 & 0.19 & 0.21 & 0.26 & 0.32\\
$F_{\mathrm{TS}}^{''}\textsuperscript{\emph{b}}$ & 0.10 & 0.19 & 0.22 & 0.28 & 0.32\\
$D_{\mathrm{TS}}$ & 4.5 & 6.0 & 6.5 & 8.0 & 9.0\\
    \hline
  \end{tabular}

\textsuperscript{\emph{a}} from the polynomial approximation.\\
\textsuperscript{\emph{b}} calculated as the three-point finite difference.
\end{table}

\begin{table}
  \caption{Comparison of folding times}
  \label{tbl:times}
  \begin{tabular}{llllll}
    \hline
     $T$ & $0.1$ & $0.15$ & $0.2$ & $0.25$ & $0.3$\\
    \hline
${\langle t_{\mathrm{\mathrm{U} \rightarrow \mathrm{NL}}}\rangle}\textsuperscript{\emph{a}}$ & 93.0 & 63.0 & 59.0 & 63.0 & 63.0\\
${\langle t_{\mathrm{\mathrm{U} \rightarrow \mathrm{NL}}}\rangle}\textsuperscript{\emph{b}}$ & 48.7 & 24.9 & 21.0 & 15.2 & 8.4\\
$\langle t_{\mathrm{\mathrm{U} \rightarrow \mathrm{NL}}}\rangle\textsuperscript{\emph{c}}$ & 144.6 & 103.4 & 92.5 & 89.9 & 85.0 \\
$\langle t_{\mathrm{\mathrm{NL} \rightarrow \mathrm{N}}}\rangle\textsuperscript{\emph{c}}$ & 105.8 & 81.0 & 92.0 & 133.8 & 235.3\\
$\langle t_{\mathrm{\mathrm{U} \rightarrow \mathrm{N}}}\rangle\textsuperscript{\emph{c}}$ & 250.4 & 184.4 & 184.5 & 223.7 & 320.3\\
    \hline
  \end{tabular}

\textsuperscript{\emph{a}} calculated from the slope of the simulated $\mathrm{U} \rightarrow \mathrm{NL}$ decay curve.\\
\textsuperscript{\emph{b}} Kramers formula [Eq. (2), the main text] for the average values of $F_{\mathrm{U}}^{''}$ and $F_{\mathrm{TS}}^{''}$ (Table \ref{tbl:parameters}).\\
\textsuperscript{\emph{c}} simulated times.
\end{table}

\clearpage

{\centering \section{Friction Constant $\gamma=3M/\tau $}}

\begin{figure}\centering%
\resizebox{0.49\linewidth}{!}{ \includegraphics*{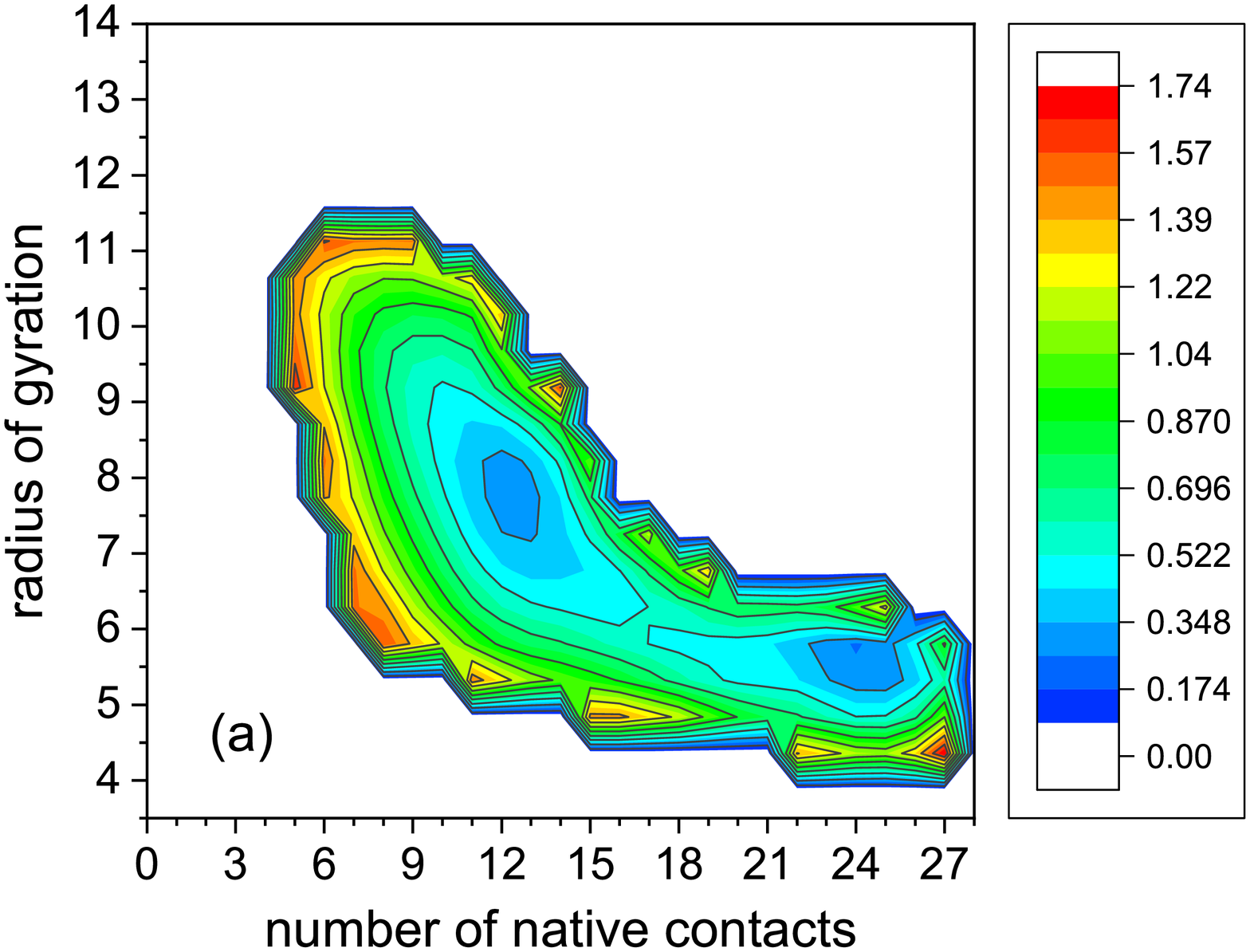}}%
\hfill
\resizebox{0.49\linewidth}{!}{ \includegraphics*{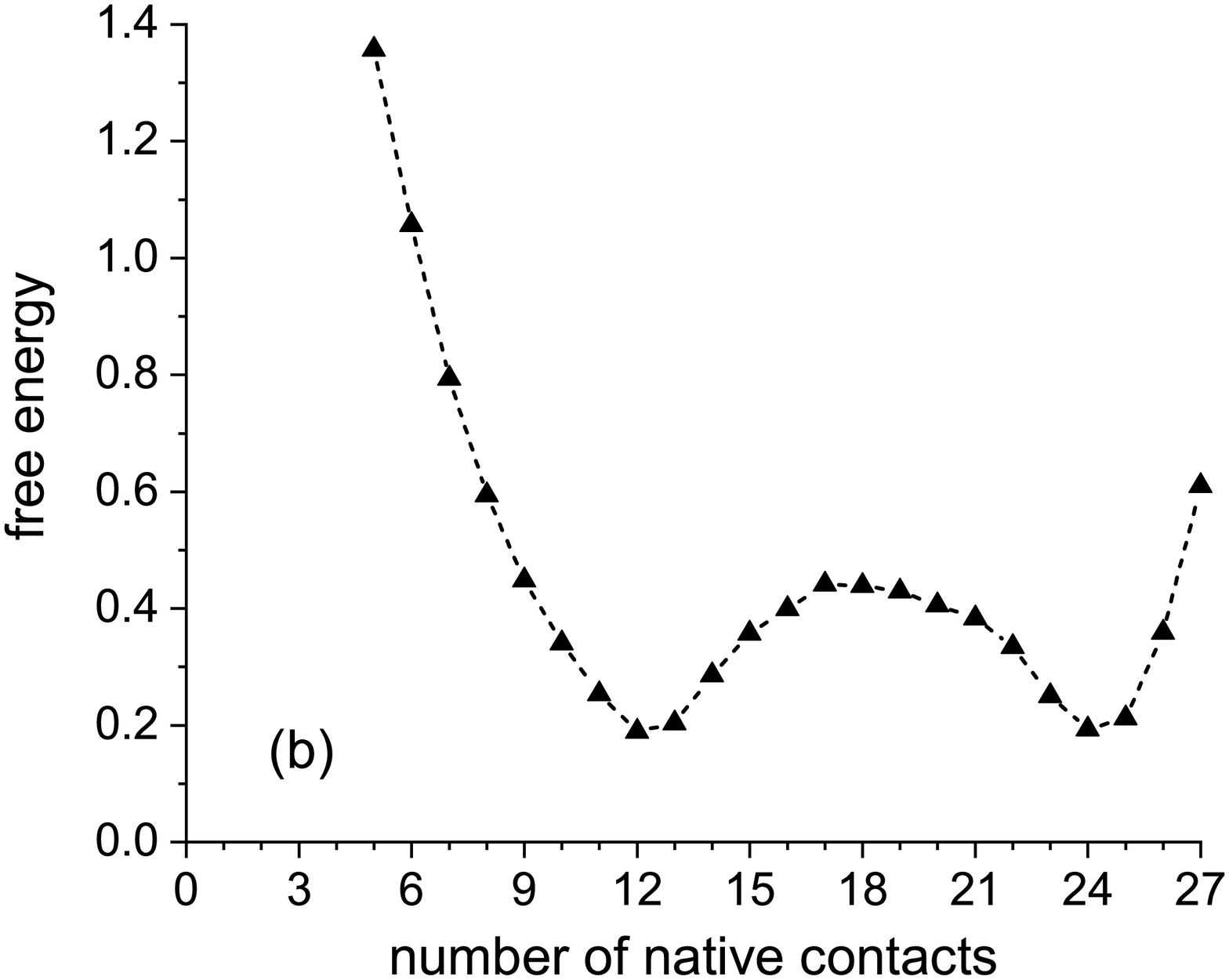}}%
\hfill
\resizebox{0.49\linewidth}{!}{ \includegraphics*{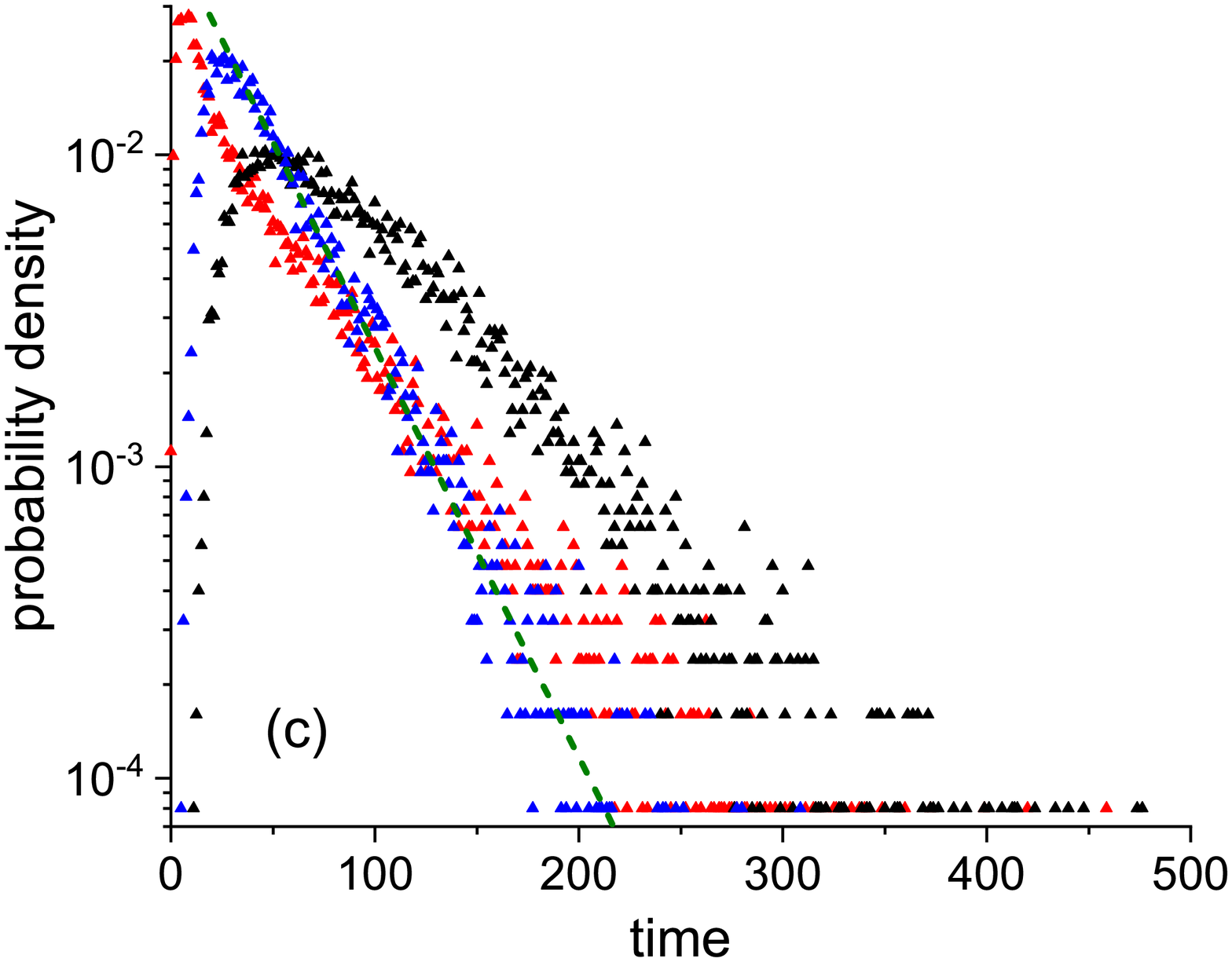}}%
\hfill
\resizebox{0.49\linewidth}{!}{ \includegraphics*{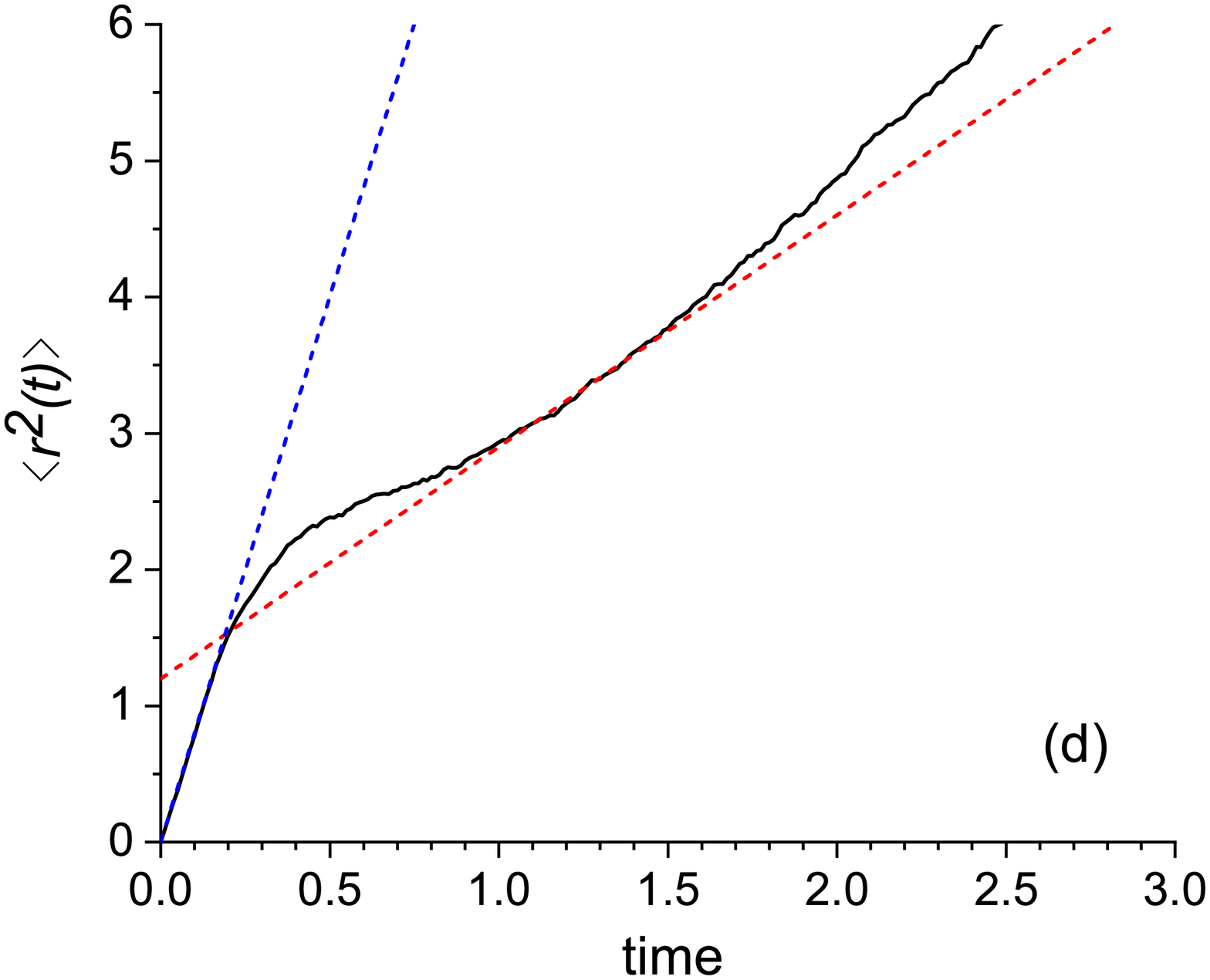}}%
\caption{$T=0.1$. ({\bf{a}}) The free energy surface $F(N_{\mathrm{nat}},R_{\mathrm{g}})$, and ({\bf{b}}) free energy profile $F(N_{\mathrm{nat}})$. ({\bf{c}}) First-passage time distributions: the $\mathrm{U} \rightarrow \mathrm{NL}$ trajectories (blue), the $\mathrm{NL} \rightarrow \mathrm{N}$ trajectories (red), and the U-N trajectories (black); the dashed green line denotes an exponential fit to the $\mathrm{U} \rightarrow \mathrm{NL}$ distribution. ({\bf{d}}) The time-dependent mean-square deviation from the transition state in the number of native contacts (black curve); the blue and red dashed lines are the linear fits to the curve for short and long times, respectively.}   
\label{g3_all_0_1}
\end{figure}

\begin{figure}\centering%
\resizebox{0.49\linewidth}{!}{ \includegraphics*{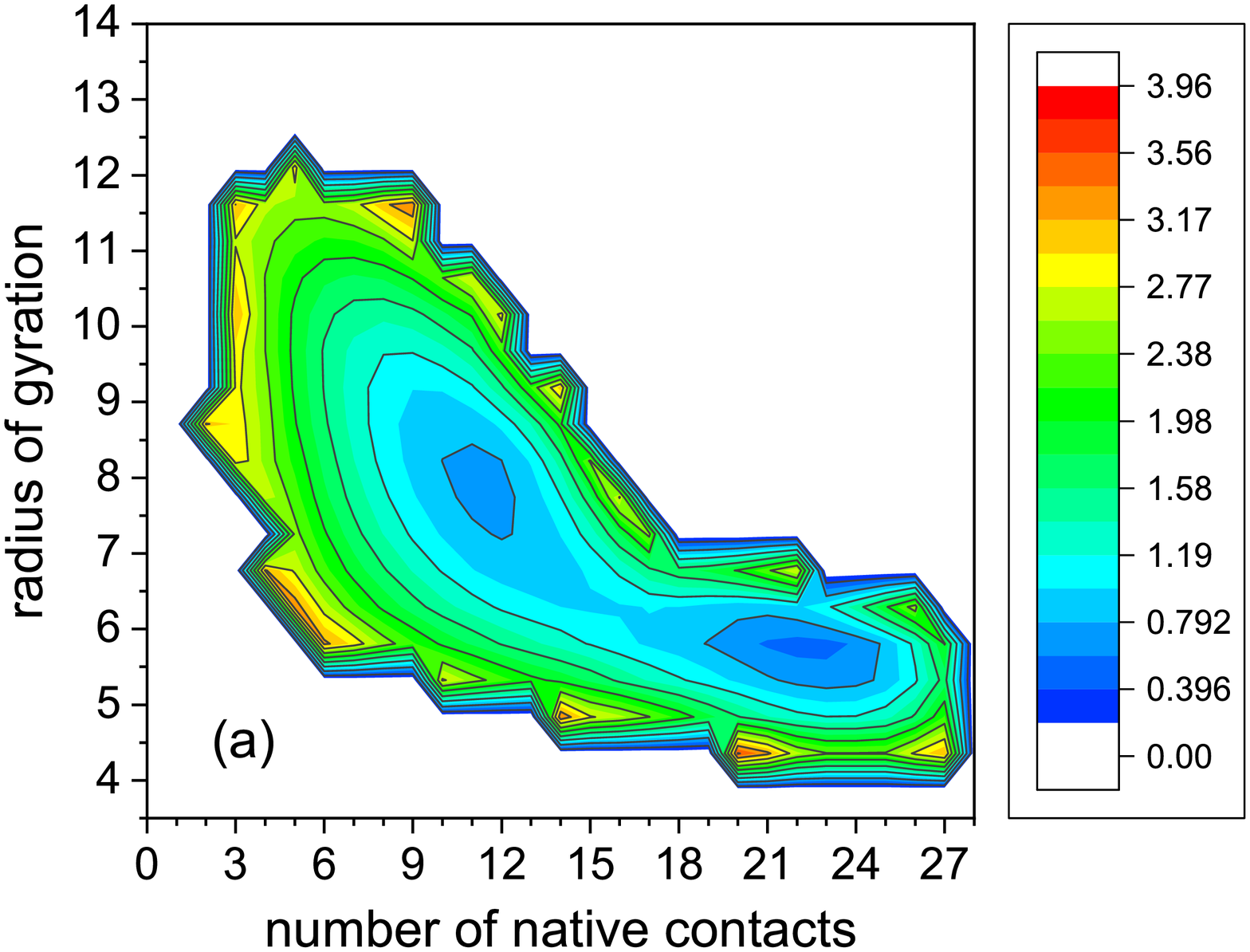}}%
\hfill
\resizebox{0.49\linewidth}{!}{ \includegraphics*{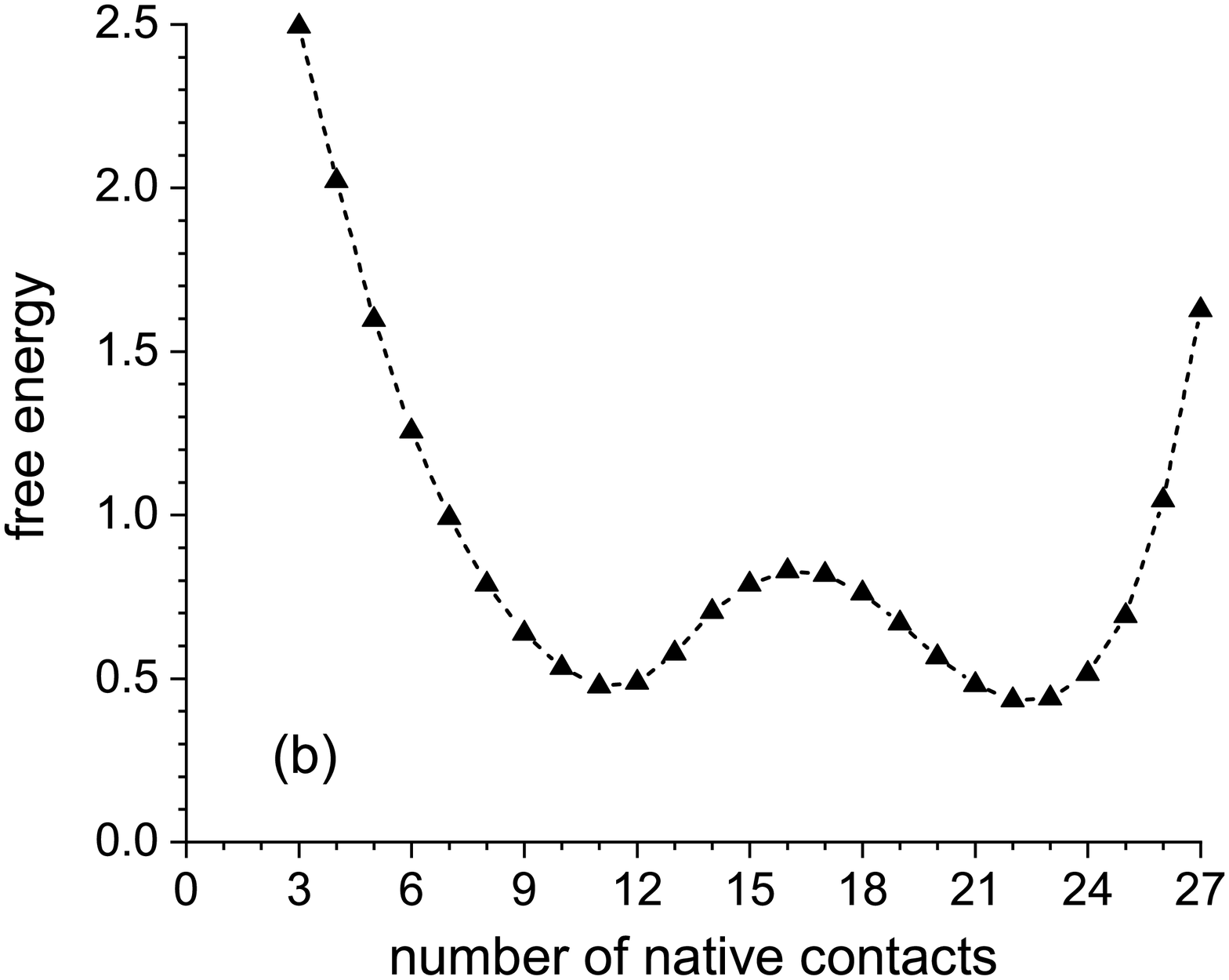}}%
\hfill
\resizebox{0.49\linewidth}{!}{ \includegraphics*{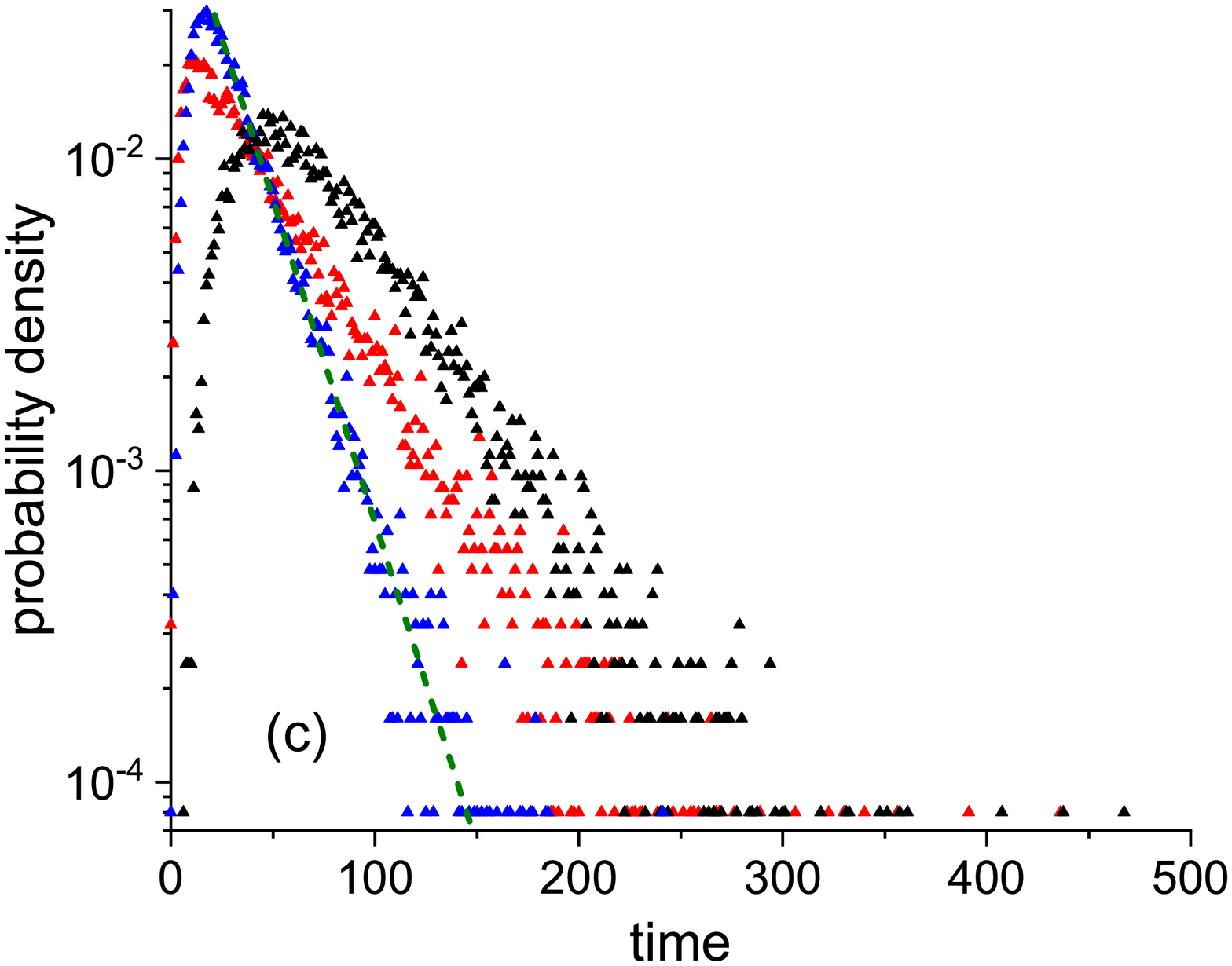}}%
\hfill
\resizebox{0.49\linewidth}{!}{ \includegraphics*{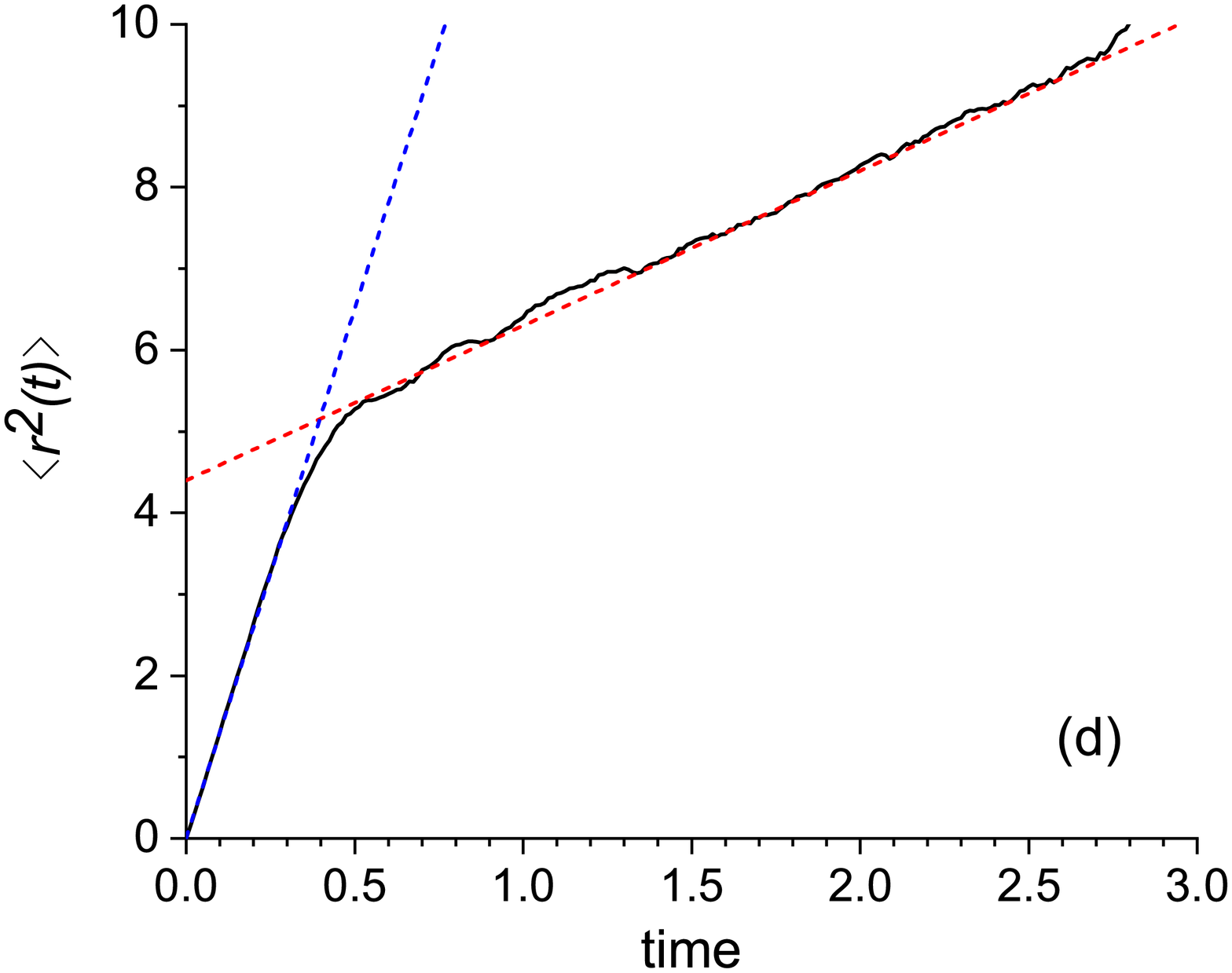}}%
\caption{$T=0.2$. The notations are as in Fig. \ref{g3_all_0_1}.}   
\label{g3_all_0_2}
\end{figure}

\begin{figure}\centering%
\resizebox{0.49\linewidth}{!}{ \includegraphics*{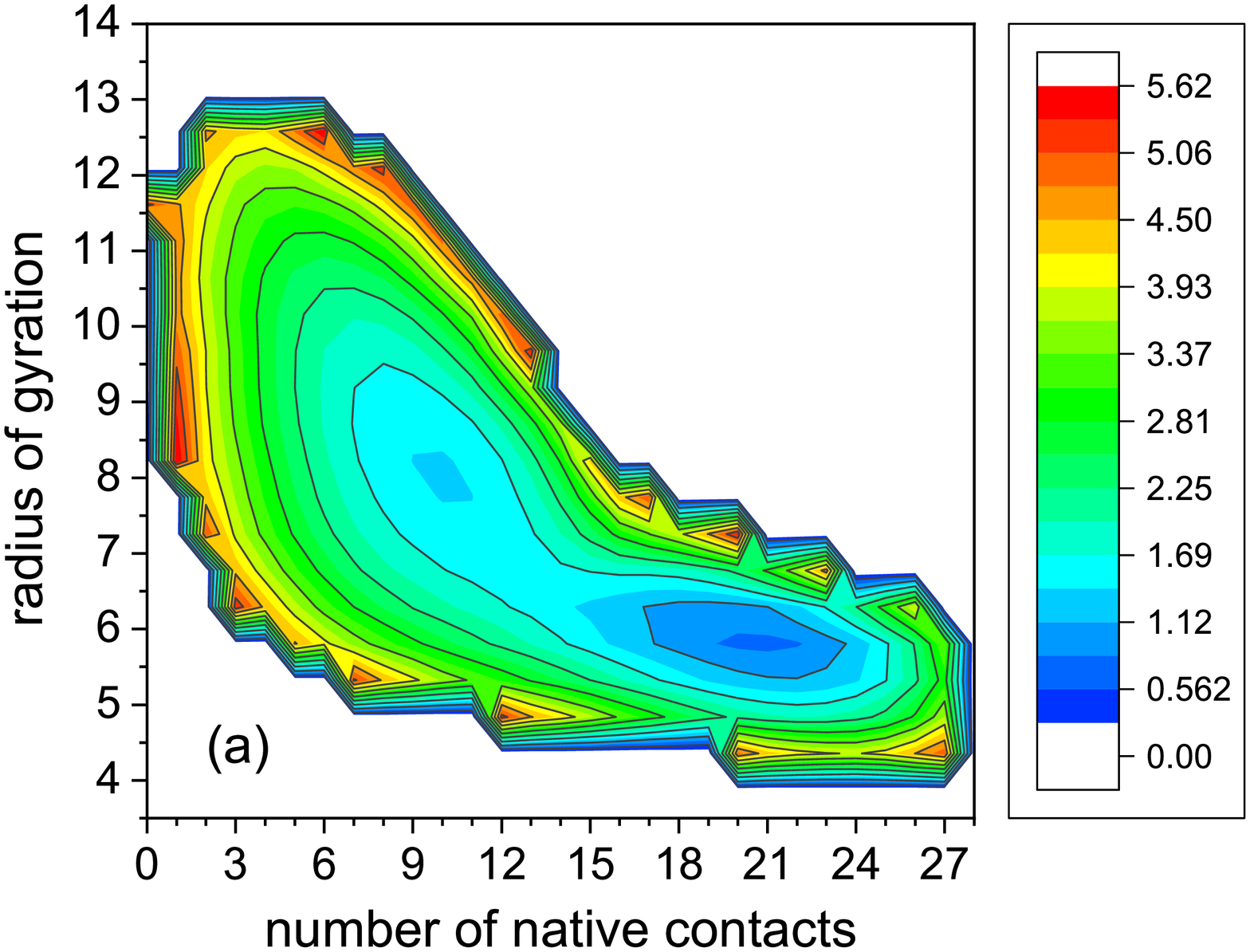}}%
\hfill
\resizebox{0.49\linewidth}{!}{ \includegraphics*{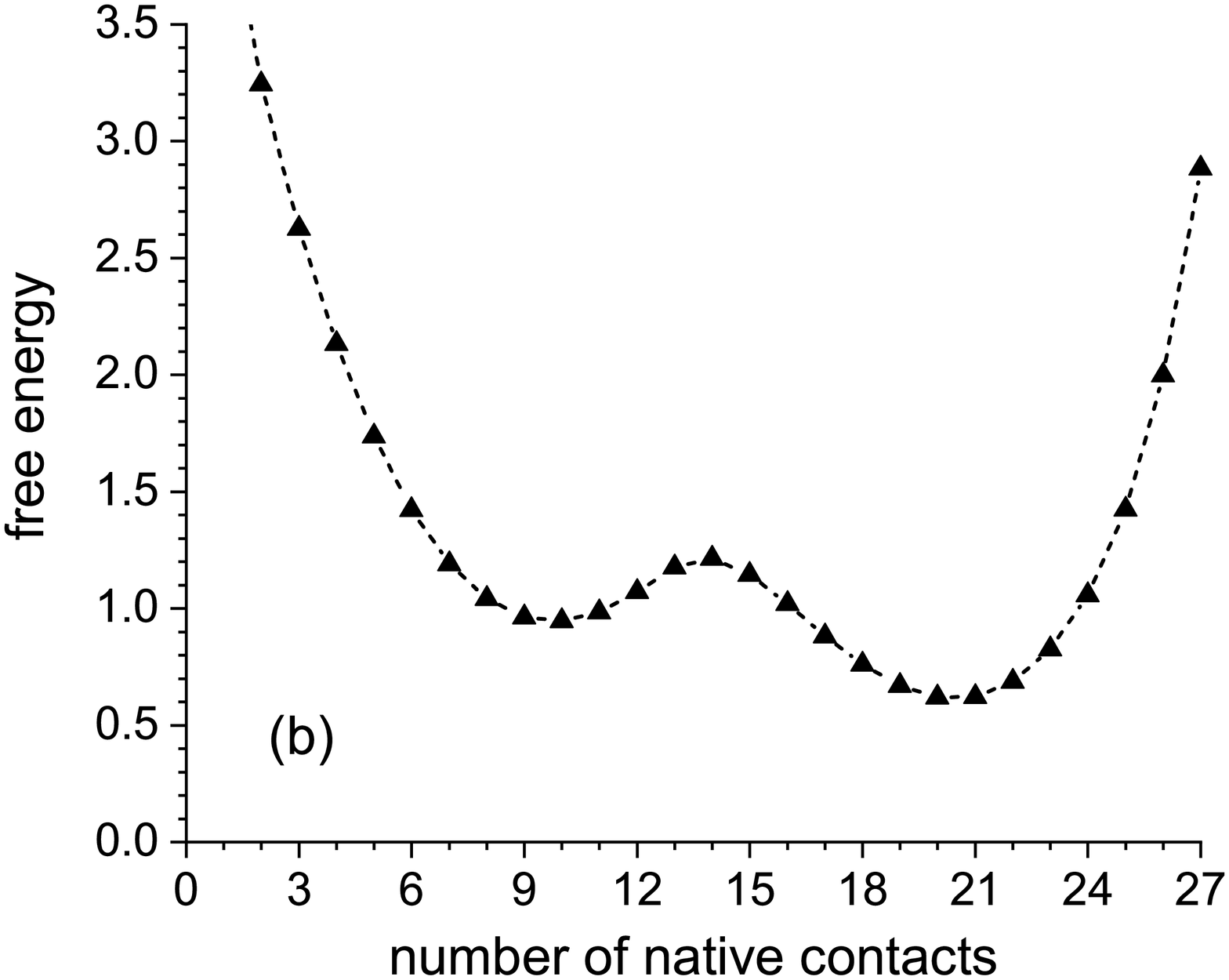}}%
\hfill
\resizebox{0.49\linewidth}{!}{ \includegraphics*{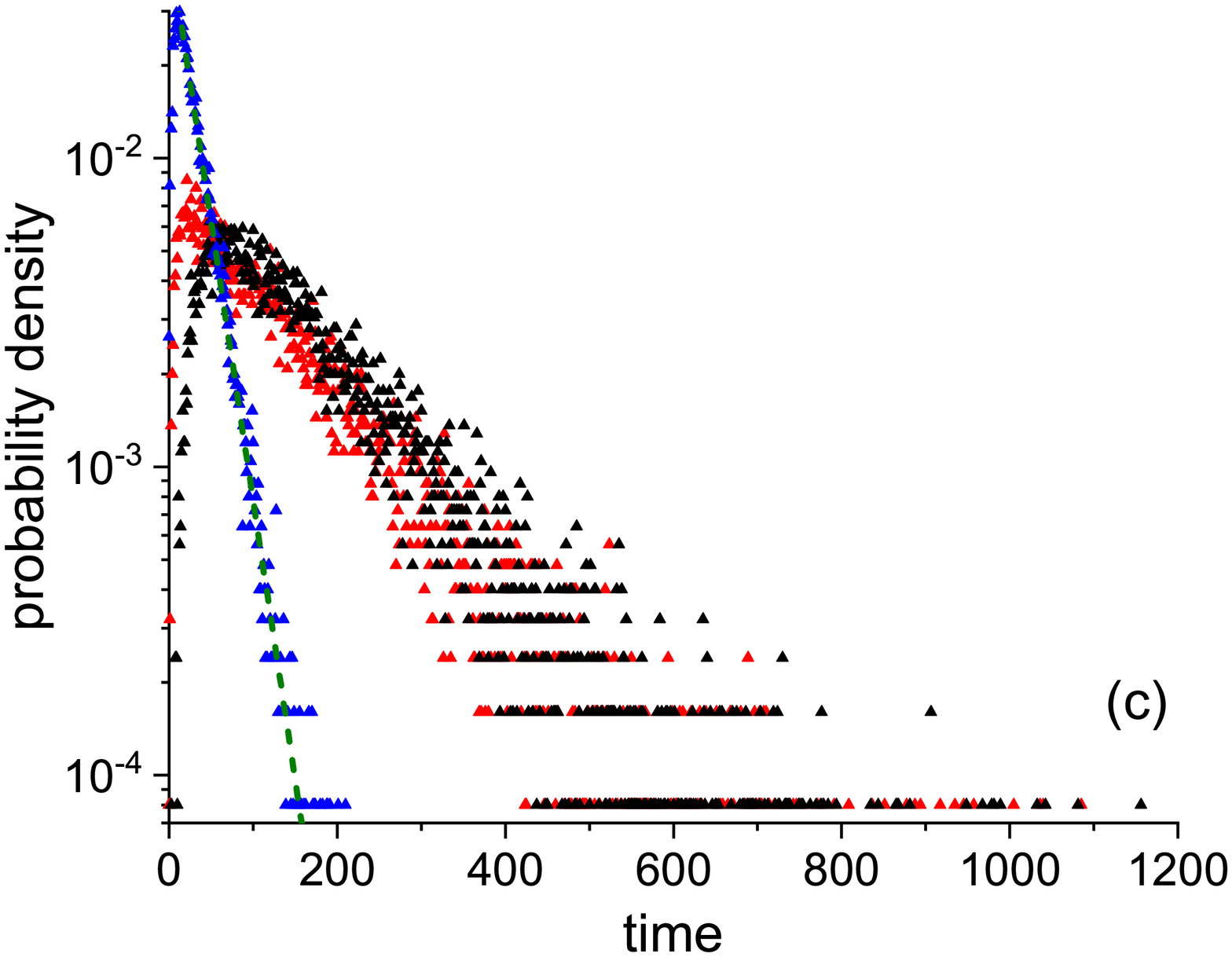}}%
\hfill
\resizebox{0.49\linewidth}{!}{ \includegraphics*{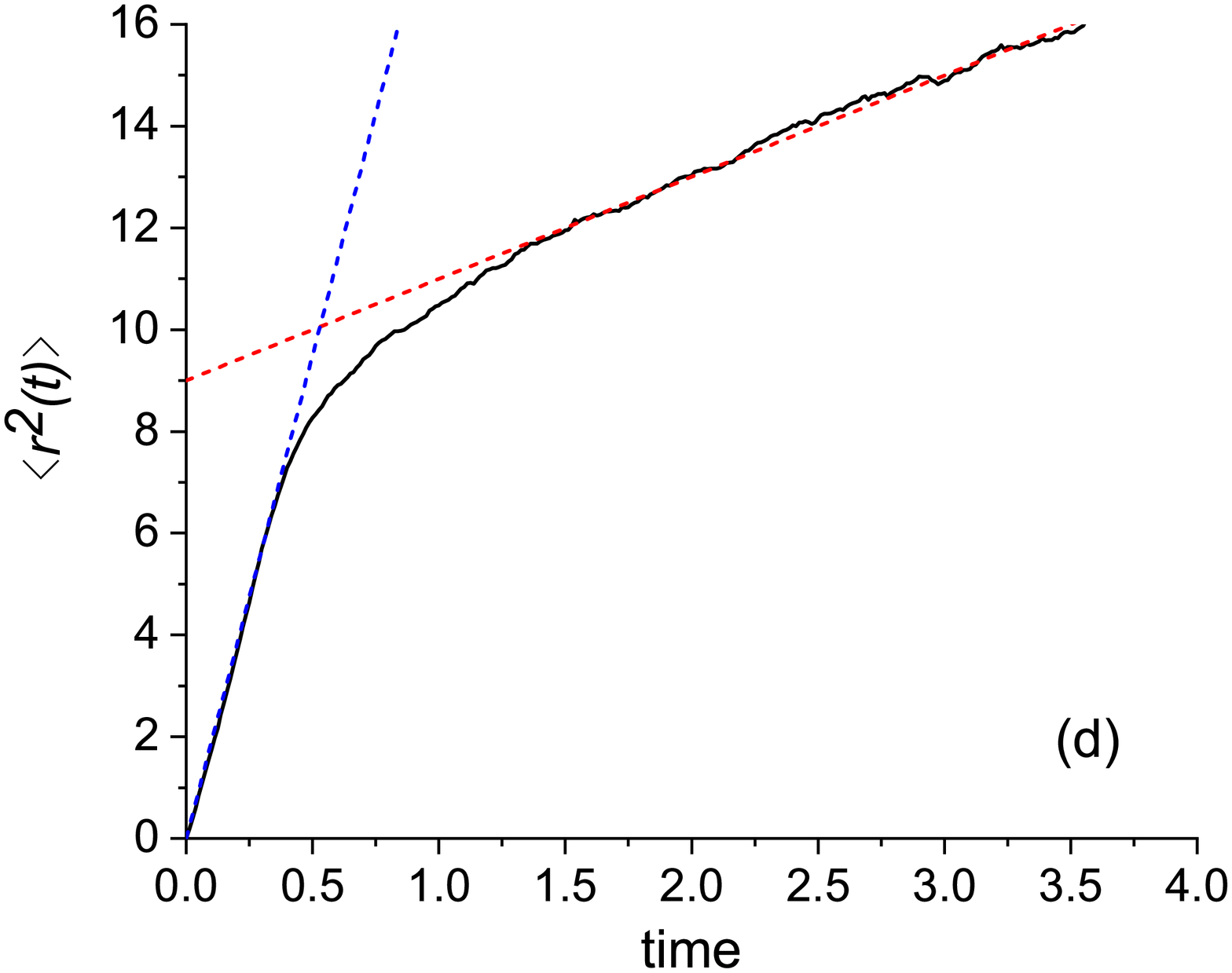}}%
\caption{$T=0.3$. The notations are as in Fig. \ref{g3_all_0_1}.}   
\label{g3_all_0_3}
\end{figure}

\clearpage

\begin{figure}\centering%
\resizebox{0.7\linewidth}{!}{ \includegraphics*{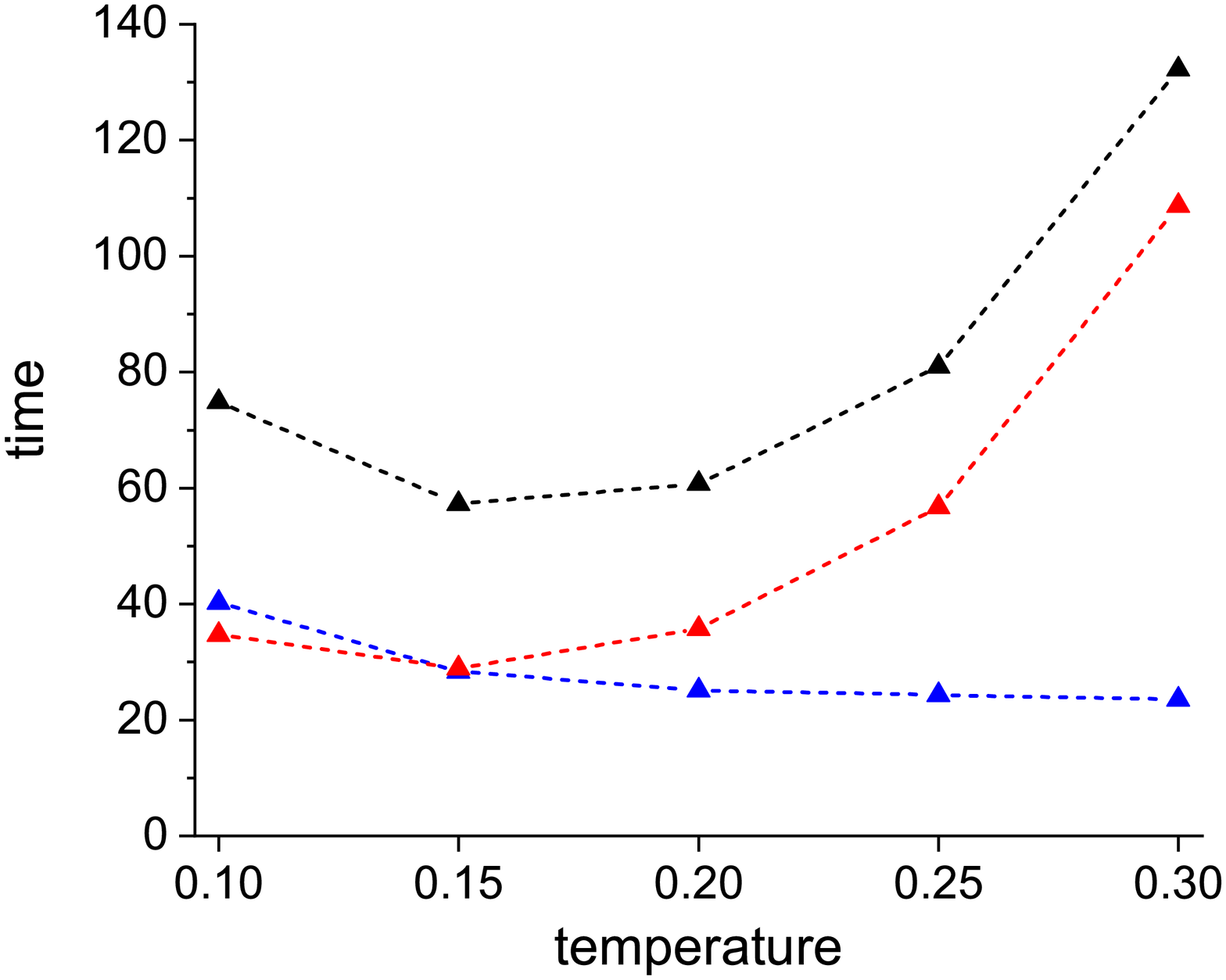}}%
\caption{The simulated MFPT times: the $\mathrm{U} \rightarrow \mathrm{NL}$ times (blue), $\mathrm{NL} \rightarrow \mathrm{N}$ times (red), and $\mathrm{U} \rightarrow \mathrm{N}$ times (black). The dashed lines are to guide the eye.}  
\label{g3_u_shape}
\end{figure}

\begin{table}
  \caption{Parameters to calculate the $\mathrm{U} \rightarrow \mathrm{NL}$ transition time with the Kramers rate formula}
  \label{tbl:g3_parameters}
  \begin{tabular}{llllll}
    \hline
     $T$ & $0.1$ & $0.15$ & $0.2$ & $0.25$ & $0.3$\\
    \hline
$\Delta F$ & 0.25 & 0.32 & 0.35 & 0.35 & 0.27\\
$F_{\mathrm{U}}^{''}\textsuperscript{\emph{a}}$ & 0.28 & 0.30 & 0.27 & 0.24 & 0.23\\
$F_{\mathrm{U}}^{''}\textsuperscript{\emph{b}}$ & 0.28 & 0.29 & 0.26 & 0.25 & 0.23\\
$F_{\mathrm{TS}}^{''}\textsuperscript{\emph{a}}$ & 0.13 & 0.17 & 0.23 & 0.28 & 0.33\\
$F_{\mathrm{TS}}^{''}\textsuperscript{\emph{b}}$ & 0.21 & 0.18 & 0.23 & 0.29 & 0.33\\
$D_{\mathrm{TS}}$ & 4.0 & 5.8 & 6.5 & 8.5 & 9.5\\
    \hline
  \end{tabular}

\textsuperscript{\emph{a}} from the polynomial approximation.\\
\textsuperscript{\emph{b}} calculated as the three-point finite difference.
\end{table}

\begin{table}
  \caption{Comparison of Folding Times}
  \label{tbl:g3_times}
  \begin{tabular}{llllll}
    \hline
     $T$ & $0.1$ & $0.15$ & $0.2$ & $0.25$ & $0.3$\\
    \hline
${\langle t_{\mathrm{\mathrm{U} \rightarrow \mathrm{NL}}}\rangle}\textsuperscript{\emph{a}}$ & 33.0 & 23.0 & 21.0 & 23.0 & 24.0\\
${\langle t_{\mathrm{\mathrm{U} \rightarrow \mathrm{NL}}}\rangle}\textsuperscript{\emph{b}}$ & 42.6 & 26.8 & 18.5 & 10.7 & 6.5\\
$\langle t_{\mathrm{\mathrm{U} \rightarrow \mathrm{NL}}}\rangle{\emph{c}}$ & 40.2 & 28.4 & 25.1 & 24.3 & 23.5 \\
$\langle t_{\mathrm{\mathrm{NL} \rightarrow \mathrm{N}}}\rangle{\emph{c}}$ & 34.7 & 28.9 & 35.7 & 56.7 & 108.7\\
$\langle t_{\mathrm{\mathrm{U} \rightarrow \mathrm{N}}}\rangle{\emph{c}}$ & 74.9 & 57.3 & 60.8 & 81.0 & 132.2\\
    \hline
  \end{tabular}

\textsuperscript{\emph{a}} calculated from the slope of the simulated $\mathrm{U} \rightarrow \mathrm{NL}$ decay curve.\\
\textsuperscript{\emph{b}} Kramers formula [Eq. (2), the main text] for the average values of $F_{\mathrm{U}}^{''}$ and $F_{\mathrm{TS}}^{''}$ (Table {\ref{tbl:g3_parameters}}).\\
\textsuperscript{\emph{c}} simulated times.
\end{table}

\begin{figure}\centering%
\resizebox{0.7\linewidth}{!}{ \includegraphics*{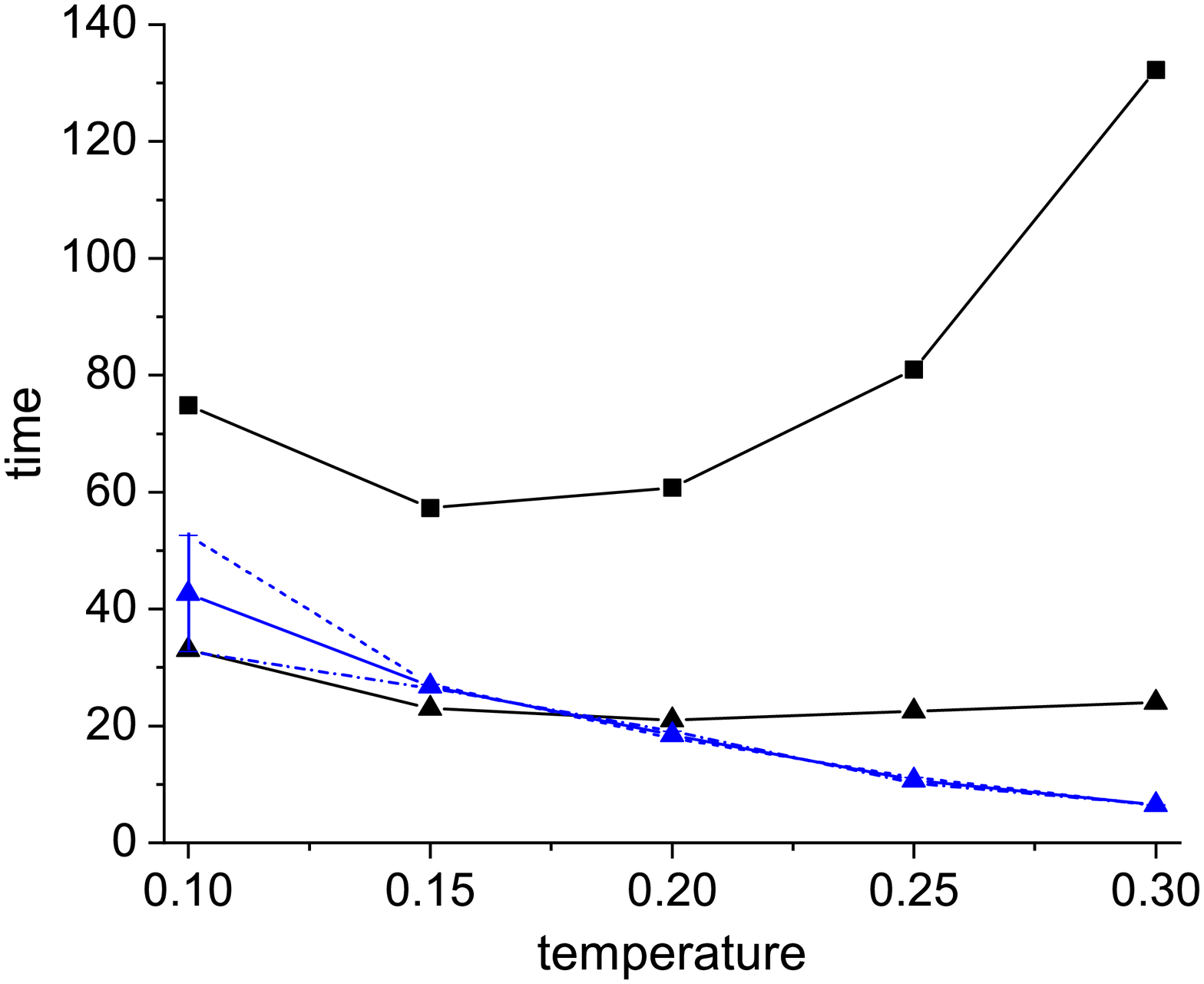}}%
\caption{The black squares are for the $\langle t_{\mathrm{\mathrm{U} \rightarrow \mathrm{N}}} \rangle$ times from simulations, the black triangles denote the $\langle t_{{\mathrm{U} \rightarrow \mathrm{NL}}} \rangle$ times calculated from the slopes of the simulated $\mathrm{U} \rightarrow \mathrm{NL}$ decay curves, and the blue triangles are for  $\langle t_{\mathrm{\mathrm{U} \rightarrow \mathrm{NL}}} \rangle$ times from Eq. (2) of the main text with the average values of $F_{\mathrm{U}}^{''}$ and $F_{\mathrm{TS}}^{''}$ (the dashed and dash-dotted blue lines indicate the results for $F_{\mathrm{U}}^{''}$ and $F_{\mathrm{TS}}^{''}$ obtained by the polynomial approximation of the FEP and calculated by finite-differences, respectively). In all cases, the lines are to guide the eye.}  
\label{times_cmp}
\end{figure}

\clearpage

{\centering \section{Friction Constant $\gamma=50M/\tau $}}

\begin{figure}\centering%
\resizebox{0.49\linewidth}{!}{ \includegraphics*{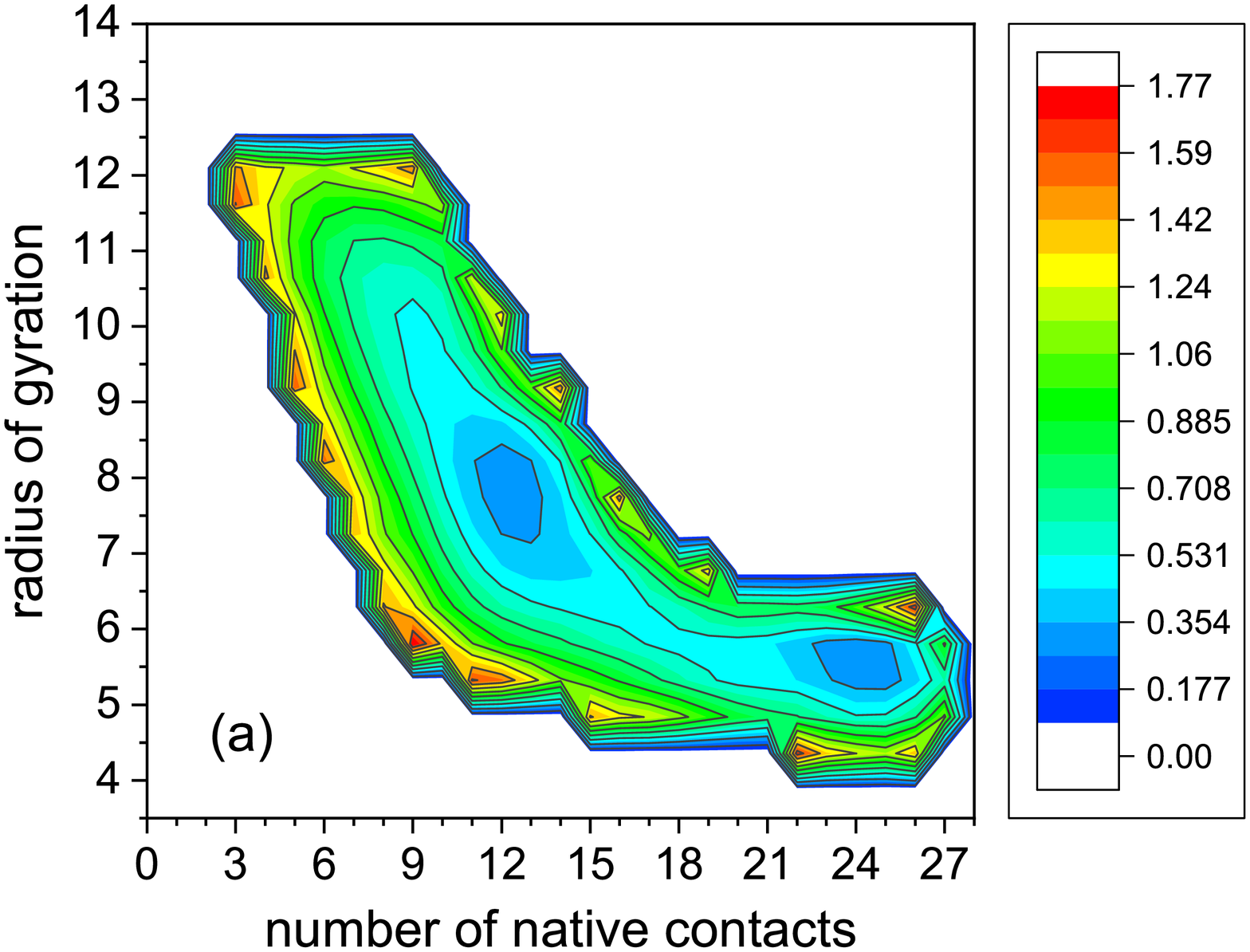}}%
\hfill
\resizebox{0.49\linewidth}{!}{ \includegraphics*{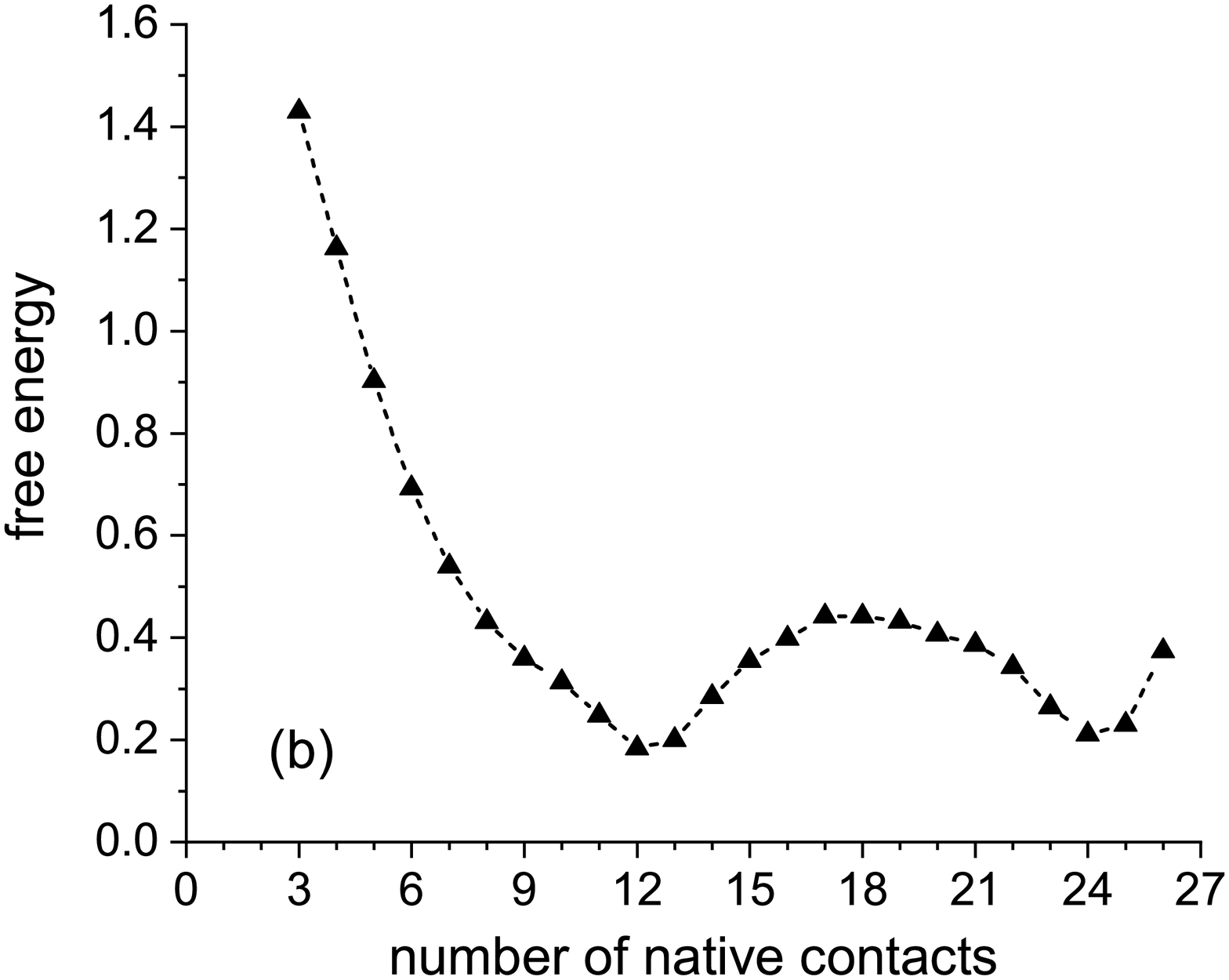}}%
\hfill
\resizebox{0.49\linewidth}{!}{ \includegraphics*{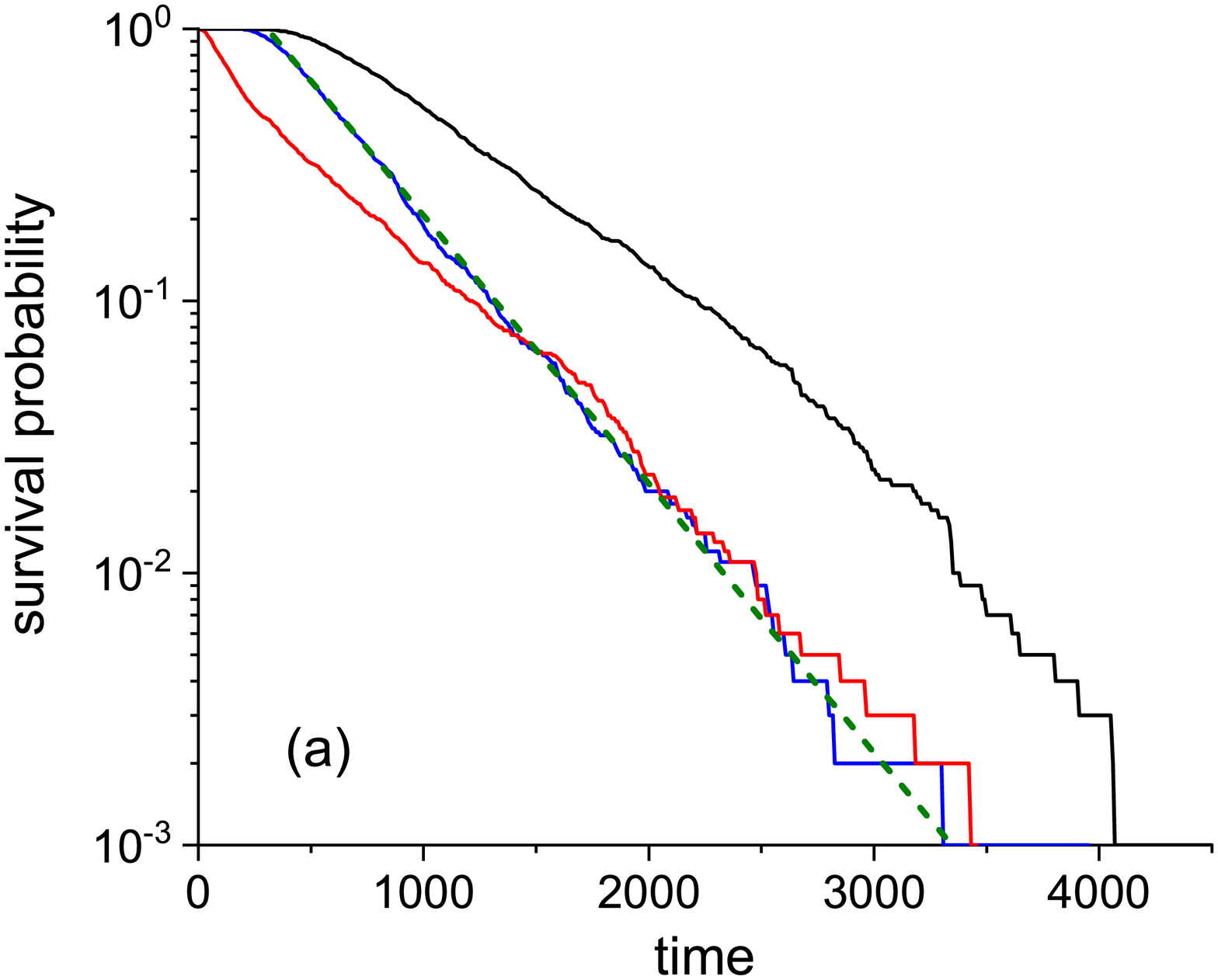}}%
\hfill
\resizebox{0.49\linewidth}{!}{ \includegraphics*{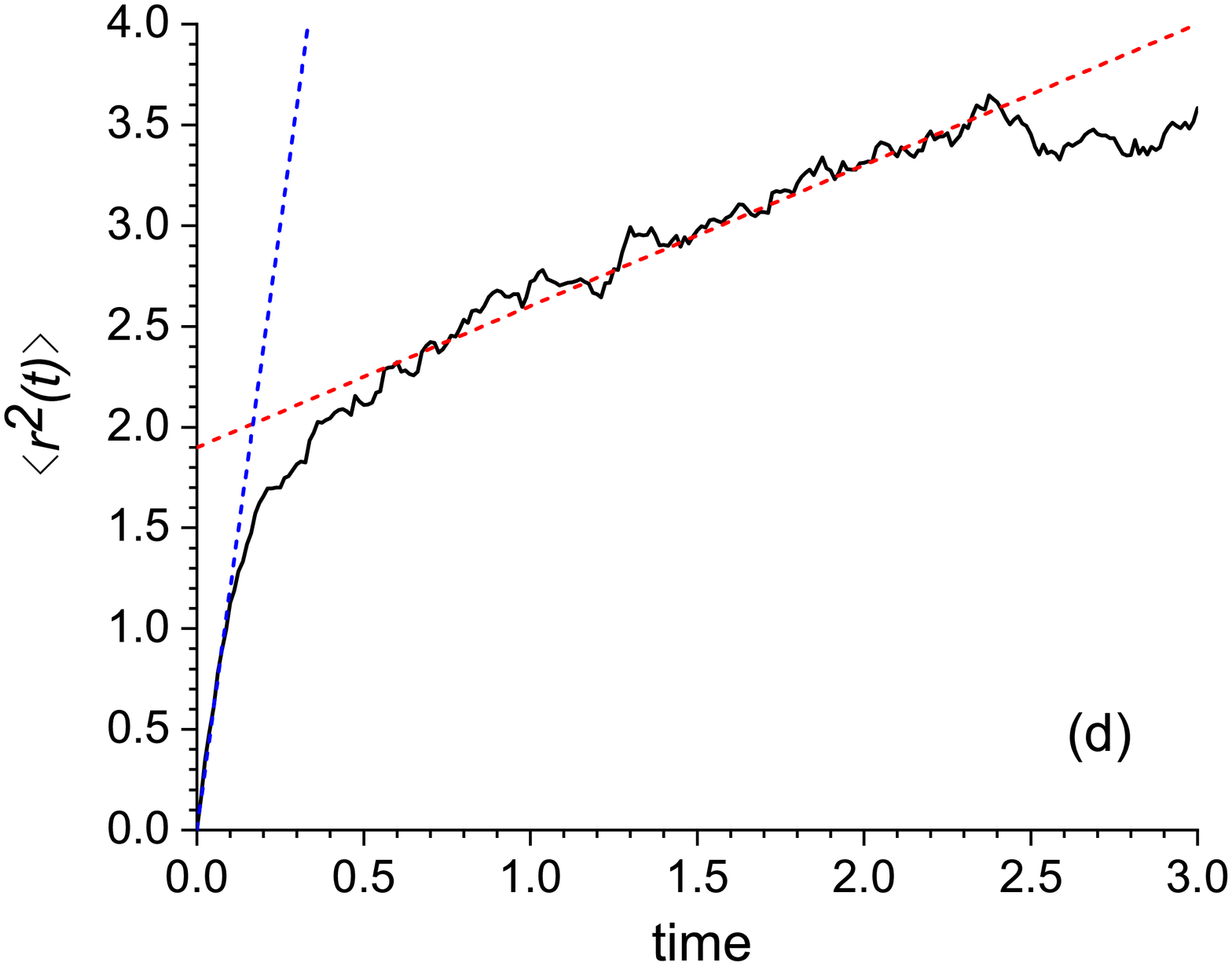}}%
\caption{$T=0.1$. ({\bf{a}}) The free energy surface $F(N_{\mathrm{nat}},R_{\mathrm{g}})$, and ({\bf{b}}) free energy profile $F(N_{\mathrm{nat}})$. ({\bf{c}}) First-passage time distributions in the form of survivaal probabilities: the $\mathrm{U} \rightarrow \mathrm{NL}$ trajectories (blue), the $\mathrm{NL} \rightarrow \mathrm{N}$ trajectories (red), and the U-N trajectories (black); the dashed green line denotes an exponential fit to the $\mathrm{U} \rightarrow \mathrm{NL}$ distribution. ({\bf{d}}) The time-dependent mean-square deviation from the transition state in the number of native contacts (black curve); the blue and red dashed lines are the linear fits to the curve for short and long times, respectively.}   
\label{g50_all_0_1}
\end{figure}

\begin{figure}\centering%
\resizebox{0.49\linewidth}{!}{ \includegraphics*{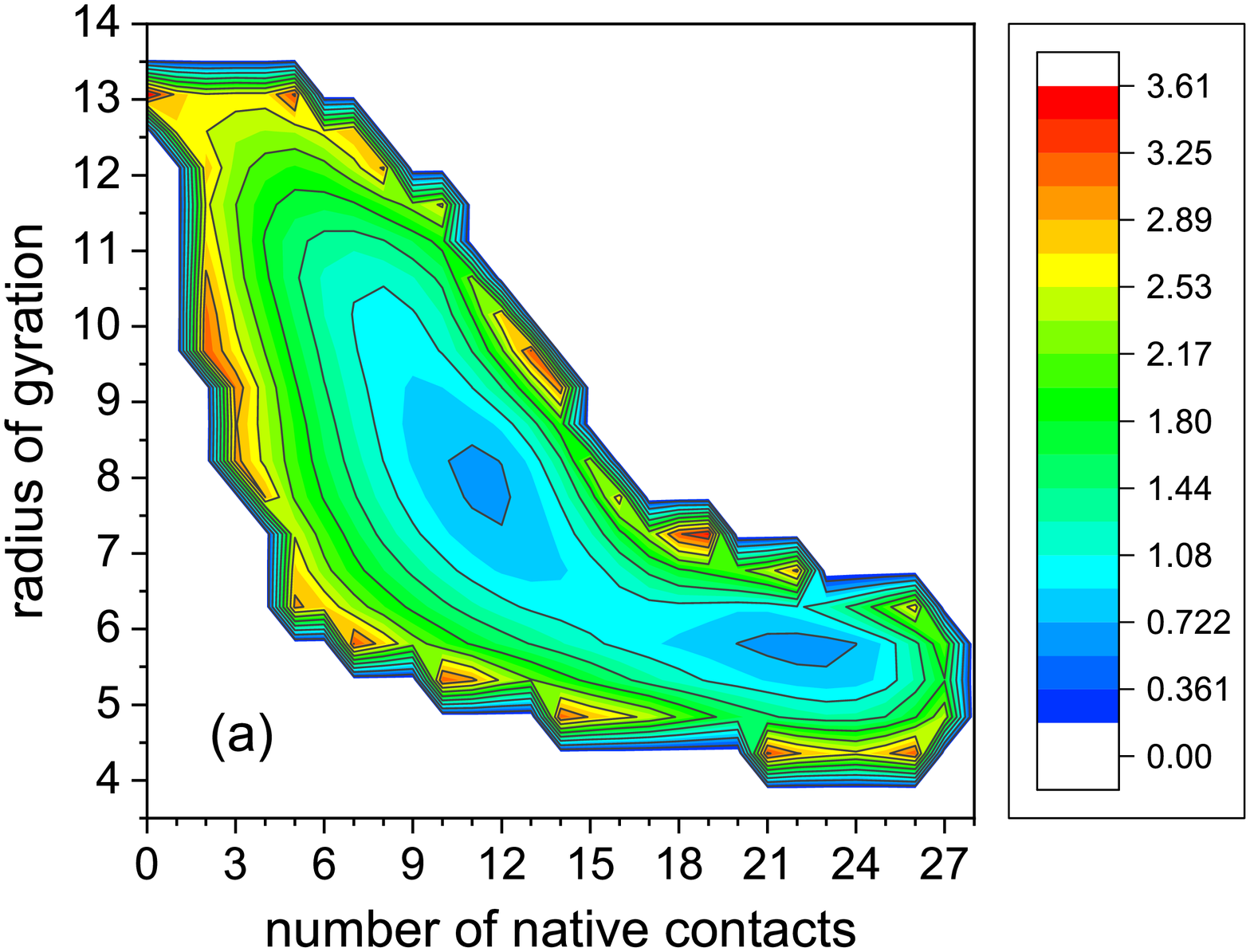}}%
\hfill
\resizebox{0.49\linewidth}{!}{ \includegraphics*{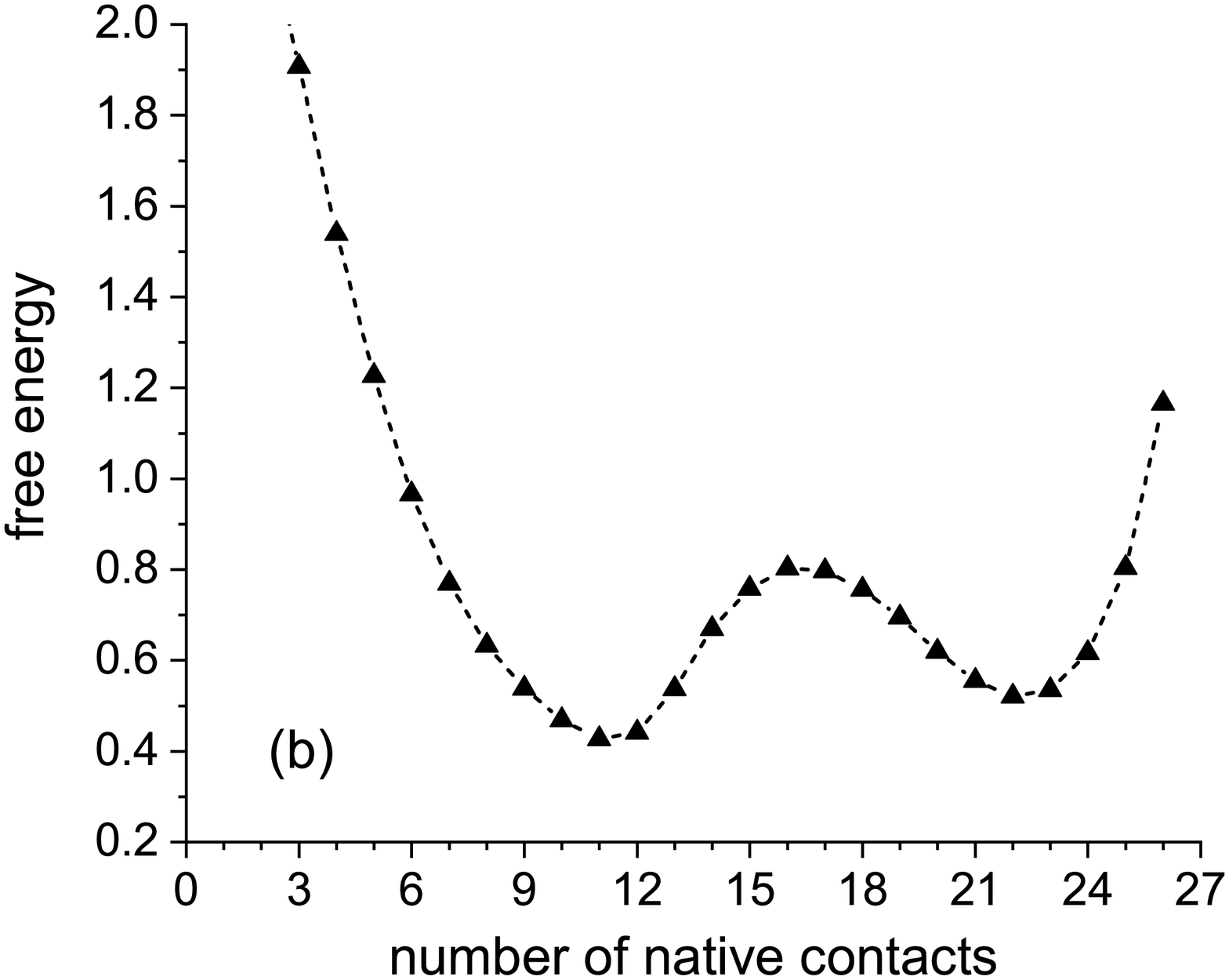}}%
\hfill
\resizebox{0.49\linewidth}{!}{ \includegraphics*{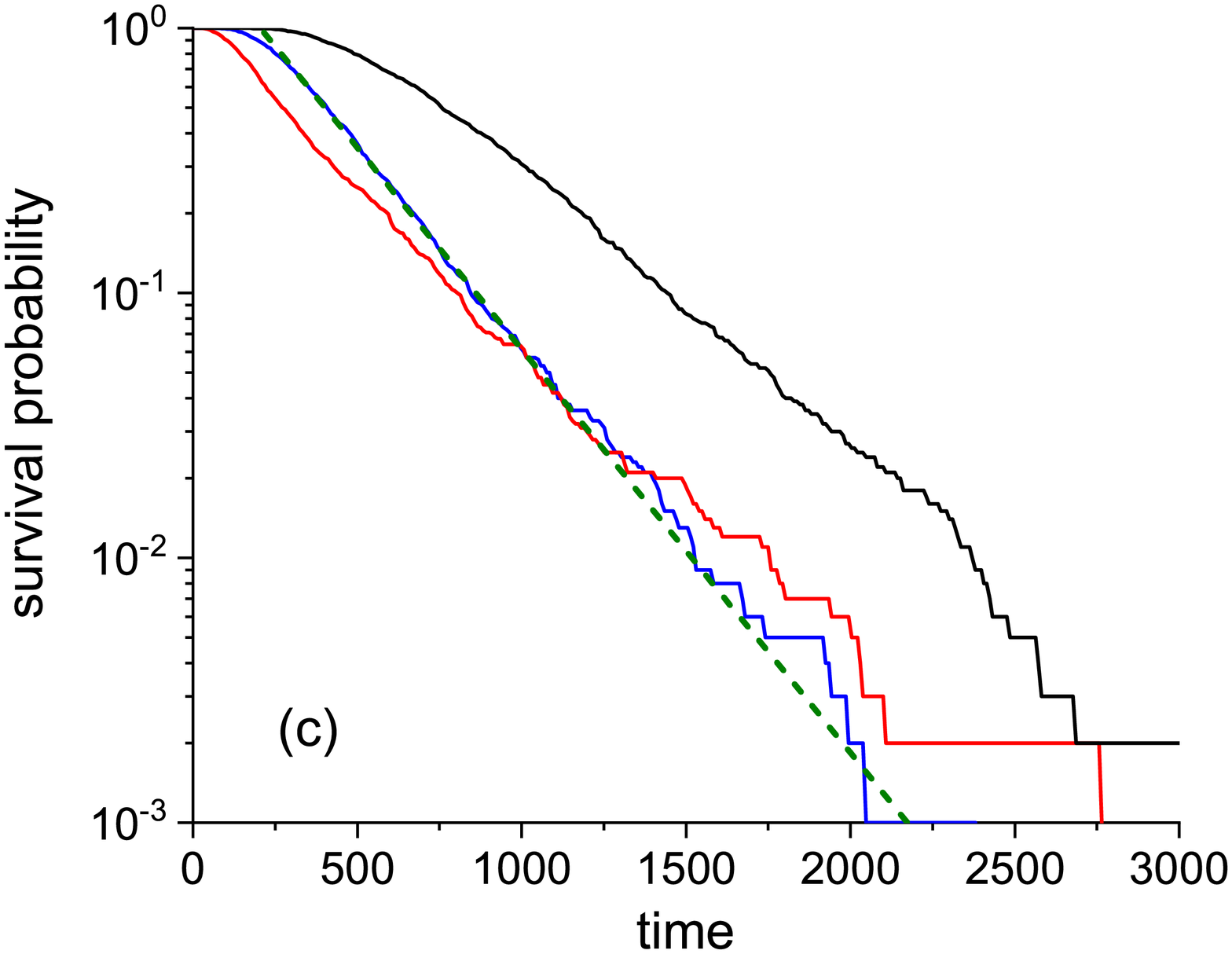}}%
\hfill
\resizebox{0.49\linewidth}{!}{ \includegraphics*{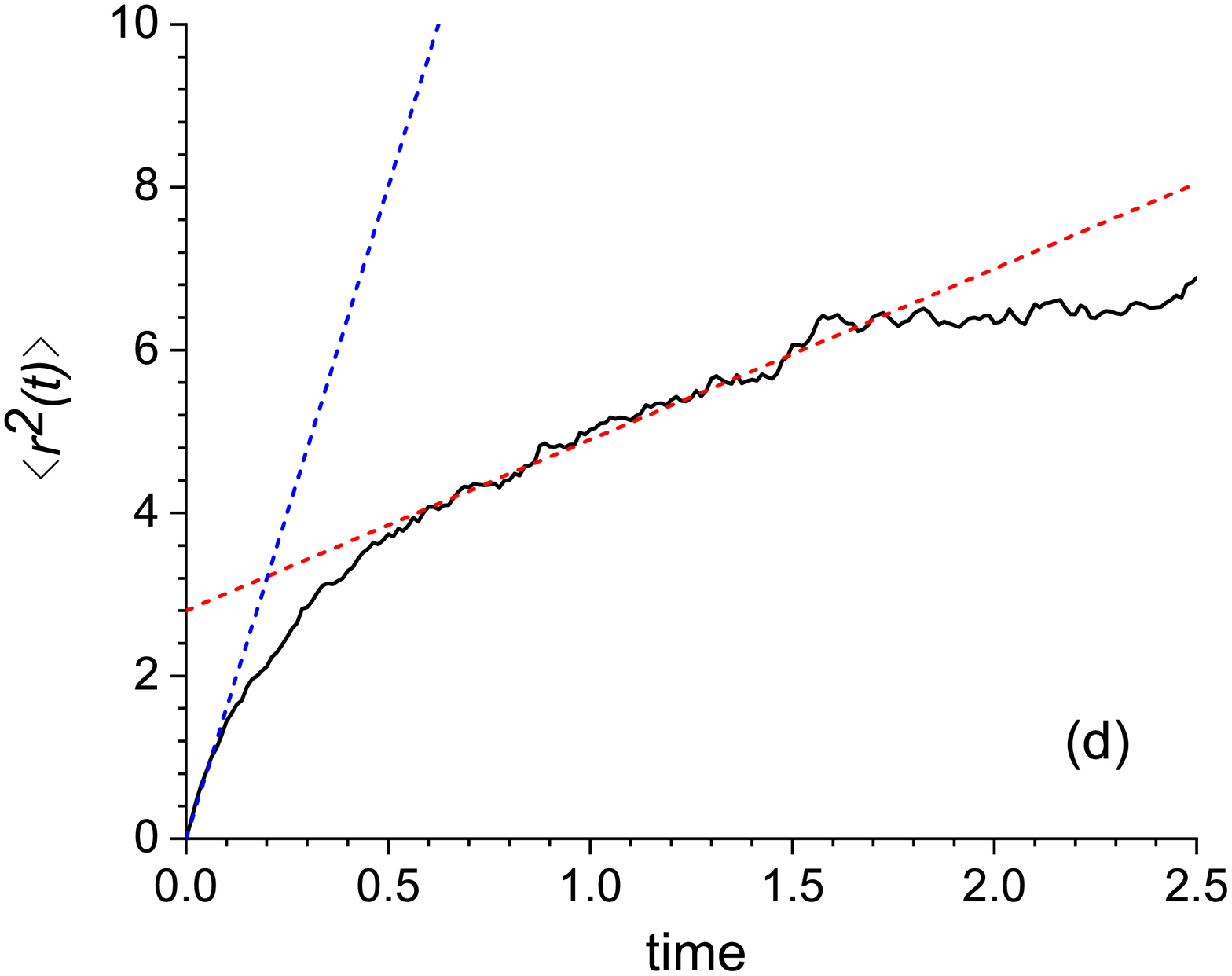}}%
\caption{$T=0.2$. The notations are as in Fig. \ref{g50_all_0_1}.}   
\label{g50_all_0_2}
\end{figure}

\begin{figure}\centering%
\resizebox{0.49\linewidth}{!}{ \includegraphics*{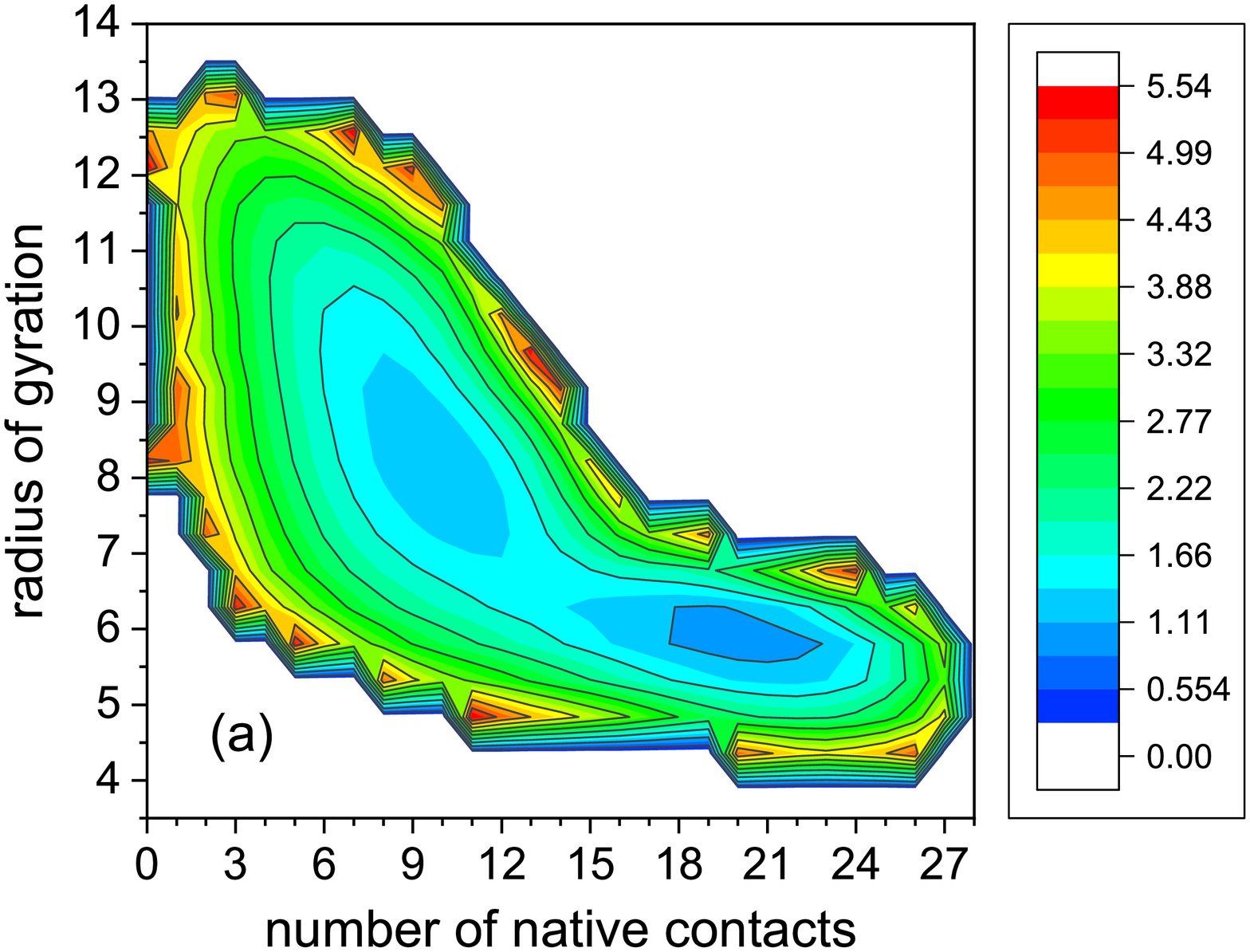}}%
\hfill
\resizebox{0.49\linewidth}{!}{ \includegraphics*{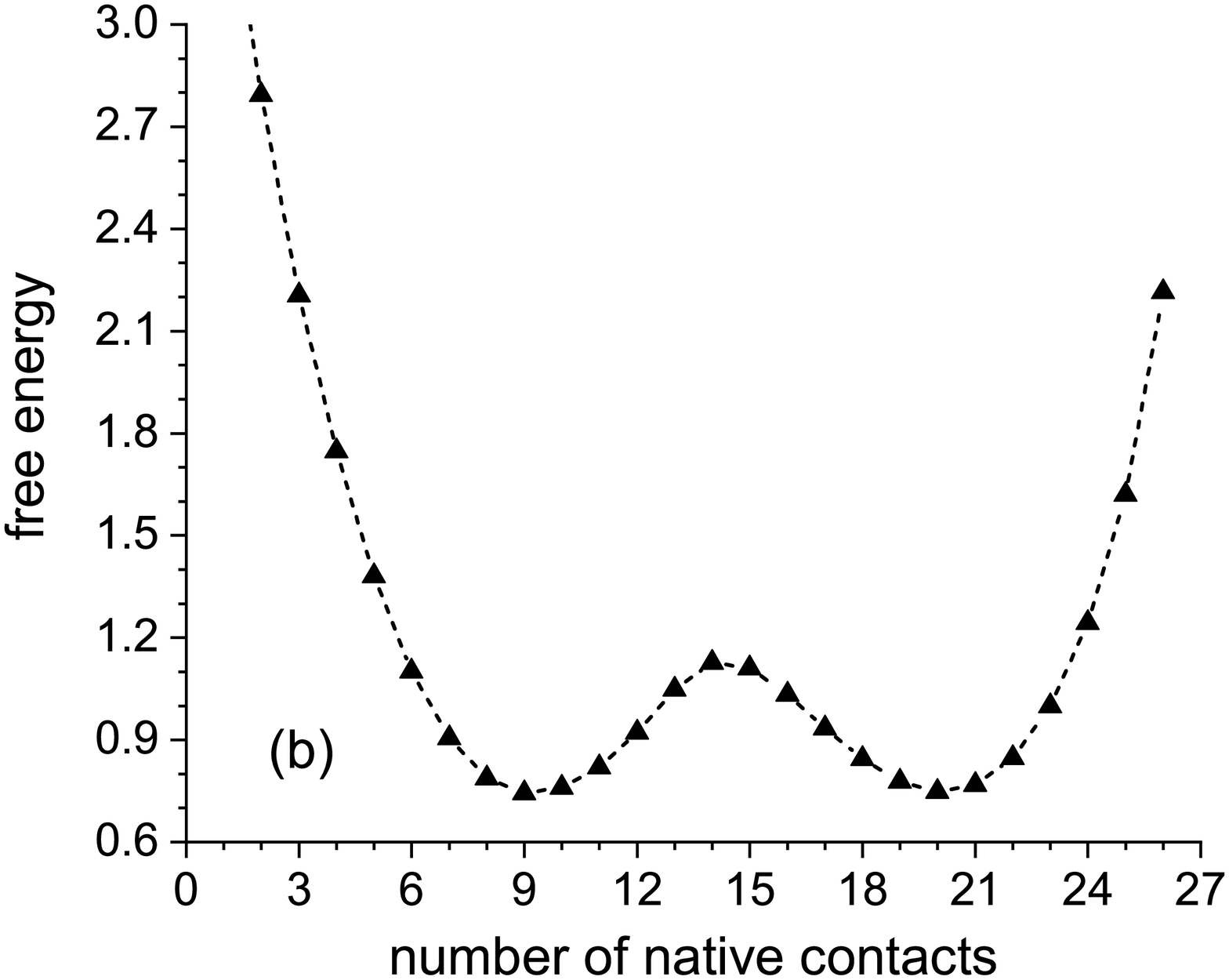}}%
\hfill
\resizebox{0.49\linewidth}{!}{ \includegraphics*{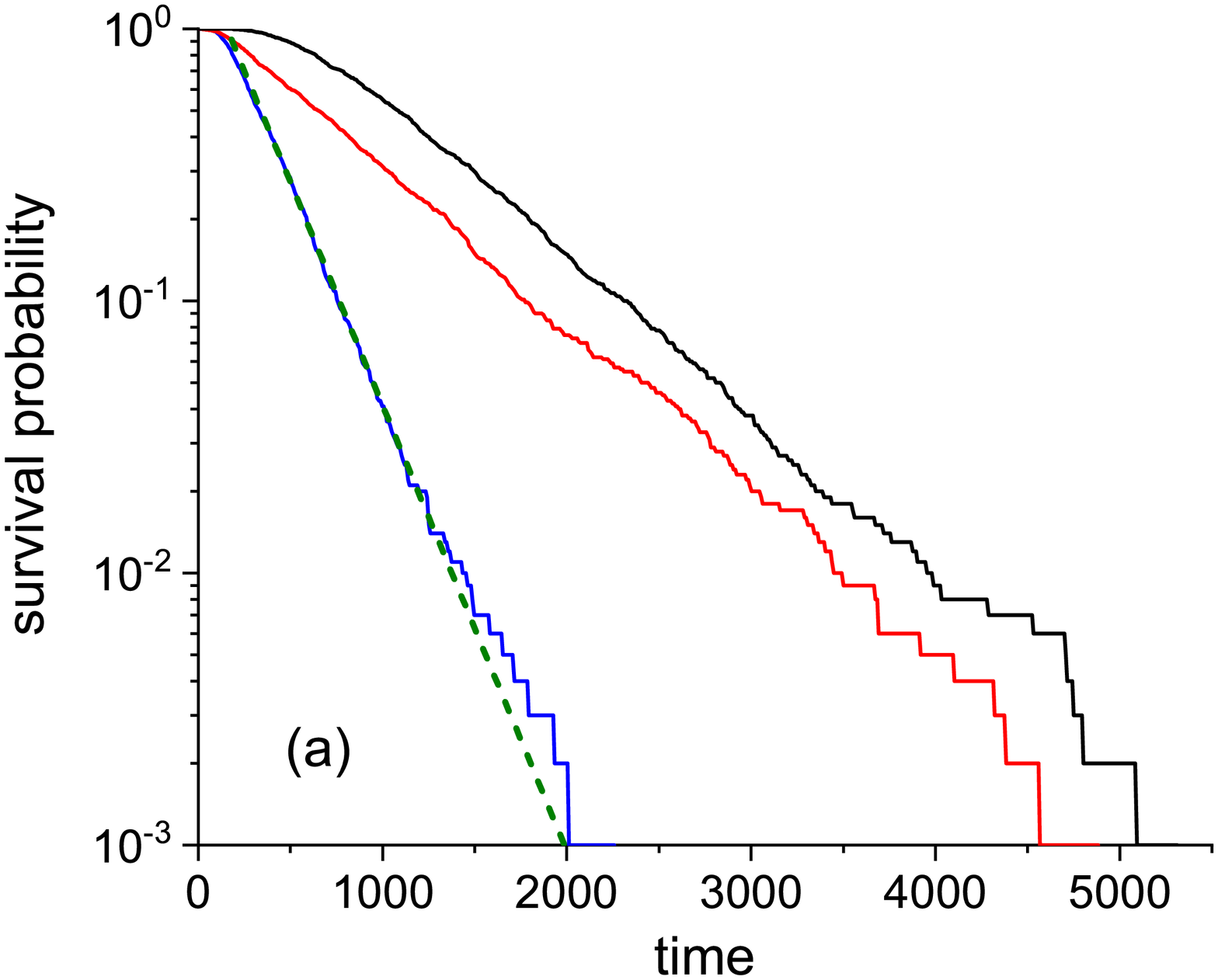}}%
\hfill
\resizebox{0.49\linewidth}{!}{ \includegraphics*{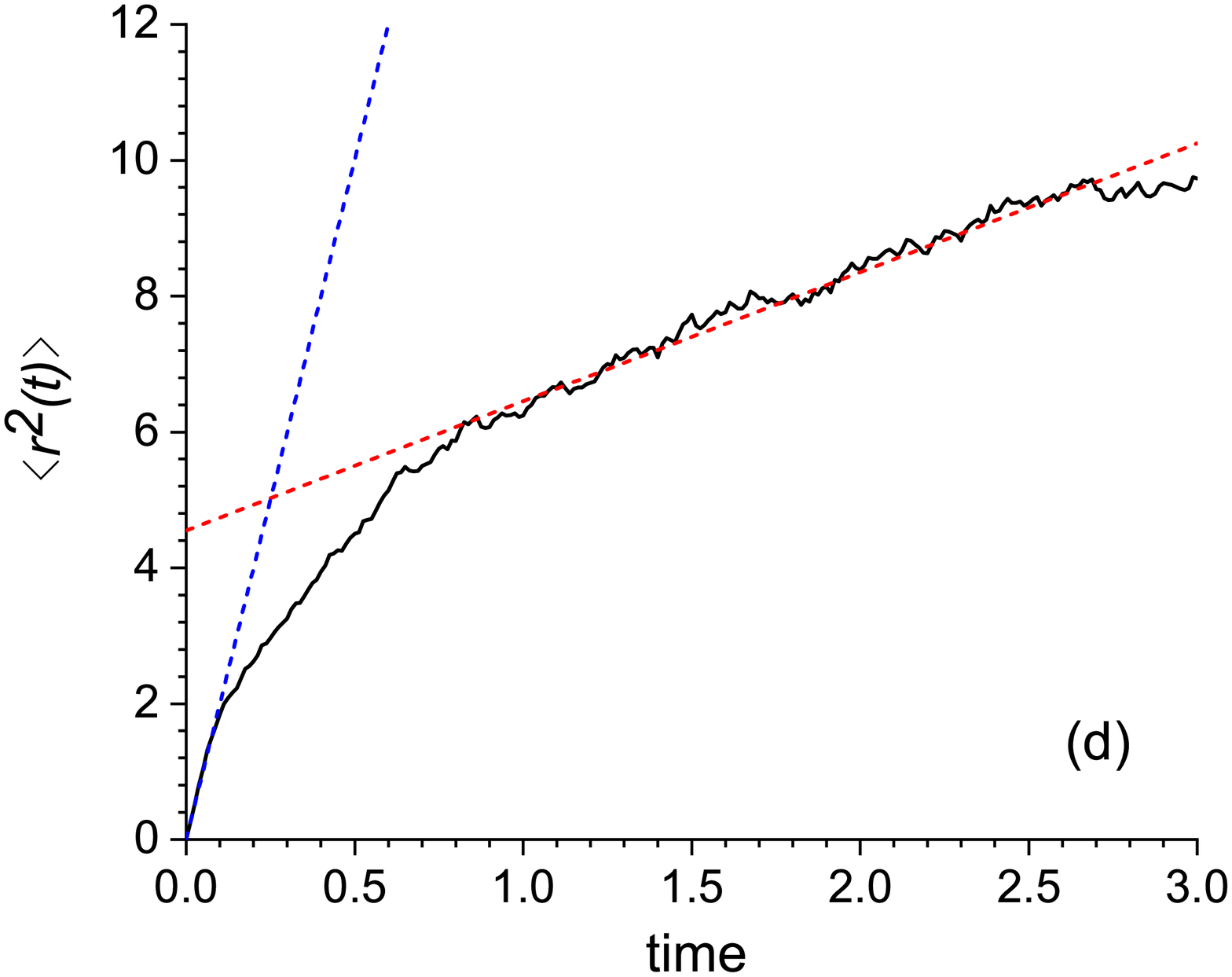}}%
\caption{$T=0.3$. The notations are as in Fig. \ref{g50_all_0_1}.}   
\label{g50_all_0_3}
\end{figure}

\clearpage

\begin{figure}\centering%
\resizebox{0.7\linewidth}{!}{ \includegraphics*{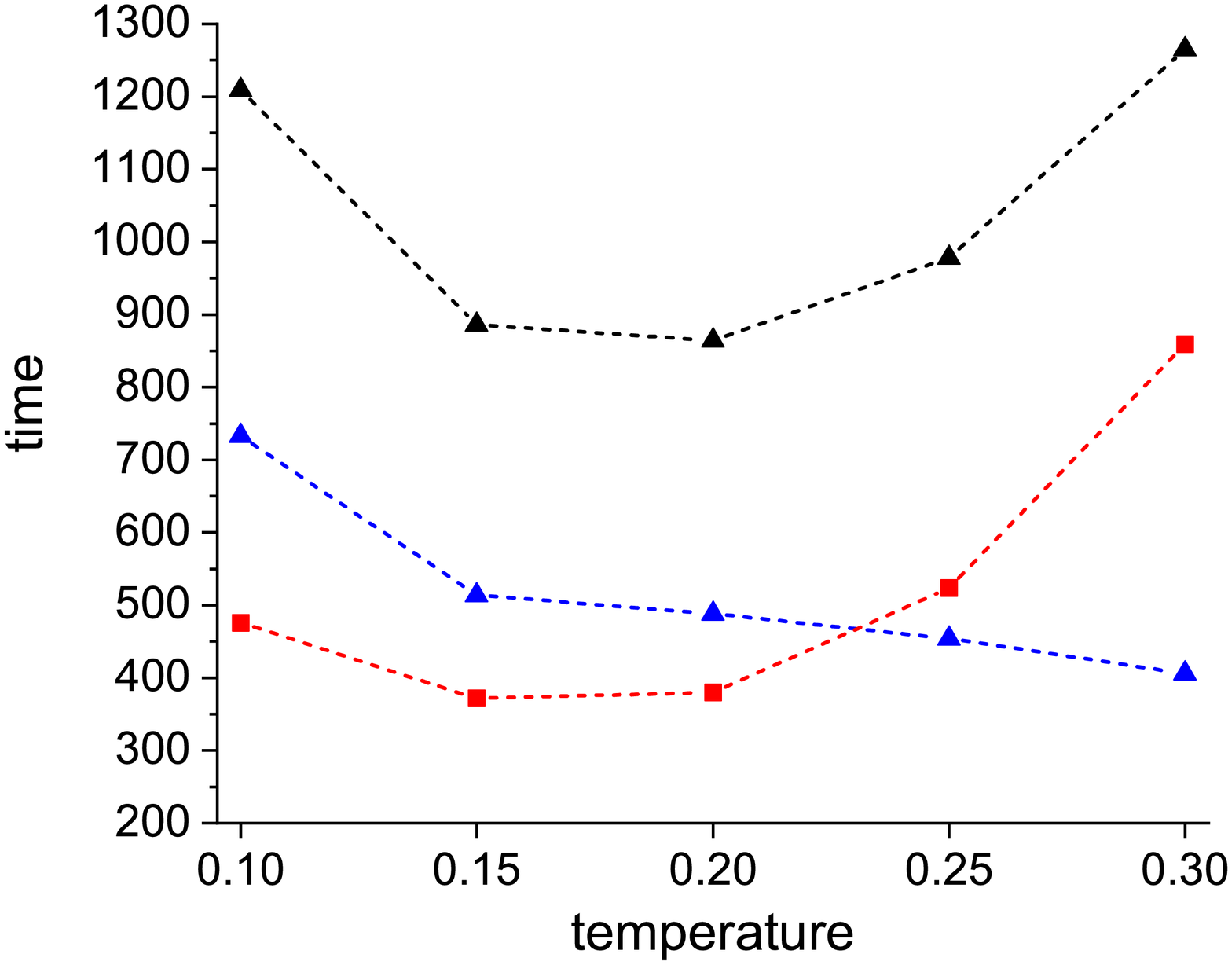}}%
\caption{The simulated MFPT times: the $\mathrm{U} \rightarrow \mathrm{NL}$ times (blue), $\mathrm{NL} \rightarrow \mathrm{N}$ times (red), and $\mathrm{U} \rightarrow \mathrm{N}$ times (black). The dashed lines are to guide the eye.}  
\label{g50_u_shape}
\end{figure}

\begin{table}
  \caption{Parameters to calculate the $\mathrm{U} \rightarrow \mathrm{NL}$ transition time with the Kramers rate formula}
  \label{tbl:g50_parameters}
  \begin{tabular}{llllll}
    \hline
     $T$ & $0.1$ & $0.15$ & $0.2$ & $0.25$ & $0.3$\\
    \hline
$\Delta F$ & 0.25 & 0.32 & 0.38 & 0.41 & 0.38\\
$F_{\mathrm{U}}^{''}\textsuperscript{\emph{a}}$ & 0.27 & 0.25 & 0.24 & 0.22 & 0.24\\
$F_{\mathrm{U}}^{''}\textsuperscript{\emph{b}}$ & 0.29 & 0.29 & 0.24 & 0.21 & 0.24\\
$F_{\mathrm{TS}}^{''}\textsuperscript{\emph{a}}$ & 0.14 & 0.16 & 0.21 & 0.26 & 0.28\\
$F_{\mathrm{TS}}^{''}\textsuperscript{\emph{b}}$ & 0.21 & 0.18 & 0.23 & 0.27 & 0.31\\
$D_{\mathrm{TS}}$ & 0.35 & 0.50 & 1.1 & 0.90 & 0.95\\
    \hline
  \end{tabular}

\textsuperscript{\emph{a}} from the polynomial approximation.\\
\textsuperscript{\emph{b}} calculated as the three-point finite difference.
\end{table}

\begin{table}
  \caption{Comparison of folding times}
  \label{tbl:g50_times}
  \begin{tabular}{llllll}
    \hline
     $T$ & $0.1$ & $0.15$ & $0.2$ & $0.25$ & $0.3$\\
    \hline
${\langle t_{\mathrm{\mathrm{U} \rightarrow \mathrm{NL}}}\rangle}\textsuperscript{\emph{a}}$ & 440 & 280 & 285 & 310 & 265\\
${\langle t_{\mathrm{\mathrm{U} \rightarrow \mathrm{NL}}}\rangle}\textsuperscript{\emph{b}}$ & 472 & 358 & 151 & 159 & 100\\
$\langle t_{\mathrm{\mathrm{U} \rightarrow \mathrm{NL}}}\rangle{\emph{c}}$ & 733 & 514 & 484 & 454 & 406 \\
$\langle t_{\mathrm{\mathrm{NL} \rightarrow \mathrm{N}}}\rangle{\emph{c}}$ & 476 & 372 & 380 & 524 & 859\\
$\langle t_{\mathrm{\mathrm{U} \rightarrow \mathrm{N}}}\rangle{\emph{c}}$ & 1209 & 886 & 864 & 978 & 1265\\
    \hline
  \end{tabular}

\textsuperscript{\emph{a}} calculated from the slope of the simulated $\mathrm{U} \rightarrow \mathrm{NL}$ decay curve.\\
\textsuperscript{\emph{b}} Kramers formula [Eq. (2), the main text] for the average values of $F_{\mathrm{U}}^{''}$ and $F_{\mathrm{TS}}^{''}$ (Table {\ref{tbl:g50_parameters}}).\\
\textsuperscript{\emph{c}} simulated times.
\end{table}
\clearpage

\begin{figure}\centering%
\resizebox{0.7\linewidth}{!}{ \includegraphics*{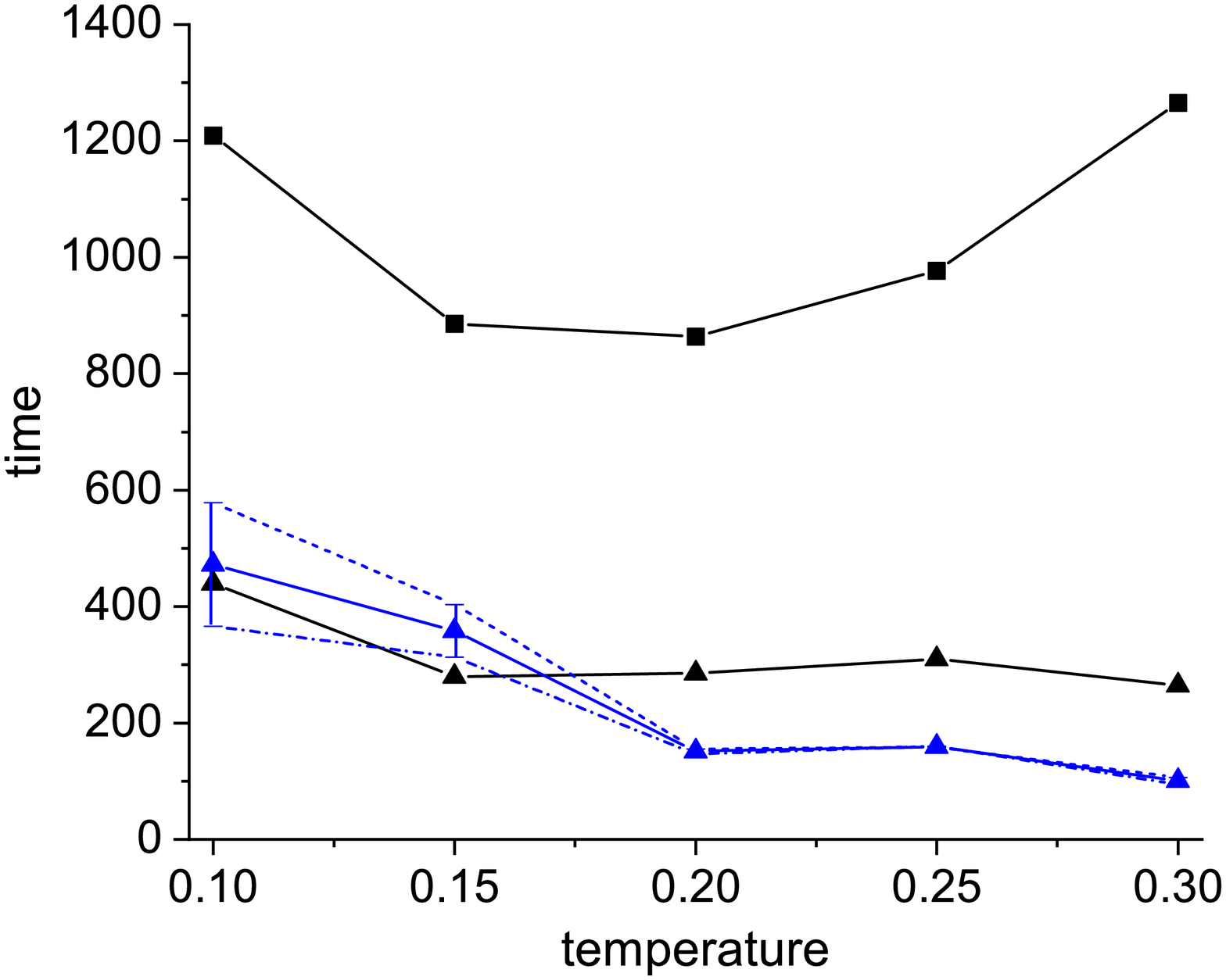}}%
\caption{The black squares are for the $\langle t_{\mathrm{\mathrm{U} \rightarrow \mathrm{N}}} \rangle$ times from simulations, the black triangles denote the $\langle t_{\mathrm{\mathrm{U} \rightarrow \mathrm{NL}}} \rangle$ times calculated from the slopes of the simulated $\mathrm{U} \rightarrow \mathrm{NL}$ decay curves, and the blue triangles are for  $\langle t_{\mathrm{\mathrm{U} \rightarrow \mathrm{NL}}} \rangle$ times from Eq. (2) of the main text with the average values of $F_{\mathrm{U}}^{''}$ and $F_{\mathrm{TS}}^{''}$ (the dashed and dash-dotted blue lines indicate the results for $F_{\mathrm{U}}^{''}$ and $F_{\mathrm{TS}}^{''}$ obtained by the polynomial approximation of the FEP and calculated by finite-differences, respectively). In contrast to the cases of $\gamma=3M/\tau $ and $\gamma=10M/\tau $, where the diffusion coefficient was calculated from $\langle R^{2} (t) \rangle$ at short times, in the given case it was calculated at longer times where $\langle R^{2} (t) \rangle \sim t$ (the red dashed curves in Figs. \ref{g50_all_0_1} - \ref{g50_all_0_3}). If the approximation of $\langle R^{2} (t) \rangle$ at short times is used (the blue dashed curves), the Kramers times are one order smaller.  In all cases, the lines are to guide the eye.}  
\label{times_cmp}
\end{figure}

\clearpage

{\centering \section{Different Thresholds to Terminate the MD Trajectories}}

\begin{figure}\centering%
\resizebox{0.49\linewidth}{!}{ \includegraphics*{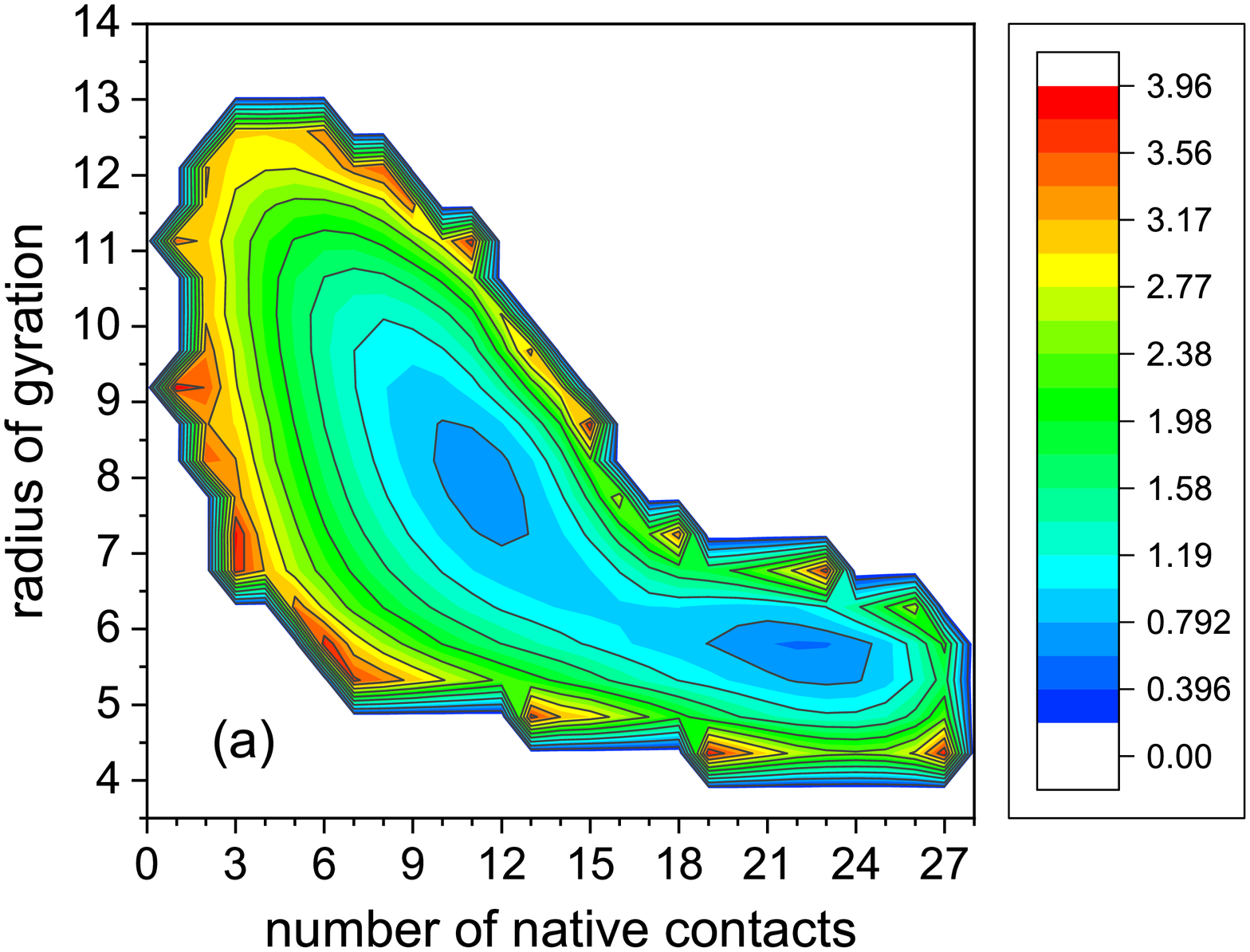}}%
\hfill
\resizebox{0.49\linewidth}{!}{ \includegraphics*{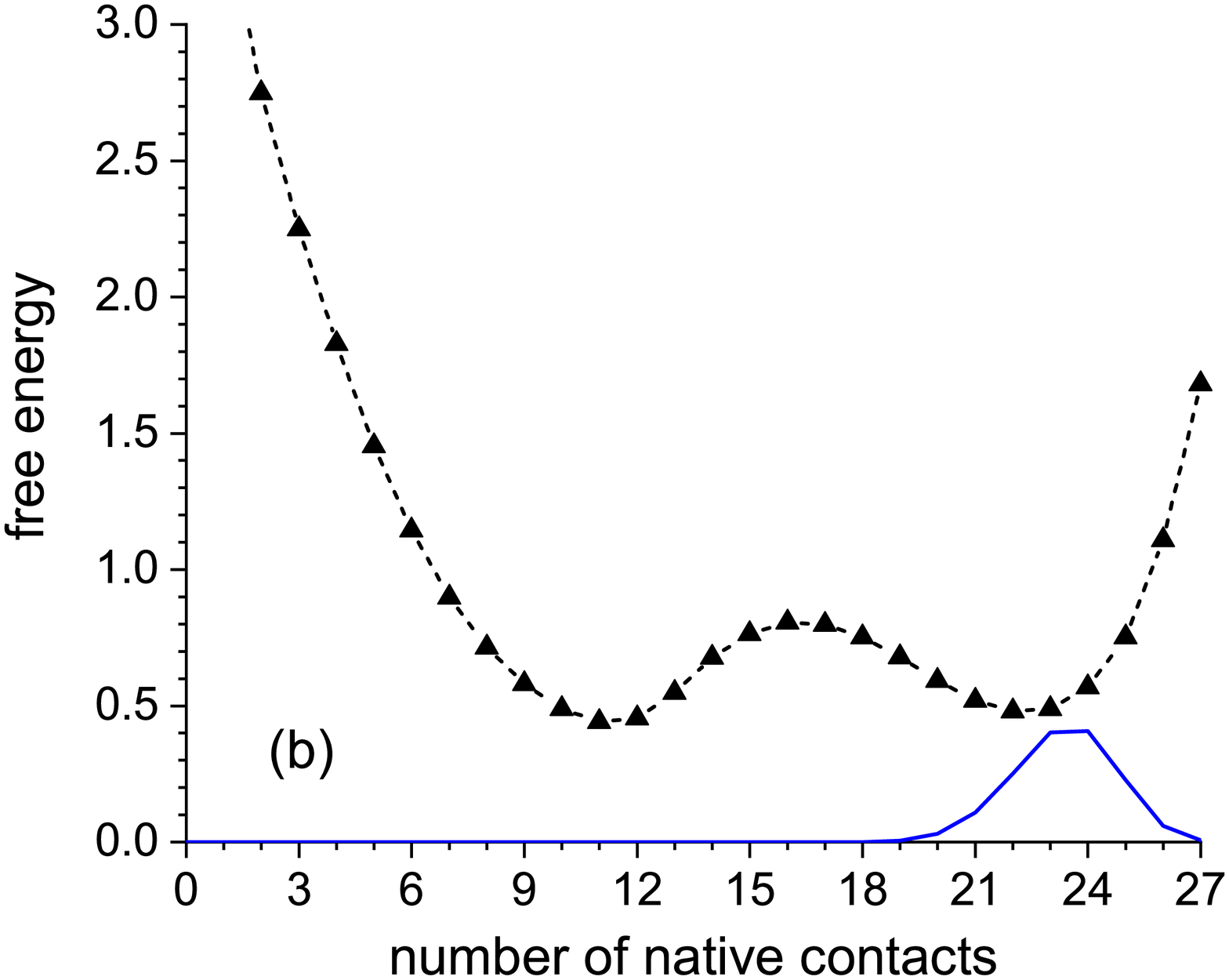}}%
\hfill
\resizebox{0.49\linewidth}{!}{ \includegraphics*{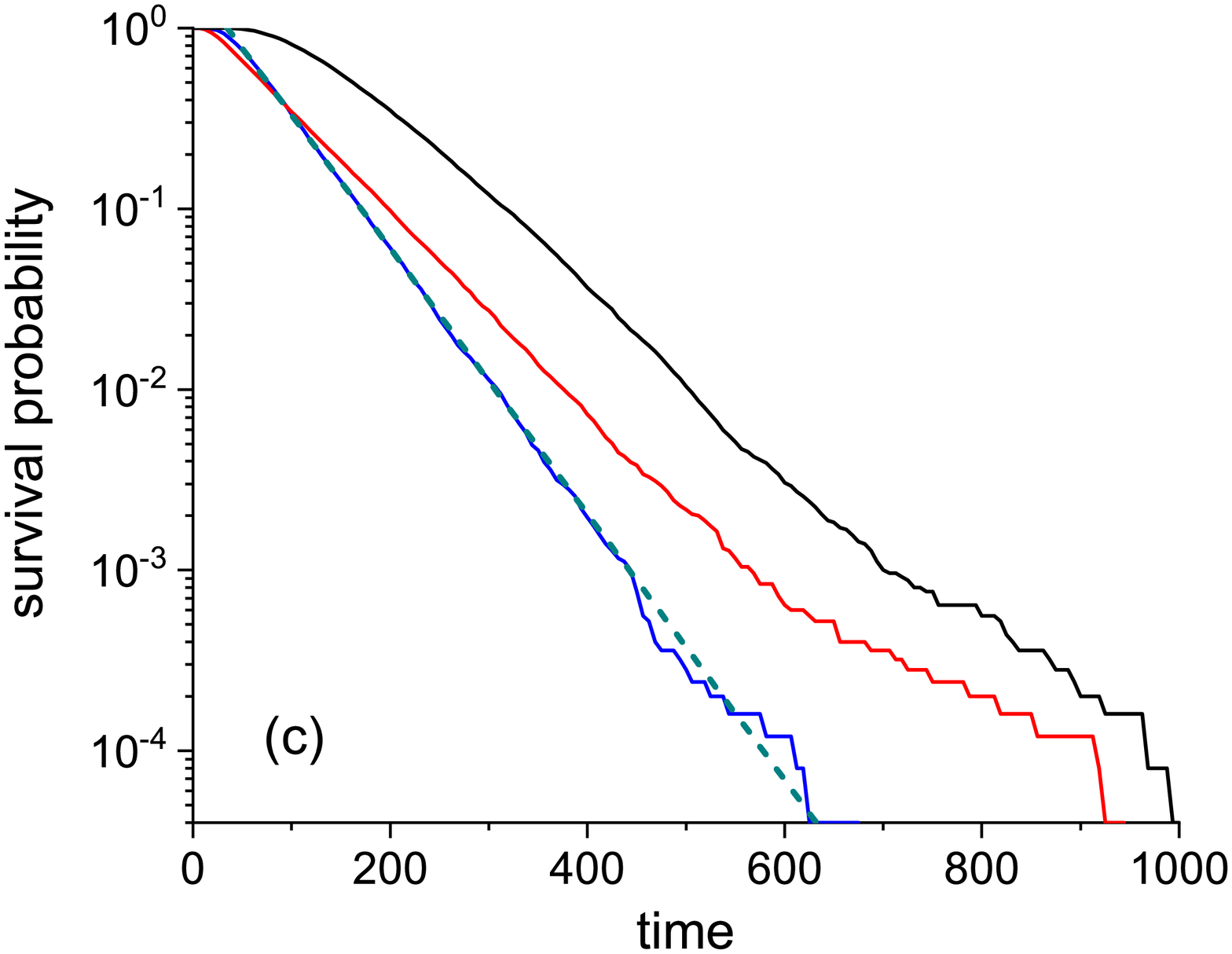}}%
\hfill
\resizebox{0.49\linewidth}{!}{ \includegraphics*{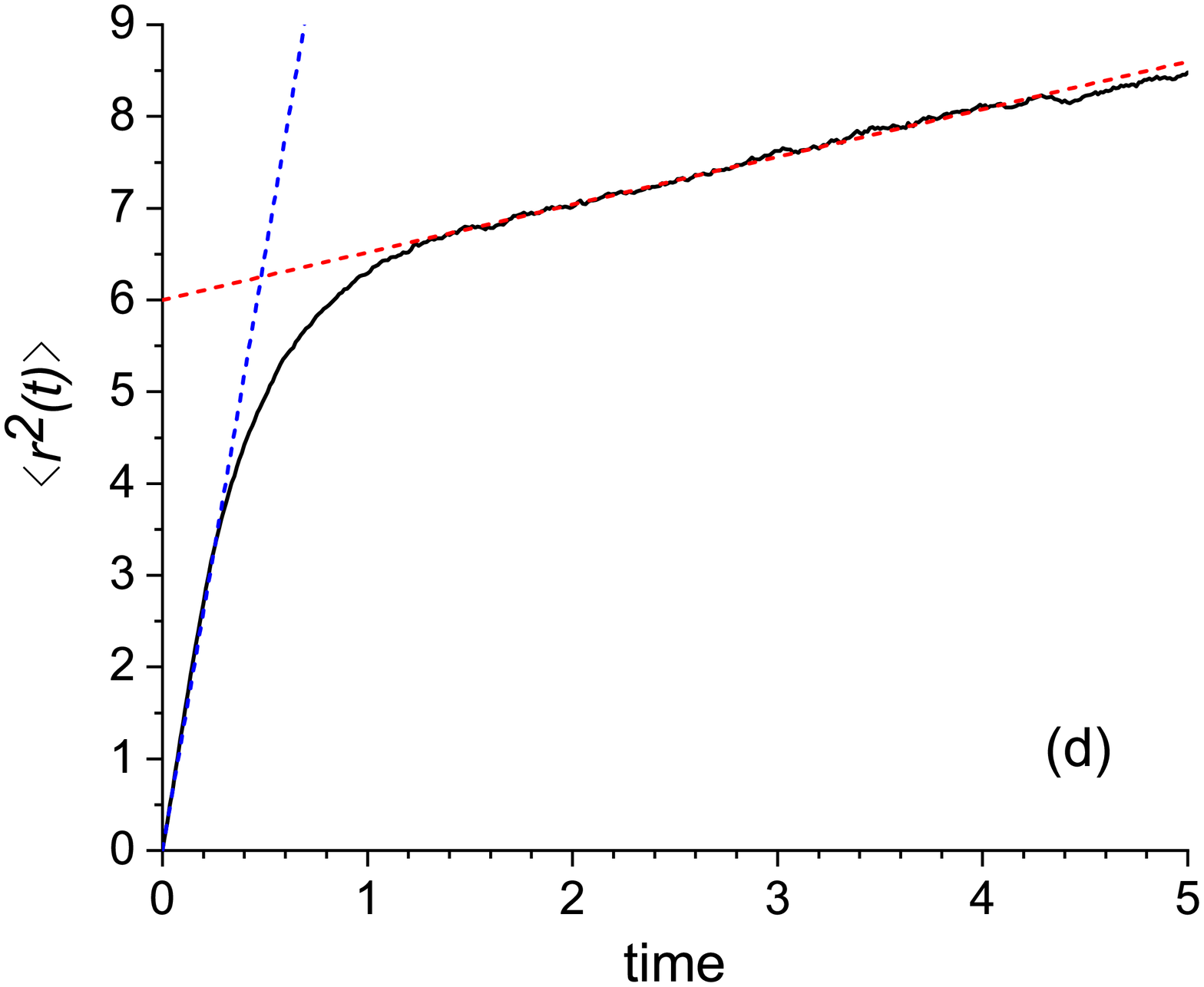}}%
\caption{The trajectories were terminated as the RMSD from the native state was less than 1.0 {\AA}; $T=0.2$. ({\bf{a}}) The free energy surface, and ({\bf{b}}) free energy profile (black curve) with the normalized distributions of the protein states in the native-state ensemble (blue curve). ({\bf{c}}) First-passage time distributions in the form of survival probabilities: the $\mathrm{U} \rightarrow \mathrm{NL}$ trajectories (blue), the $\mathrm{NL} \rightarrow \mathrm{N}$ trajectories (red), and the U-N trajectories (black); the dashed green line denotes an exponential fit to the $\mathrm{U} \rightarrow \mathrm{NL}$ distribution. ({\bf{d}}) The time-dependent mean-square deviation from the transition state in the number of native contacts (black curve); the blue and red dashed lines are the linear fits to the curve for short and long times, respectively.}   
\label{g3_all_0_2_rmsd}
\end{figure}

\begin{figure}\centering%
\resizebox{0.49\linewidth}{!}{ \includegraphics*{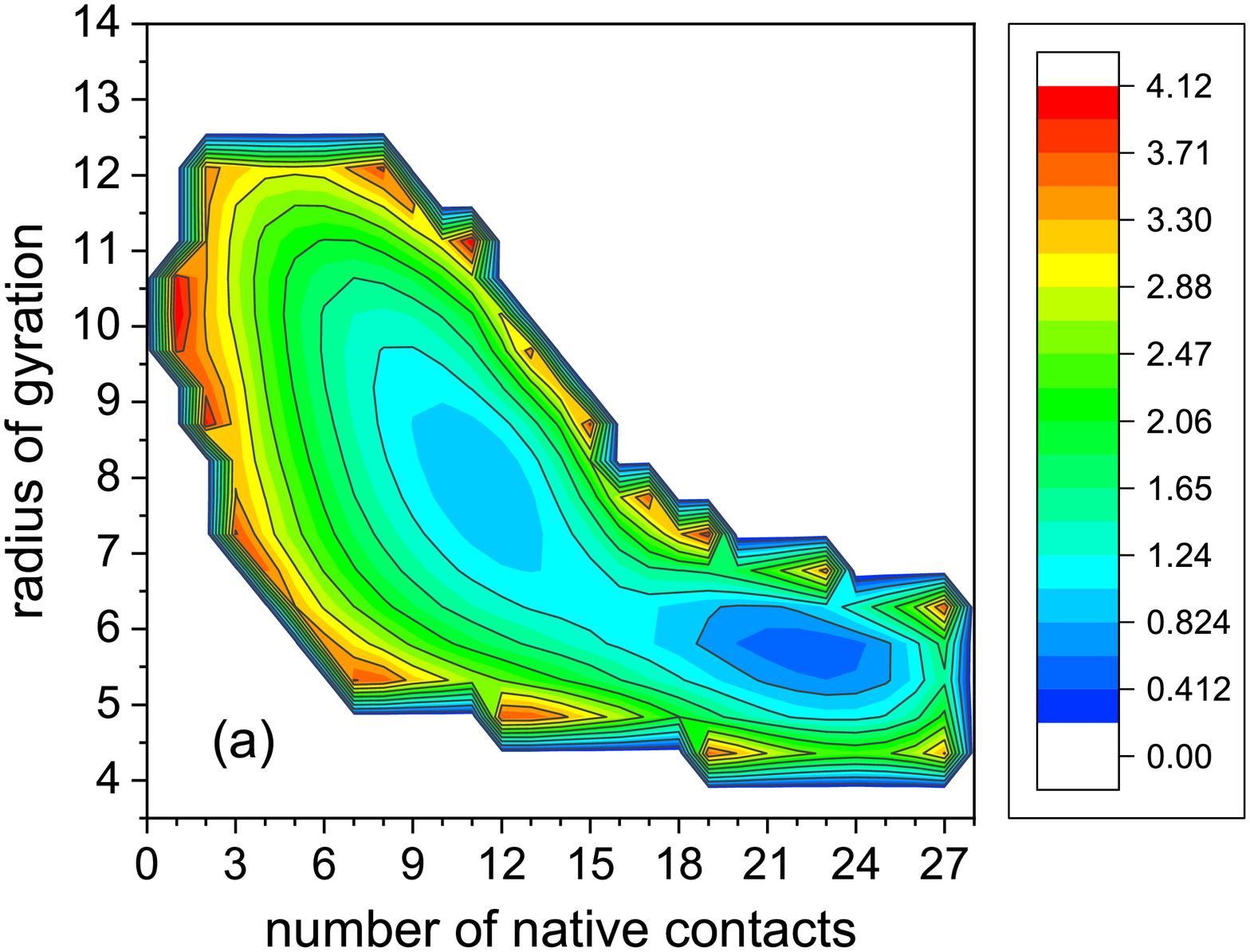}}%
\hfill
\resizebox{0.49\linewidth}{!}{ \includegraphics*{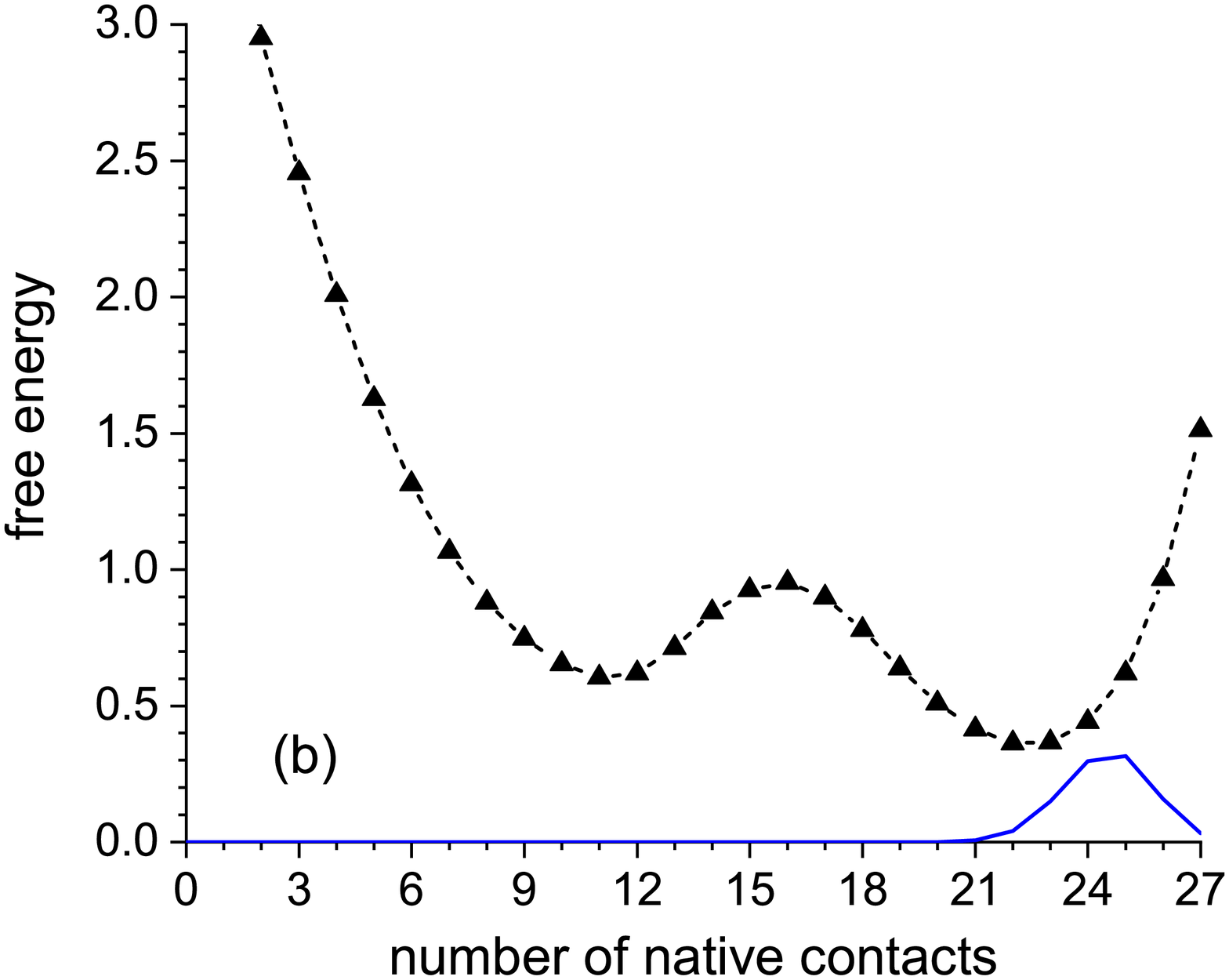}}%
\hfill
\resizebox{0.49\linewidth}{!}{ \includegraphics*{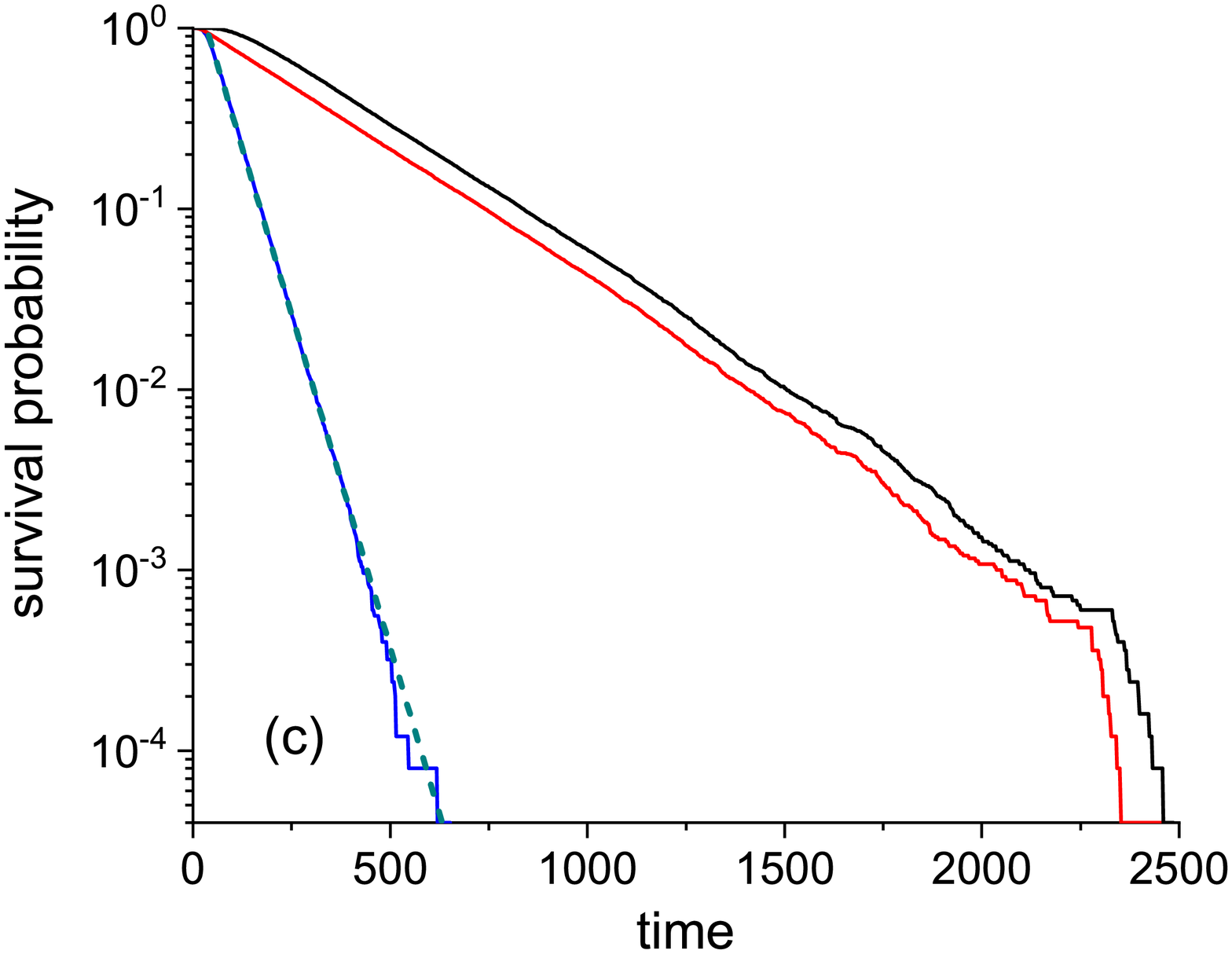}}%
\hfill
\resizebox{0.49\linewidth}{!}{ \includegraphics*{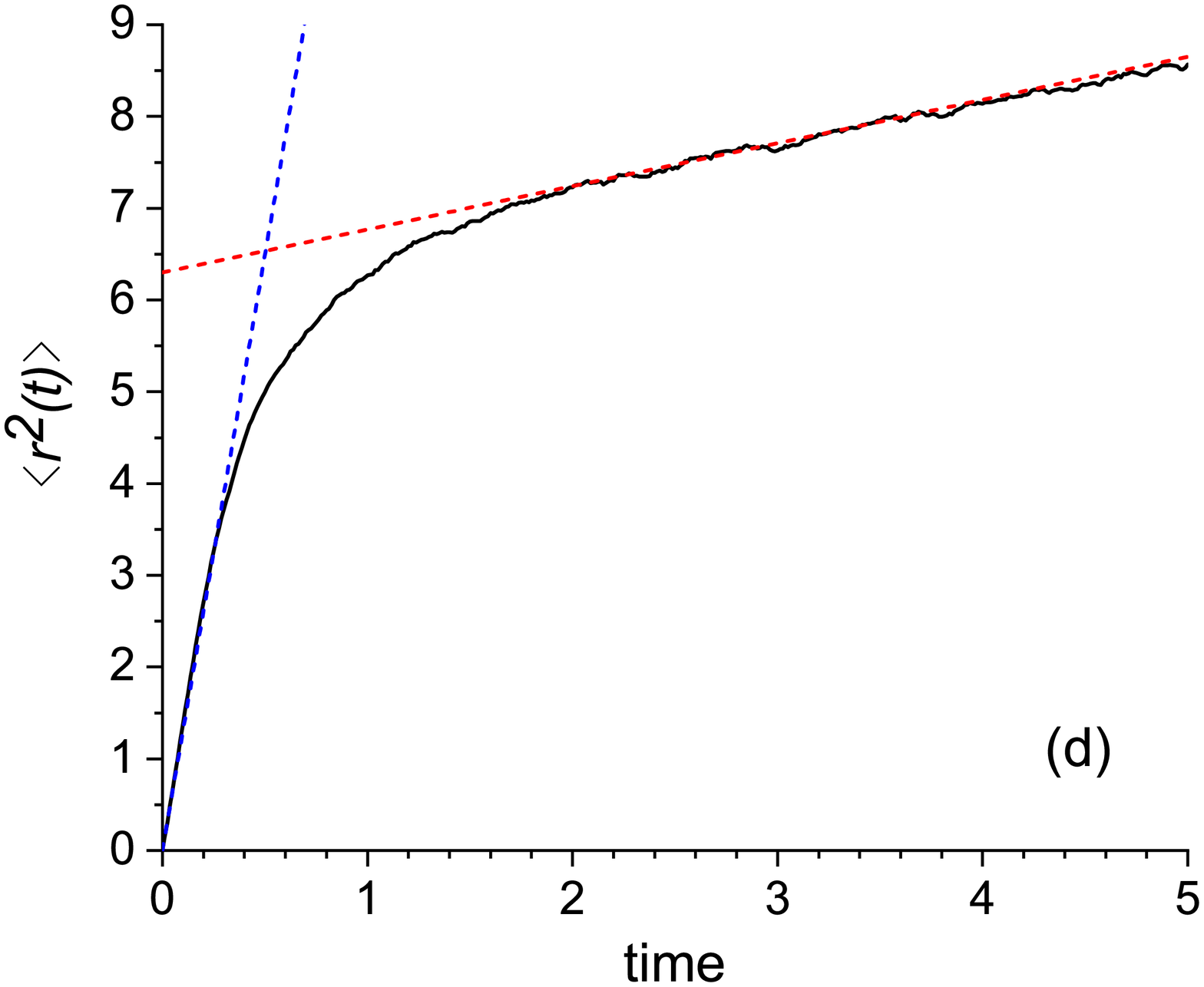}}%
\caption{The trajectories were terminated as the RMSD from the native state was less than 0.65 {\AA}; $T=0.2$. The notations are as in Fig. \ref{g3_all_0_2_rmsd}.}   
\label{g3_all_0_2_nnat}
\end{figure}

\begin{figure}\centering%
\resizebox{0.49\linewidth}{!}{ \includegraphics*{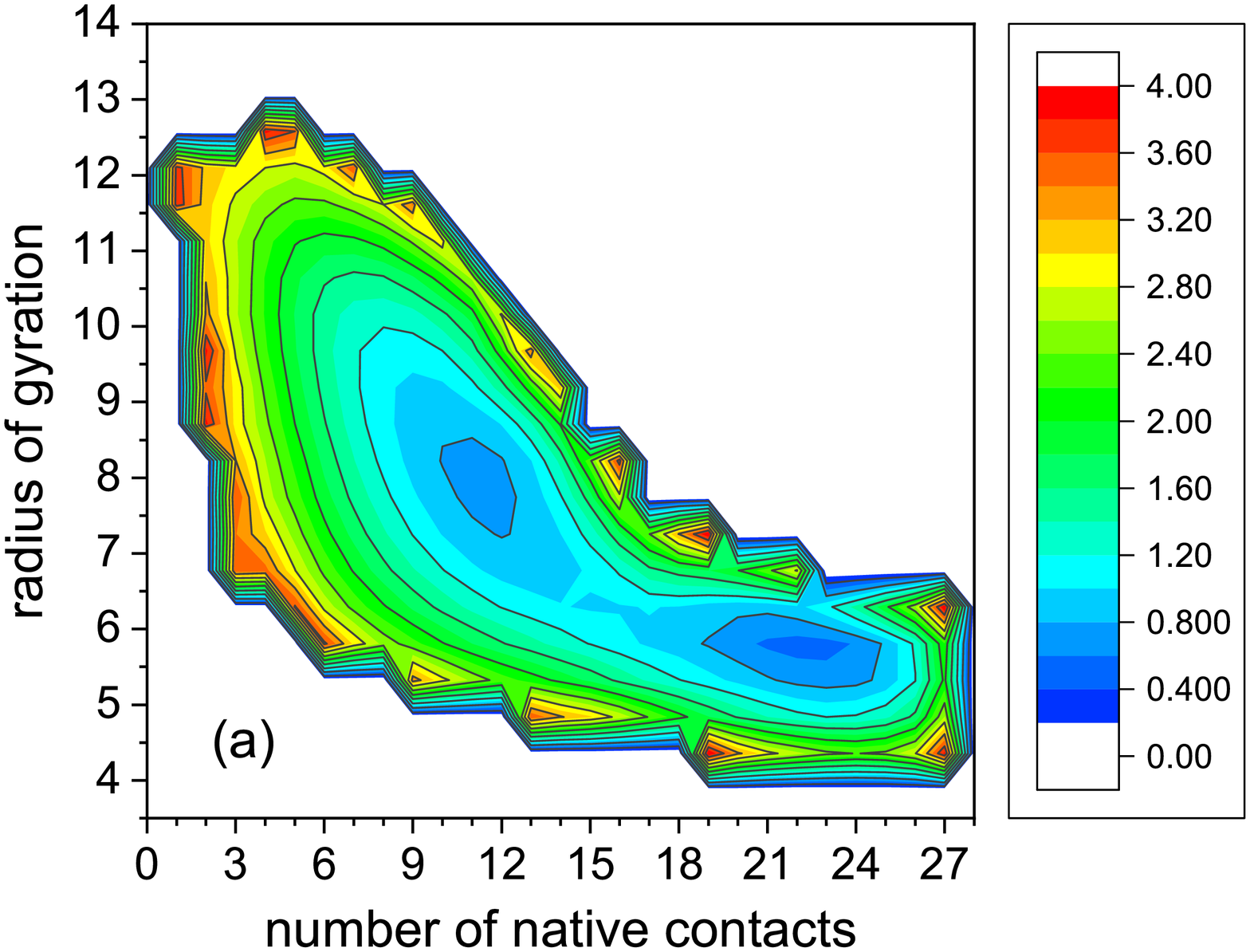}}%
\hfill
\resizebox{0.49\linewidth}{!}{ \includegraphics*{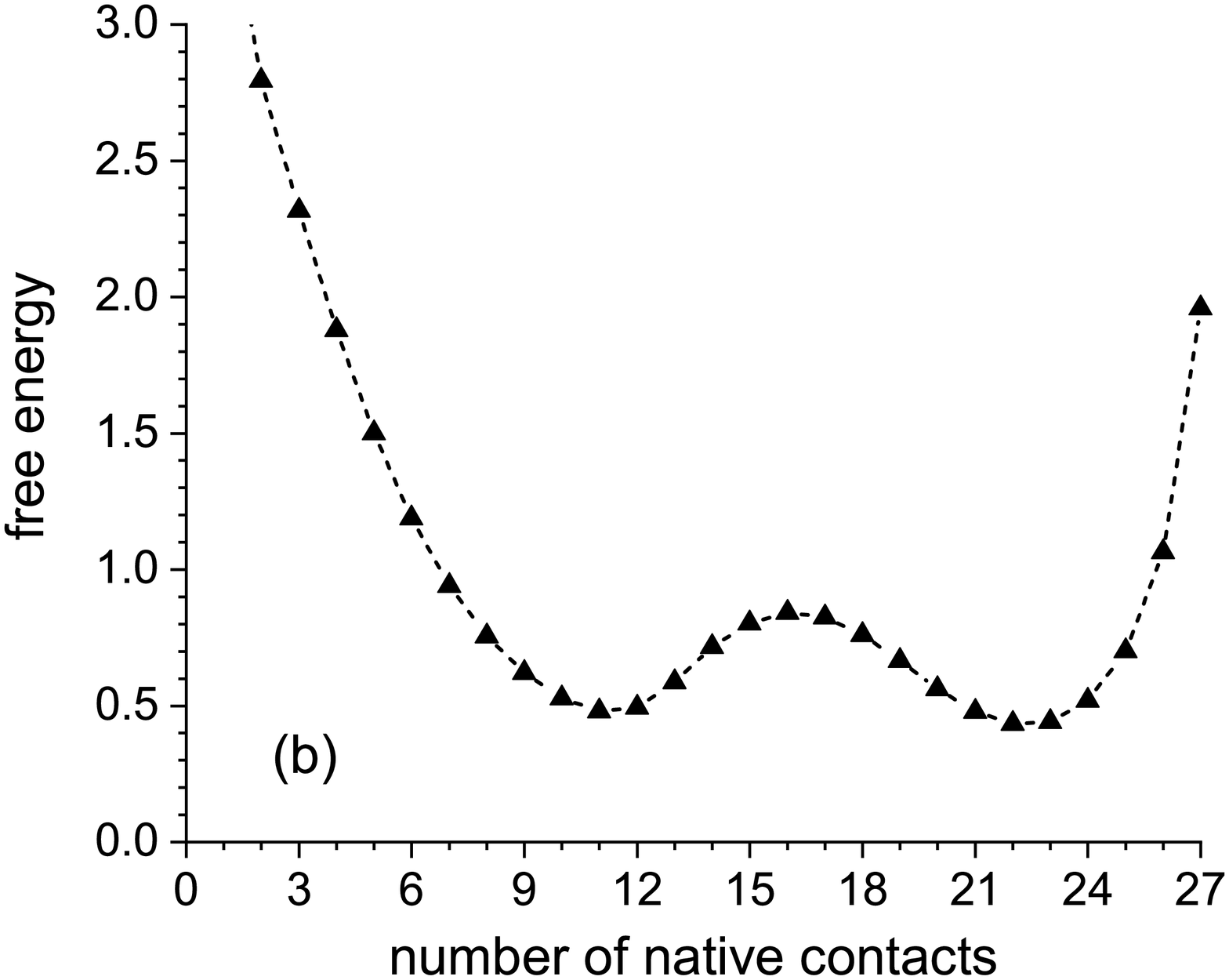}}%
\hfill
\resizebox{0.49\linewidth}{!}{ \includegraphics*{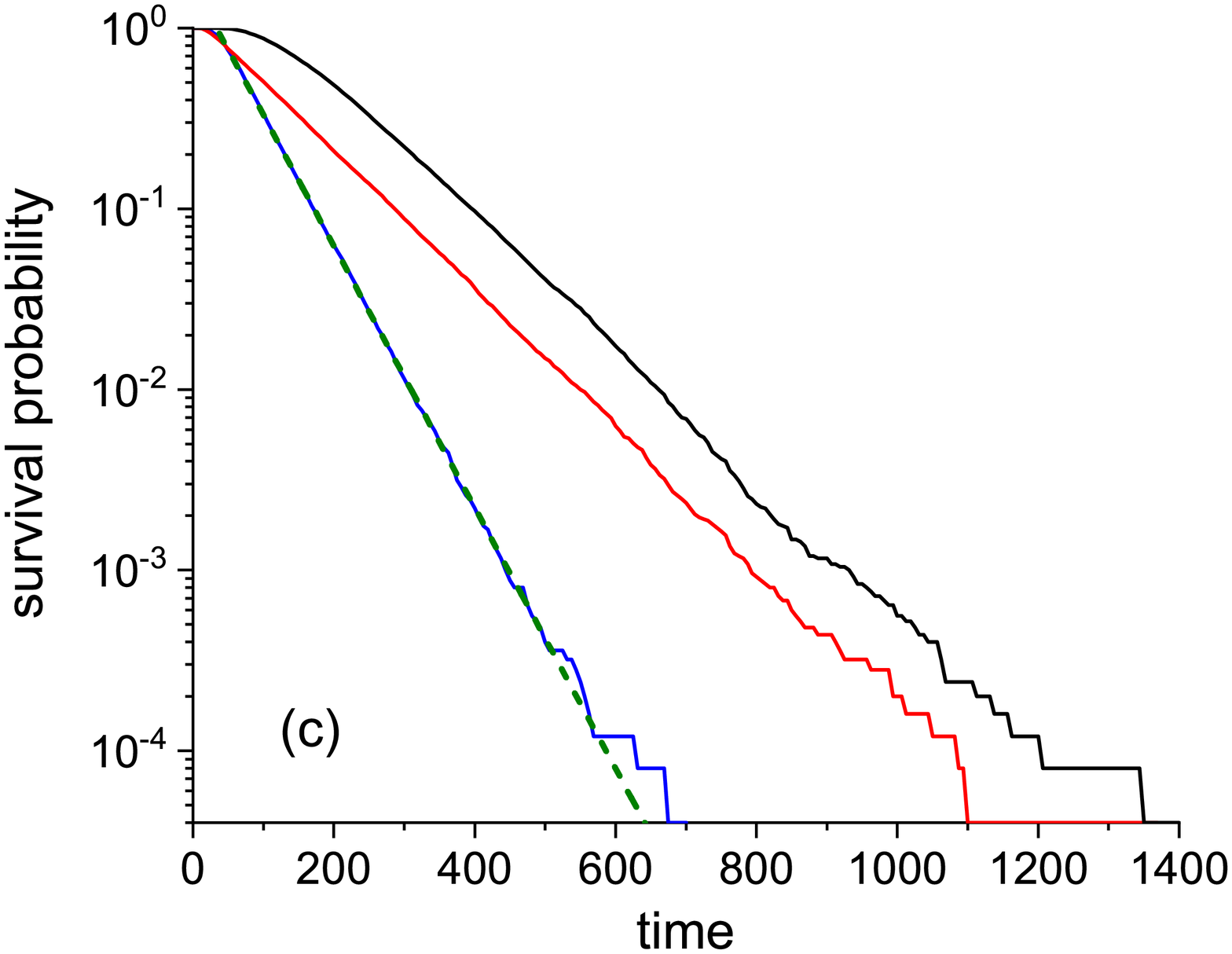}}%
\hfill
\resizebox{0.49\linewidth}{!}{ \includegraphics*{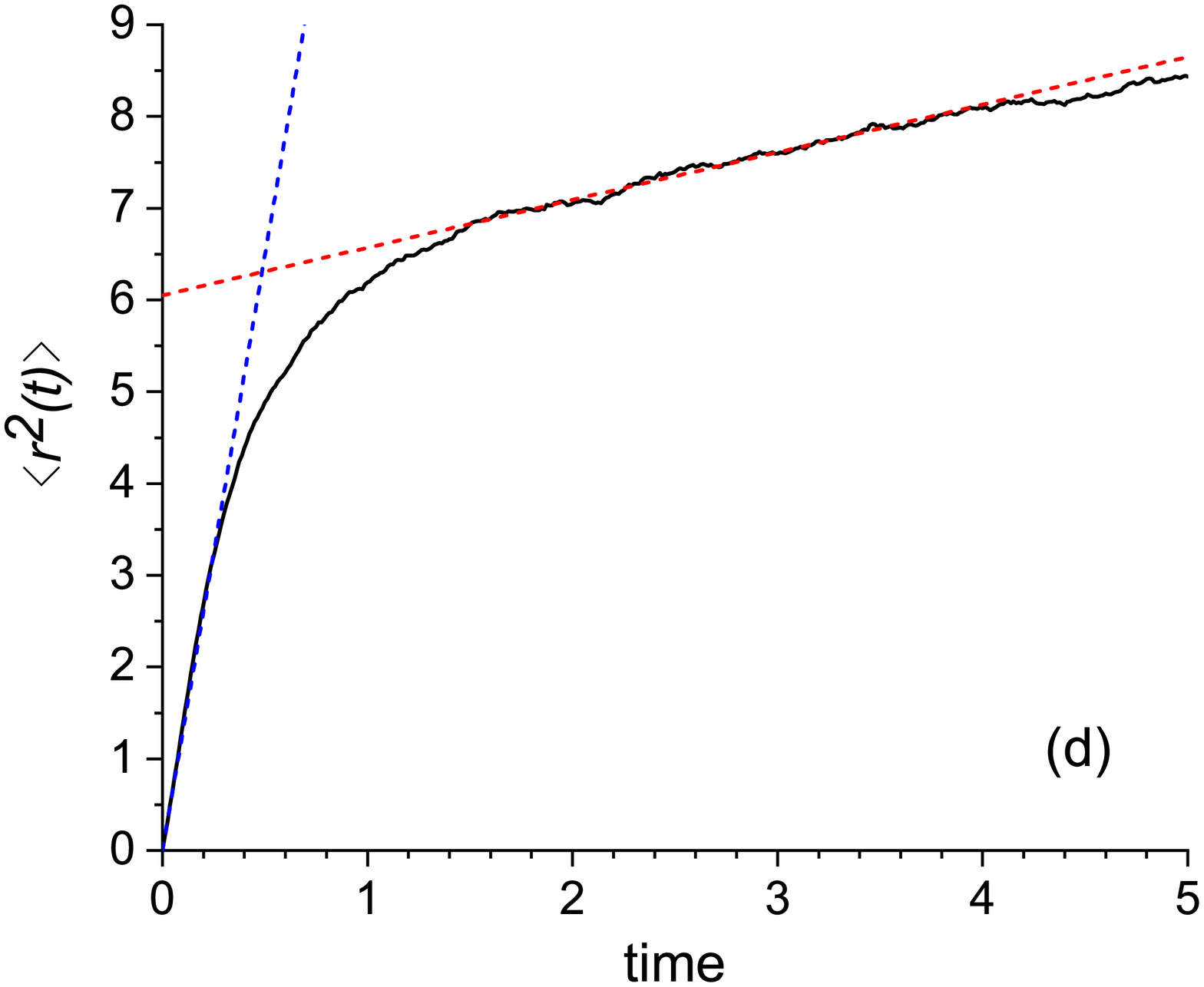}}%
\caption{The trajectories were terminated as the number of native contacts $N_{\mathrm{nat}}$ was equal to the number of native contacts  in the native state $N_{\mathrm{nat}}^{\mathrm{NAT}}=27$; $T=0.2$. The notations are as in Fig. \ref{g3_all_0_2_rmsd}.}   
\label{g3_all_0_2_nnat}
\end{figure}